%% file: Feature_Paper_ESIT_SPM_REVISED_2nd_ARXIV.tex
\documentclass[a4paper, draftcls, 11pt, onecolumn, twoside]{IEEEtran}
\IEEEoverridecommandlockouts
\usepackage{soul} 
\usepackage[normalem]{ulem}
\usepackage{graphicx,cite,acronym,amsmath,amssymb, algorithmic,amsfonts,color,floatflt,stfloats,bm,textgreek,multirow}
\usepackage[ruled,vlined]{algorithm2e}
\usepackage[caption=false,labelformat=empty,font=footnotesize]{subfig}
\usepackage{xcolor}
\usepackage{bm}
\usepackage[cmintegrals]{newtxmath}
\usepackage{graphics,hhline,array,cite,bbm}
\usepackage{latexsym}
\usepackage{theorem}
\usepackage{times}
\usepackage{makeidx}
\usepackage{colortbl} 
\usepackage{comment} 
\definecolor{Red}{HTML}{FF0000}
\definecolor{CornflowerBlue}{RGB}{100,149,237}
\definecolor{GreenYellow}{RGB}{173,255,47}
\definecolor{darknavy}{RGB}{0, 0, 128}
\newcolumntype{C}[1]{>{\centering\arraybackslash}m{#1}}
\usepackage{units}
\usepackage{setspace}

\DeclareGraphicsExtensions{.png,.jpg,.eps}
\input{acronyms}

\input{def_EM_SP}
\begin{document}
\bstctlcite{IEEEexample:BSTcontrol}
\title{An Overview on Over-the-Air\\ Electromagnetic Signal Processing}
\author{
\IEEEauthorblockN{Davide Dardari,~\IEEEmembership{Fellow,~IEEE,} \\ Giulia Torcolacci,~\IEEEmembership{Graduate Student Member,~IEEE,} \\ Gianni  Pasolini,~\IEEEmembership{Member,~IEEE,} and Nicol\`o Decarli,~\IEEEmembership{Member,~IEEE}}
 \IEEEcompsocitemizethanks{\IEEEcompsocthanksitem}}

\maketitle

\begin{abstract} 

This article provides a tutorial on \emph{over-the-air electromagnetic signal processing} (ESP) for next-generation wireless networks, addressing the limitations of digital processing to enhance the efficiency and sustainability of future 6th Generation (6G) systems. It explores the integration of electromagnetism and signal processing (SP)  under a unified framework by highlighting how their convergence can drive innovations for 6G technologies. Key topics include electromagnetic (EM) wave-based processing, the application of metamaterials and advanced antennas to optimize EM field manipulation with a reduced number of radiofrequency chains, and their applications in holographic multiple-input multiple-output systems. By showcasing enabling technologies and use cases, the article illustrates how wave-based processing can minimize energy consumption, complexity, and latency, offering an effective framework for more sustainable and efficient wireless systems. This article aims to assist researchers and professionals in integrating advanced EM technologies with conventional SP methods.
\end{abstract}

\newpage

\section{Introduction}
\IEEEPARstart{A}{fter} the publication of Shannon's landmark paper in 1948, the digital revolution rapidly accelerated, shifting information processing and communication from the analog realm to the digital domain. Since then, the exponential growth in digital computing capabilities has pushed the boundary between analog and digital processing in wireless systems increasingly closer to the devices' antenna. This evolution is mainly driven by software-defined radio implementations, where analog processing is relegated to \ac{IF} and \ac{RF} circuits, mainly for basic operations such as filtering, amplification, phase shifting, and frequency conversion. In fact, except for optical systems—where analog computing has been established for several years and remains an active research area \cite{McM:23}—most signal processing in current wireless systems takes place in the digital domain at baseband. 

This approach, however, could be questioned when addressing the design of next-generation wireless systems, such as \ac{6G} networks, which are expected to impose extremely stringent requirements to deliver enhanced performance in terms of capacity, reliability, latency, and support for new functionalities such as \ac{ISAC}\cite{alsabah20216g,Pre:J24}.
An analysis of the evolution of mobile radio technologies over recent decades reveals that the growing demands of each new generation—from 1G to 5G—have been addressed through the adoption of advanced solutions such as small cell deployments, millimeter wave (mmWave) communications, and massive \ac{MIMO} systems.
These innovations have effectively boosted network capacity, enhanced performance, and supported ubiquitous connectivity \cite{BjoEldLarLozPoo:23}. Looking ahead to the next generation, significant research efforts are focused on the use of higher frequency bands—such as \ac{THz}—and electrically large antenna arrays, including metasurfaces and \ac{XL-MIMO} systems \cite{YouCaiLiuDiRDumYen:24}. Notably, their combined deployment enables operation in the radiative near-field region of the antennas, unlocking unprecedented levels of communication and sensing performance, flexibility, and spatial resolution \cite{ZhaShlGuiDarEld:J23, DarDec:J21}.

However, we have reached the point where further advancing the capabilities of already established technologies and incorporating new ones is hindered by issues related to practical implementation, hardware cost, and complexity, as well as excessive processing time and power consumption. These issues often result in the so-called ``digital bottleneck", posing a substantial challenge to the feasibility, scalability, and sustainability of future wireless networks.

Increasingly, it is becoming clear that reversing the prevailing trend toward fully digital processing is necessary, prompting the design of solutions in which at least part of the signal processing is performed in the analog domain at \ac{RF}.   
 For instance, hybrid digital-analog solutions in \ac{MIMO} systems have been extensively explored to mitigate digital processing burdens, though often at the cost of reduced flexibility \cite{BjoEldLarLozPoo:23}. Nevertheless, hybrid \ac{XL-MIMO} systems, even with analog precoding and low-resolution data converters, entail substantial power consumption. Moreover, the proliferation of antennas and RF components further complicates hardware design \cite{CasYanChaHea:25}. Similarly, the integration of \ac{ISAC} into future mobile networks, widely regarded as a key technological breakthrough, presents substantial challenges, particularly due to the requirement for transceivers to process signals with very high dynamic ranges, e.g., from strong communication signals to weak radar echoes. This necessitates high-resolution \acp{ADC}, which significantly increases power consumption in both fully digital and hybrid architectures, potentially making \acp{ADC} and digital processing units the primary sources of energy consumption, surpassing that of the analog RF front-end  \cite{Analog:23}. Although low-resolution \acp{ADC}, including extreme one-bit quantization, can offer competitive performance in some applications when paired with advanced signal processing, their practical viability depends on factors such as computational complexity, processing unit cost, and potential strict constraints on responsiveness \cite{Pre:J24}.

It is precisely the growing awareness of the limitations of current technologies, coupled with the pressing need to reduce their energy impact, that is fueling the exploration of innovative design paradigms.
A promising approach toward this goal is to transfer part of the signals' processing directly to the \ac{EM} level \cite{ZhuWanDaiDebPoo:24,DiRMig:24,BjoChaHeaMarMezSanRusCasJunDem:24, WeiGonZhaShaCheDaiDebYue:24, migliore2020cares}.
This approach is known as \emph{\acf{ESP}}, also referred to as \emph{\acf{ESIT}} or \emph{\acf{EIT}}.
Specifically, \ac{ESP} can be defined as an interdisciplinary scientific field focused on exploiting the synergies between innovative \ac{EM} technologies, \ac{SP} algorithms, and information theory. This term refers to the study and development of novel technological solutions where an \ac{EM} wave, generated by one or more radiating sources, is perturbed during its propagation, i.e., over-the-air, by reconfigurable, typically passive, \ac{EM} structures. This approach aims to perform specific \ac{SP} functions with reduced energy consumption, complexity, and latency—since processing occurs at the speed of light—compared to their digital counterparts. In fact, \ac{ESP} can be viewed as part of the broader \emph{holographic communications} paradigm, which envisions a holistic method for manipulating the \ac{EM} field generated or sensed by antennas with unprecedented flexibility~\cite{DarDec:J21}.
This paradigm involves designing reconfigurable radio propagation environments \cite{Dar:J24} that can arbitrarily control signal transmission, manipulation, and reflection through the massive deployment of peculiar \ac{EM} devices based on metamaterials. In these devices, the reflection, refraction, or scattering properties of a set of reconfigurable scatterers composing their structure are optimized to achieve specific processing and radiation characteristics. 

Admittedly, many studies on the design of innovative \ac{EM} structures and processing algorithms are still in their infancy, and significant progress is needed in terms of practical implementations and concrete applications. Nonetheless, the seamless integration of electromagnetism and \ac{SP} offers a transformative opportunity for advancing next-generation wireless networks, particularly through the emergence of \ac{ESP}. 

The primary objective of this article is to present, in a unified way, the fundamental limits and new opportunities in processing \ac{EM} waves using physically consistent models, establishing a bridge between typically fragmented frameworks such as \ac{EM}, antenna, circuit, and \ac{SP} theories. While traditional \ac{SP} techniques are effective in the digital realm, they often struggle with the complexities and constraints posed by \ac{EM} environments and emerging \ac{EM} technologies. This discrepancy underscores the necessity for specialized \ac{SP} algorithms that align with the physical properties and constraints of \ac{EM} waves. 
This article seeks to shed light on this issue and provide a survey of recent advancements in \ac{ESP}. It emphasizes the importance of bridging \ac{EM} theory with \ac{SP} and advocates for a unified strategy to enhance the sustainability of wireless communication technologies.

Targeted at readers with expertise in \ac{SP} and/or wireless communications, the article begins with a brief historical overview before revisiting fundamental \ac{EM} principles needed to establish a general framework for exploring the fundamental limits and practical algorithms of \ac{ESP}. Key technologies will be presented as specific cases, supported by numerical examples that underscore the need for electromagnetically consistent optimization algorithms.

By addressing the limitations, opportunities, and challenges of this integrated \ac{ESP} approach, we aim to inspire further research and provide valuable insights. We hope this work serves as a useful resource for researchers and practitioners, highlighting the essential role of integrating these disciplines in driving innovative solutions and applications for future \ac{6G} networks.

\section{``Over-the-Air" EM-Based Signal Processing: An Old Idea Made New} 

\begin{figure}[!t]
\centering\includegraphics[width=0.85\columnwidth]{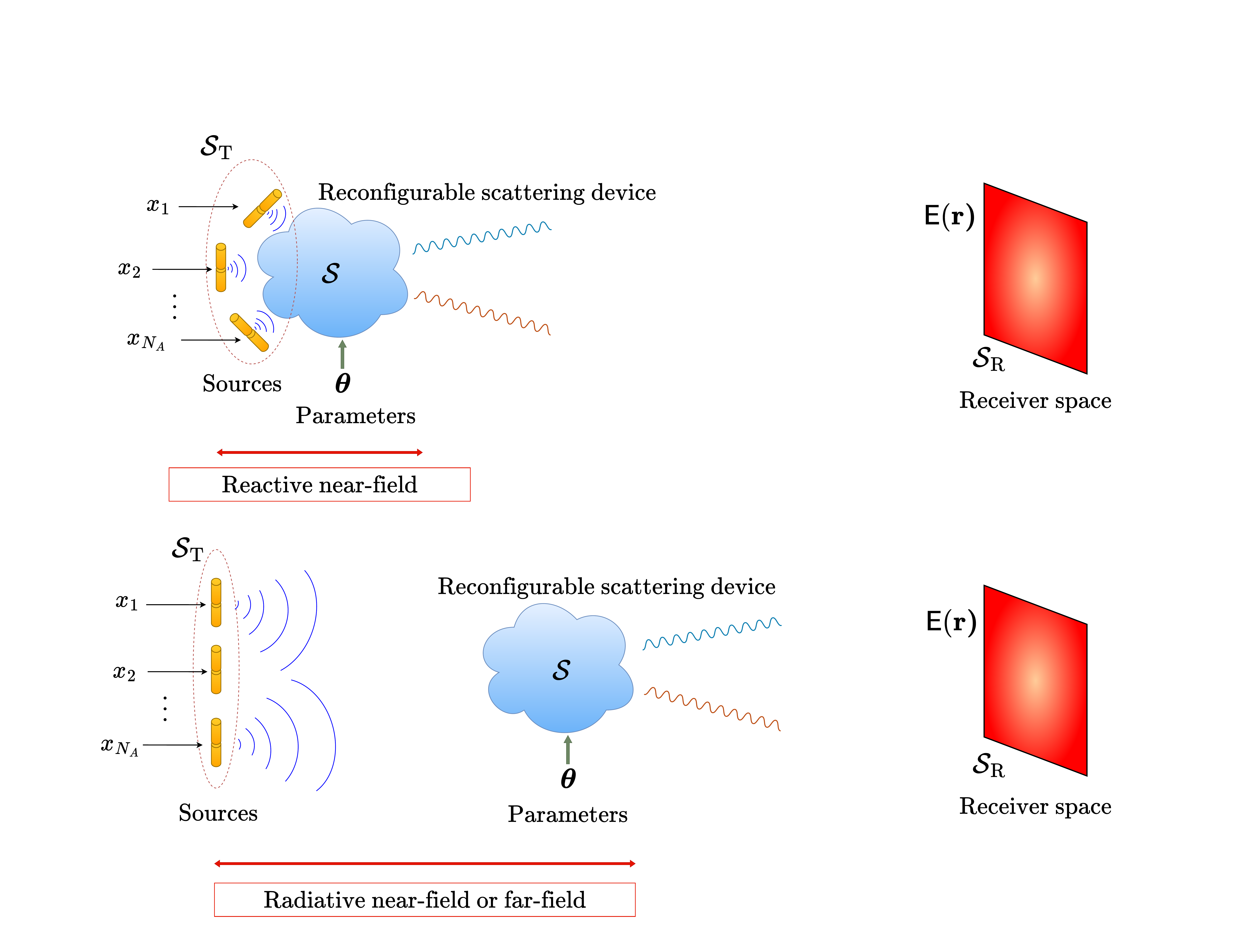}
\caption{\ac{EM} signal processing performed in the reactive near-field (top) or in the radiative near-field (or far-field) (bottom) regions of the sources using a reconfigurable scattering device.} 
\label{Fig:Scenarios}
\end{figure}

In a general setting, \ac{ESP} refers to the capability of processing \ac{EM} waves over the air by introducing one (or many) reconfigurable scattering devices between the source and receiver spaces, denoted as $\St$ and $\Sr$, respectively (see Fig.~\ref{Fig:Scenarios}). The source space might represent a transmitter equipped with multiple active antennas, while the receiver space may correspond to one or more receivers.  

Let $\CalS$ denote the space occupied by the reconfigurable scattering device, whose configuration can be modified through a set $\btheta$ of parameters according to the specific implementation technology. Typically, the scattering device is a linear and passive structure composed of a large number of elementary reconfigurable scattering elements, enabling wave processing with significantly reduced power consumption. As will be illustrated in Section~\ref{Sec:Technologies}, an antenna connected to a reconfigurable load provides a simple example of a reconfigurable scattering element.
The purpose of the reconfigurable scattering device is to manipulate the \ac{EM} field generated in the source space to produce a desired \ac{EM} field in the receiver space. The processing occurs over the air, as the received signal results from the coupling among the scattering elements and their interaction with the sources themselves.

The scenario shown in Fig.~\ref{Fig:Scenarios} can encompass various \ac{ESP} technologies and geometric configurations. Specifically, \ac{ESP} solutions can generally be categorized into two types:
\begin{itemize}

\item \emph{Reactive region configuration}: The scattering device is located in the reactive region of the source space (typically limited to a few wavelengths) so that inductive coupling takes place between them (see Fig.~\ref{Fig:Scenarios}, top) and it can support the implementation of signal processing in the \ac{EM} domain. Its role includes enabling large antenna arrays while minimizing the number of \ac{RF} chains, reducing power consumption, and lowering latency compared to traditional \ac{MIMO} architectures. 
The dual scheme where the scattering device is located close to the receiver space is also possible, thus moving the \ac{EM} signal processing to the receiver side.  

\item \emph{Radiative near and far field region configuration}: The reconfigurable scattering device is positioned in the radiative region of the source and receiver spaces and it is used to control the propagation environment (see Fig.~\ref{Fig:Scenarios}, bottom). This approach offers a cost-effective alternative to deploying additional access points. In some configurations, the source and receiver spaces might match (see Section~\ref{Sec:Radiative}).
\end{itemize}
 Regardless of the specific category or underlying technology, the core \ac{ESP} problem can be formulated as follows: Given a set of sources located in the source space $\St$, determine the configuration $\btheta$ of the scattering device that modifies the generated wave to obtain the desired \ac{EM} field at the receiver space $\Sr$, while satisfying the constraints imposed by \ac{EM} laws and the selected technology. For instance, one might require that the structure performs a multifunctional processing such that 
the electrical field $\Em(\boldr)$, observed in the receiver space $\Sr$, is a linear combination of different arbitrary spatial responses $f_n \left (\boldr \right )$, with $\boldr \in \Sr$, associated with the information $x_n$ transmitted by the $n$-th source. Denote by $\tilde{f}_n \left (\mathbf{r}; \boldsymbol{\theta} \right )$ the actual response of the reconfigurable scattering device associated with the $n$-th source obtained with the configuration parameters $\boldsymbol{\theta}$. The problem is to configure the scattering device such that the actual responses are a good approximation of the desired ones according to some fidelity measure, and hence
\begin{equation} 
\Em(\boldr)=\sum_{n=1}^{\Na} \tilde{f}_n \left (\mathbf{r}; \boldsymbol{\theta} \right ) \, x_n \simeq \sum_{n=1}^{\Na} f_n \left (\mathbf{r} \right ) \, x_n \, ,
\end{equation}
where $\Na$ denotes the number of active sources.
In other words, the challenge is to obtain different spatial responses simultaneously using the same configuration $\boldsymbol{\theta}$.

%

A further step involves making the set of parameters $\btheta$ dependent on data generated internally by the scattering device itself (e.g., when it incorporates a sensor in an \ac{IoT} application).
This allows the information to be directly embedded into the scattered signal, eliminating the need for additional \ac{RF} signal generation. As a result, this approach can lead to substantial energy savings on the sensor side \cite{LiaZhaWanLonZhoYan:22,DarLotDecPas:J23,IJ133_NatComm_9_4334_2018}.

Notably, the concept of reconfigurable scatterers has a long and well-established history in the field of antenna theory. A pivotal milestone and precursor to modern solutions is R. Harrington's theory of loaded scatterers \cite{MauHar:73}, which is closely linked with characteristic mode decomposition. This theory involves configuring scatterers such that the principal radiating mode of the \ac{EM} structure, when illuminated by the source wave, aligns with the desired response. Harrington later extended this theory to reactively controlled arrays, which represent one of the earliest realizations of reconfigurable scattering systems and correspond to the first category illustrated in Fig.~\ref{Fig:Scenarios} (top). In this context, passive scattering elements are positioned near a radiating antenna to exploit mutual coupling within the reactive near-field region, thereby modifying the antenna's radiation characteristics \cite{Har:78}.

In recent years, these concepts have been revisited and advanced thanks to technological and material innovations (e.g., metamaterials), enabling the development of new \ac{EM} structures that address the increasing demand for enhanced performance, flexibility, and cost efficiency in future wireless systems. These include \ac{XL-MIMO} arrays (also known as \acp{ELAA}) \cite{WanZhaDuShaAiNiyDeb:23}, \acp{LIS} or \ac{HMIMO} surfaces \cite{Dar:J20,GonGavJiHuaAleWeiZhaDebPooYue:24}, \acp{DMA} \cite{ZhaShlGuiDarEld:J23}, fluid and reconfigurable antennas \cite{WonNewHaoTonCha:23}, and the latest developments in reactively controlled arrays, such as \acp{ESPAR} and \acp{DSA} \cite{BucJuaKamSib:20,Dar:C24}.
 Significant advancements have concurrently been made in the optical community with the development of all-optical diffractive \acp{DNN} \cite{LinRivYarVelLuoJarOzc:18}, and a similar concept has recently been applied to \ac{RF} systems \cite{gu2024classification}, inspiring the development of the \ac{SIM} technology, even though this technology is currently limited to linear processing capabilities \cite{AnXuNgAleHuaYueHan:23}. Additionally, the introduction of novel volumetric metamaterials has enabled \ac{3D} \ac{EM} processing functions, such as first-derivative computation \cite{Sil:14}.

A notable example of new metamaterials is given by metasurfaces, which are \ac{2D} structures, either homogeneous or inhomogeneous, composed of interacting sub-wavelength-sized cells called meta-atoms. These structures may provide control of the reflected and transmitted \ac{EM} field \cite{BarHamLonMonRamVelAleBil:21,MarMac:22}. 
Both metasurfaces and more conventional solutions, such as backscattering antenna arrays, have facilitated the development of \acp{RIS}, which have garnered significant interest within the scientific community. \acp{RIS} are primarily used to manipulate the propagation environment and are considered a cornerstone of the emerging \emph{smart radio environments} paradigm, whose aim is to integrate the propagation environment into the system's design and optimization loop \cite{DiRDanTre:22}. Specifically, \acp{RIS} belong to the second category,  shown in Fig.~\ref{Fig:Scenarios} (bottom), where they are typically employed to control the propagation environment \cite{DiRDanTre:22,BjoWymMatPopSanCar:22}.

To build a general framework for the description of \ac{ESP}, the foundational \ac{EM} principles, on which the subsequent developments are based, are introduced in Section~III. To this end, the preliminaries on the Green's operator and the wavenumber representation are introduced in Section~III-A, and then leveraged in Section~III-B to define the concept of \textit{communication modes}, first between a transmitting and a receiving space, and then in the presence of a programmable \ac{EM} scatterer. This progression enables the definition of the mode transfer matrix for the scattering devices (Section~III-C), ultimately culminating in the formal expression of the ESP problem.
To account for realistic constraints on the types of processing that can be implemented at the \ac{EM} level, Section~IV analyzes the fundamental performance limits, building on the \ac{EM} framework developed in Section~III. Specifically, the wavenumber representation and essential signal theory concepts are used to derive the \ac{DoF} in both unbounded and bounded spaces (i.e., considering spatial constraints related to the transmitter/receiver/scatterer components), thereby linking the \ac{ESP} problem previously formulated to its physical feasibility and fundamental limitations.
The connection between electromagnetic and signal processing theory is provided through the circuit-based equivalent representation in Section~V.A. Circuit theory represents one approach to building a physically consistent model of the system on which \ac{ESP} algorithms can be designed.
Finally, concrete examples of enabling \ac{EM} technologies and over-the-air signal processing methodologies are provided in the rest of Section~V. Some of the processing functions that can be achieved include over-the-air computation of \ac{MIMO} \ac{EM} precoding, \ac{DFT} computation, and \ac{DoA} signal estimation. 

\section{A General Framework for EM Signal Processing} 
\label{Sec:Model}

\subsection{Preliminaries: The Green's Operator and the Wavenumber Representation}

We consider a generic source space $\St$, characterized by a monochromatic current density $\Jm (\boldr)$ [$\unit[]{A/m^2}$], where $\boldr = r_x \versorx + r_y \versory + r_z \versorz \in \St$ denotes the position vector, and $\versorx$, $\versory$, and $\versorz$ are the orthonormal unit vectors in 3D space. The time-harmonic dependence $e^{\jmath \omega t}$, with $\omega$ denoting the angular frequency, is omitted for brevity. This current distribution acts as the source of the electromagnetic field, which propagates through space and interacts with the surrounding environment, including any scattering or receiving structures.

In general, the space $\St$ represents a volume, although it may also correspond to a surface or a wire source. The current density $\Jm (\boldr)$ can be an impressed current in the case of an active source, such as a transmitting antenna, or an induced/equivalent current in the case of a scattering device.  
The electrical field $\Em(\boldr)$ [$\unit[]{V/m}$] generated by  $\Jm(\boldr)$ can be computed under the free-space condition by considering the Helmholtz equation, whose solution is given by \cite{BalB:24} 
\begin{equation} \label{eq:Er}
 \Em(\boldr) = \int_{\St} \GreenE{\boldr - \bolds} \, \Jm(\bolds) \, \mathrm{d}\bolds =
 \left (\GOperator \, \Jm \right ) (\boldr)  \, .
\end{equation}
Equation \eqref{eq:Er} implicitly defines  the Green's tensor operator $\GOperator$, with $\GreenE{\boldr}$ being the dyadic Green's function defined as \cite{BalB:24} 
\begin{align} \label{eq:Gej}
& \GreenE{\boldr} = -\jmath \frac{ \eta \, e^{-\jmath \kappazero \, |\boldr|}}{2 \lambda |\boldr|}  \cdot \left [\left ( \boldI - \hat{\boldr} \, \hat{\boldr}^\transpose \right ) + \frac{\jmath \lambda }{2 \pi |\boldr|} \left  ( \boldI - 3 \hat{\boldr} \, \hat{\boldr}^\transpose \right ) - \frac{ \lambda^2 }{(2 \pi |\boldr|)^2}  \left ( \boldI - 3 \hat{\boldr} \, \hat{\boldr}^\transpose \right )\right ]\, ,  
\end{align}
where $\kappazero=2\pi / \lambda$ is the wavenumber, $\lambda$ is the wavelength, $\hat{\boldr}=\boldr/|\boldr|$, $\jmath$ denotes the imaginary unit, and $\eta$ is the free-space impedance. 
 $\GreenE{\boldr}$ is the solution of the Helmholtz equation for an impulsive current density $\Jm(\boldr)=\IIm \cdot \delta(\boldr)$, with $\IIm$ being the unit dyadic,\footnote{The unit dyadic, also known as the identity dyadic or unit tensor, is a mathematical construct in vector and tensor analysis. It is the second-order tensor equivalent of the scalar number 1 in scalar algebra or the identity matrix in linear algebra.} and $\delta(\boldr)$ is the delta Dirac pseudo-function. When $|\boldr| \gg \lambda$, only the first term in \eqref{eq:Gej} contributes to the \ac{EM} radiation. 
The electrical field in the current-free portion of the space must satisfy the homogeneous wave equation \cite{BalB:24} 
\begin{equation} \label{eq:Homogeneous}
\nabla^2 \Em(\boldr) + \kappazero^2 \, \Em(\boldr)=0\, ,   
\end{equation}
where $\nabla^2$ represents the Laplacian operator. 

For what follows, it is convenient to introduce the representation of the fields in the wavenumber domain $\wavefreq = \wavefreqx \versorx + \wavefreqy \versory + \wavefreqz \versorz$ through the \ac{3D} Fourier transform. Specifically, given a generic field $\Am(\boldr)$, we define 
\begin{align}
  \TildeAm (\wavefreq) &=\fourierthree{\Am(\boldr)}= \intthree \Am(\boldr) \, e^{-\jmath \, \wavefreq \scalprod \boldr} \, \mathrm{d}\boldr\, ,   
  \end{align}
with $\scalprod$ denoting the scalar product. 

By applying the Fourier transform to both sides of \eqref{eq:Homogeneous} we obtain
\begin{equation} \label{eq:Helmholtzk}
  \left (\wavefreqx^2+\wavefreqy^2+\wavefreqz^2- \kappazero^2 \right ) \TildeEm(\wavefreq) = 0\, ,  
\end{equation}
where we have considered that
$\fourierthree{\nabla^2 \Am(\boldr)}=- |\wavefreq|^2 \, \TildeAm(\wavefreq)$.
Equation \eqref{eq:Helmholtzk} reveals that $\TildeEm(\wavefreq)=\fourierthree{\Em(\boldr)}$ must vanish everywhere except on the wavenumber support 
\begin{equation}  \label{eq:wavecondition}
\CalE=\left \{(\wavefreqx,\wavefreqy,\wavefreqz) \in \mathbb{R}^3 :  \wavefreqx^2+\wavefreqy^2+\wavefreqz^2=\kappazero^2  \right \} \, .
\end{equation} 
In other words, the \ac{EM} propagation phenomenon acts as a filter in the wavenumber domain with support in $\CalE$. Since $\TildeEm(\wavefreq)$ is defined over a support with zero measure in the waveform domain, $\TildeEm(\wavefreq)$ and  \eqref{eq:Helmholtzk} have to be intended in the distributional sense\footnote{ This means that for all test functions \(\varphi(\wavefreq)\),
$
\intthree \left (\wavefreqx^2+\wavefreqy^2+\wavefreqz^2- \kappazero^2 \right ) \TildeEm(\wavefreq)\varphi(\wavefreq) \, \mathrm{d}\wavefreq = 0$.}
and $\TildeEm(\wavefreq)$
takes the form \cite{Mar:18}
\begin{align} \label{eq:TildeEm}
  \TildeEm(\wavefreq) = \TildeEm_0(\wavefreq) \,  \delta \left (\wavefreqx^2+\wavefreqy^2+\wavefreqz^2- \kappazero^2 \right )\, ,   
  \end{align}
where $\TildeEm_0(\wavefreq)$ defines the angular dependence (direction) of the wave.
When applied to the \ac{EM} field $\Em(\boldr)$ in the presence of current sources, the inverse Fourier representation expresses it in terms of mathematical plane waves \cite{BalB:24,Dar:J24}  
\begin{align} 
 \Em(\boldr) &= \invfourierthree{\TildeEm (\wavefreq)}= \frac{1}{(2 \pi)^3}  \intthree  \TildeEm (\wavefreq)  \, e^{\jmath \, \wavefreq \scalprod \boldr} \, \text{d} \wavefreq = \frac{1}{(2 \pi)^3}  \intthree \TildeGm(\wavefreq)\, \TildeJm(\wavefreq) \, e^{\jmath \, \wavefreq \scalprod \boldr} \, \text{d} \wavefreq \nonumber \\
 &=\frac{\jmath \eta}{\kappazero (2 \pi)^3}  \intthree  \frac{\crossprod{\wavefreq}{  {\crossprod{\wavefreq}{  \TildeJm(\wavefreq) }} } }{  |\wavefreq|^2-\kappazero^2 } \, e^{\jmath \, \wavefreq \scalprod \boldr} \, \text{d} \wavefreq\, , 
 \label{eq:invfourier}
\end{align}
where $\TildeGm(\wavefreq)=\fourierthree{\GreenE{\boldr)}}$, $\TildeJm(\wavefreq)=\fourierthree{\Jm(\boldr)}$, and $\crossprod{}{}$ denotes the cross product.
In fact, for a fixed wavenumber $\wavefreq$,  $\TildeEm (\wavefreq)  \, e^{\jmath \, \wavefreq \scalprod \boldr}$ in \eqref{eq:invfourier} represents a plane wave with direction $\wavefreq/\kappazero$.

It is interesting to emphasize that the generic plane wave $\TildeEm (\wavefreq)  \, e^{\jmath \, \wavefreq \scalprod \boldr}=\TildeEm (\wavefreq)  \, e^{\jmath \, \wavefreqx  \, r_x}\,  e^{\jmath \, \wavefreqy \, r_y} \,  e^{\jmath \, \wavefreqz \, r_z}$ in \eqref{eq:invfourier} radiates $\forall \boldr$, i.e., in all the \ac{3D} space, only when $\wavefreqx$, $\wavefreqy$, and $\wavefreqz$ are all real (\emph{visible region}). Vice versa, if at least one of them is complex,\footnote{For example, this might happen if $\wavefreqx^2+\wavefreqy^2>\kappazero^2$ so that to satisfy \eqref{eq:wavecondition} it should be $\wavefreqz=\pm \jmath \sqrt{\wavefreqx^2+\wavefreqy^2-\kappazero^2} \in \mathbb{C}$.} let us say $\wavefreqz$,  it is   $e^{\jmath \, \wavefreqz \, r_z}=e^{-| \wavefreqz  \, r_z|} \rightarrow 0$ for  $|r_z|$ larger than a few wavelengths and the wave becomes evanescent, i.e., it does not radiate (\emph{invisible region}). In summary, only the spectral components of $\TildeEm(\wavefreq)$ for $\wavefreq \in \CalE$ represent \ac{EM} waves, i.e., are physically consistent, and only the subset having real components radiates.  
For each wavenumber, i.e., plane wave, because of the $\crossprod{\wavefreq}{  {\crossprod{\wavefreq}{  \TildeJm(\wavefreq) }} }$ term in \eqref{eq:invfourier},  
$ \TildeEm (\wavefreq)$ is subject to the relation $\TildeEm (\wavefreq) \scalprod \wavefreq =0$, limiting to two the \ac{DoF} achievable with different polarizations \cite{BalB:24}.
Incidentally, for locations $\boldr$ in the far-field region of the source, \eqref{eq:invfourier} can be approximated as
\cite{BalB:24}
%
\begin{align} \label{eq:farfield}
\Em(\boldr) &\simeq \jmath \kappazero \eta   \frac{e^{-\jmath \, \kappazero \, |\boldr|}}{4 \pi |\boldr|}   \,  \crossprod{\versorr}{ \, {\crossprod{\versorr}{  \TildeJm(\kappazero \, \versorr) }} }\, ,   
\end{align}
so that the \ac{EM} field becomes directly related to the Fourier transform of the source evaluated at direction $\kappazero \, \versorr \in \CalE$.

\subsection{Communication Modes}\label{sec:CommModes}
\label{Sec:CommModes}

For convenience, we introduce the inner product between vector functions $\Am(\boldr)$ and $\Bm(\boldr)$, defined on the generic space $\CalS$, as 
\begin{align}
  \innerprod{ \Am (\boldr) }{ \Bm(\boldr)}= \int_{\CalS} \Am (\boldr) \scalprod \Bm^* (\boldr) \, \text{d}\boldr  \, .
\end{align}
Two vector functions are said to be orthogonal if $ \innerprod{ \Am (\boldr) }{ \Bm(\boldr)}=0$.

\begin{figure}[!t]
\centering\includegraphics[width=0.8\columnwidth]{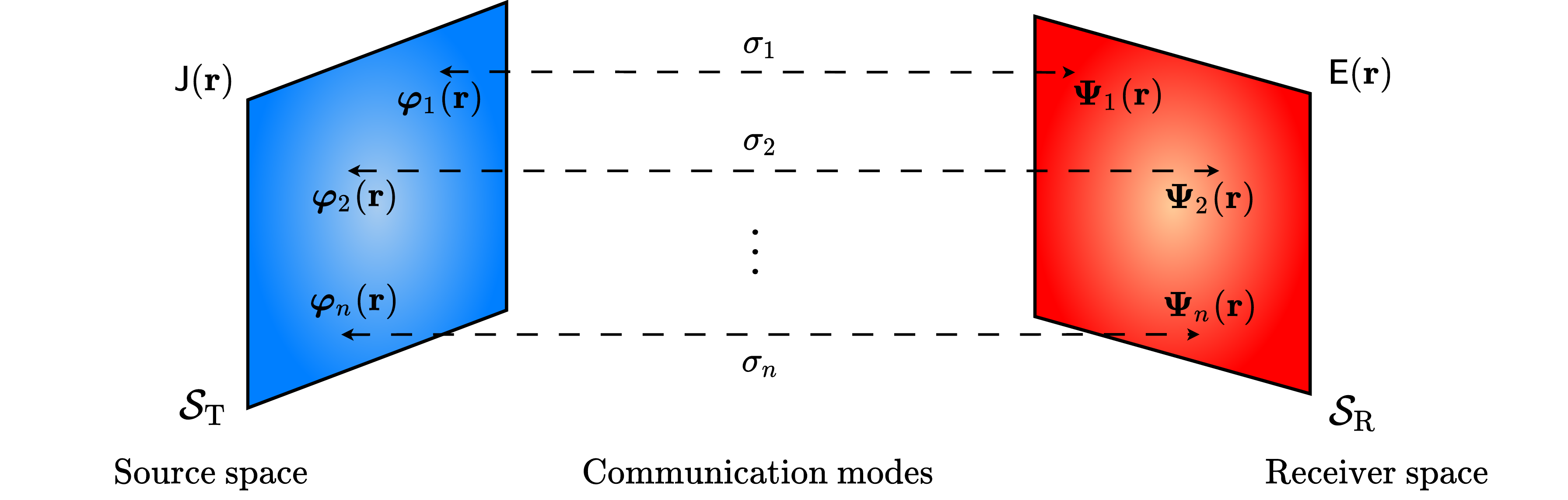}
\caption{Communication modes between two spaces.} 
\label{Fig:CommMode}
\end{figure}

A fundamental question is determining how many orthogonal communication channels—referred to as \emph{communication modes}—can be exploited between a source space $\St$ and a receiver space $\Sr$ while maximizing the coupling intensity (see Fig.~\ref{Fig:CommMode}). 
This is associated with the optimal approximation of every element in the image space of a Hilbert-Schmidt operator in terms of singular functions. 
Specifically, define $\mathcal{X}={\mathbb{L}}^2(\St)$ and $\mathcal{Y}={\mathbb{L}}^2(\Sr)$ the Hilbert spaces corresponding to the square-integrable functions defined in $\St$ and $\Sr$, respectively. The function $\Em(\boldr) \in \mathcal{Y}$ can be seen as the image of $\Jm(\boldr) \in \mathcal{X}$ through the Hilbert-Schmidt kernel $\GreenE{\boldr,\bolds}=\GreenE{\boldr - \bolds}$  on $\St \times \Sr$, which induces the operator $\GOperator:\mathcal{X} \rightarrow \mathcal{Y}$ such that, for any $\Jm(\boldr) \in \mathcal{X}$, we can write \eqref{eq:Er}.
Define the following self-adjoint Hilbert-Schmidt operators $\GOperator \, \GOperator^*$ and $\GOperator^* \, \GOperator$ and let $\{\VarPhim_n(\boldr) \}$  and $\{\Psim_n(\boldr)\}$ be the  eigenfunctions, associated with eigenvalues $\sigma_n^2$, of the following coupled eigenfunction problems  \cite{FranceschettiBook:2018} 
\begin{align} \label{eq:Eigen}
\left ( \GOperator \, \GOperator^* \, \VarPhim_n \right ) (\boldr) =\sigma_n^2 \, \VarPhim_n(\boldr) \quad \quad \quad \quad\quad\quad
\left ( \GOperator^* \, \GOperator \,  \Psim_n \right ) (\boldr) =\sigma_n^2 \, \Psim_n(\boldr) \, .
\end{align}
An important property of Hilbert-Schmidt operators is that they are compact and admit either a finite or countably infinite orthonormal basis.
In particular, two  sets of orthonormal eigenfunctions $\{\VarPhim_n(\boldr) \}$, $\{\Psim_n(\boldr)\}$ exist,  which are solutions, respectively, of the coupled eigenfunction problems in \eqref{eq:Eigen}  
with the same real eigenvalues $\sigma_1^2\ge \sigma_2^2 \ge \sigma_3^2 \ldots$ \cite{FranceschettiBook:2018,Miller:19}.
Note that $\left \{ \VarPhim_n(\boldr) \right \}$ and $\left \{ \Psim_n(\boldr) \right \}$ are two sets of orthonormal (vector) functions that are complete, respectively, in $\St$ and $\Sr$. The orthonormal condition implies that $\innerprod{\VarPhim_k(\boldr)}{ \VarPhim_n(\boldr) }= \delta_{k-n}$, and similarly, the same condition holds for the set $\{\Psim_n(\boldr)\}$. It holds that $\left (  \GOperator \,  \VarPhim_n \right ) (\boldr) =\sigma_n \, \Psim_n(\boldr)$, that is, a current excitation $\VarPhim_n(\boldr)$ in $\St$ produces an electrical field in $\Sr$ equal to $\sigma_n \, \Psim_n(\boldr)$. In general, the eigenvalues have significant magnitudes up to a certain value $\sigma_N^2$, i.e., the sum of the $\sigma_n^2$'s for $n>N$ is negligible with respect to the sum of the first $N$ eigenvalues, so that we can consider up to $N$ well-coupled communication modes between the two spaces. The number $N$ is often called the number of \ac{DoF} available in the link. Detailed discussion on optimal approximation and the definition of \ac{DoF} can be found in \cite{FranceschettiBook:2018}.

In summary, the finite coupling operator $\GOperator$ in \eqref{eq:Er} between finite spaces $\St$ and $\Sr$ is a Hilbert-Schmidt operator, and up to $N$ communication modes (i.e., orthogonal channels)  can be identified for a given level of approximation accuracy, each of them represented by the triad $\left (  \VarPhim_n(\boldr),  \Psim_n(\boldr),  \sigma_n  \right )$. 
It follows that the current density $\Jm(\boldr)$ on the space $\St$, i.e., for $\boldr \in \St$, can be well approximated  as a linear combination of the basis functions 
\begin{align}
\Jm(\boldr) & \simeq \sum_{n=1}^N j_n \, \VarPhim_n(\boldr)\, , 
\end{align}
where the complex coefficients $j_n$  are given by $j_n= \left < \Jm(\boldr) , \VarPhim_n(\boldr)  \right > $, for $ n=1,2, \ldots, N$.
Similarly, the electric field $\Em(\boldr)$ in $\Sr$ can be approximated through the basis set $\left \{ \Psim_n(\boldr) \right \}$ as \cite{Miller:19}
 \begin{align}
\Em(\boldr) & \simeq \sum_{n=1}^N  e_n \, \Psim_n(\boldr)=\sum_{n=1}^N \sigma_n \, j_n \, \Psim_n(\boldr) \, .
\end{align}
 
It is worth noting that, while in general a basis set for all functions defined within a bounded domain, depending solely on its geometry, can always be identified (for instance, the harmonic functions of the Fourier series \cite{Dar:J24}), the optimal approximating basis sets $\left( \VarPhim_n(\boldr), \Psim_n(\boldr) \right)$ obtained as the solution of \eqref{eq:Eigen} also depend on the surrounding environment through the Green’s coupling operator $\mathbb{G}$, in addition to $\St$ and $\Sr$.

\begin{figure}[!t]
\centering 
\centering\includegraphics[width=0.85\columnwidth]{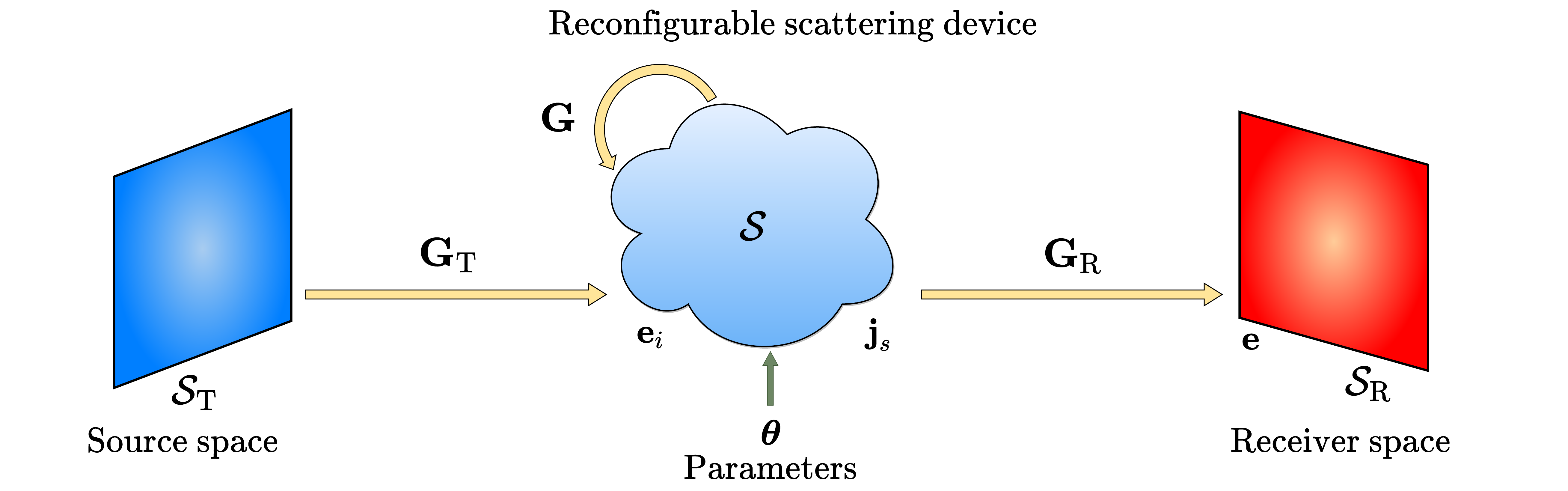}
\caption{A source and receiver space interacting through a reconfigurable scattering device acting as an EM signal processor.}
\label{Fig:Scenario}
\end{figure}

\subsection{Reconfigurable Scattering Device: Mode Transfer Matrix}\label{sec:scatterdevice}

Consider now the scenario illustrated in Fig.~\ref{Fig:Scenario}, where a source space $\St$ (e.g., a transmitter) and a receiver space $\Sr$ (the receiver) interact through a reconfigurable scattering device.
We assume the scattering device to be enclosed within the finite space $\CalS$. The impinging \ac{EM} field generated by currents located in the source space $\St$ induces a current distribution $\Jms (\boldr)$ on the scattering device, with $\boldr  \in \CalS$,  that radiates the scattered field $\Ems(\boldr)$ \cite{BalB:24}.\footnote{A more general treatment is based on the equivalent principle where fictitious electric and magnetic currents are introduced as a function of the tangent electric and magnetic fields \cite{Dar:J24}. 
For clarity, hereby we focus on a more limited case valid for dielectric and non-perfect conductors.} 

For any linear time-invariant scattering object, the induced current density $\Jms(\boldr)$, $\boldr \in \CalS$, is a linear functional of the \ac{EM} field $\Em(\boldr)$ that can be described as 
\begin{align} \label{eq:D}
\Jms(\boldr)=\int_{\CalS} \Dm(\boldr,\bolds) \, \Em(\bolds) \, d \bolds = \left ( \DOperator \, \Em \right ) (\boldr)\, , 
\end{align}
where $\Dm(\boldr,\bolds)$ is the dyadic impulse response that fully describes the relationship between the electric field and the induced current at the scatterer \cite{Dar:J24}. In \eqref{eq:D}, $\Em(\boldr)$ denotes the total electrical field, which is $\Em(\boldr)=\Emi(\boldr)+\Ems(\boldr)$, having indicated with $\Emi(\boldr)$ the electrical field incident to the scatterer generated by the source in the absence of the scatterer and with $\Ems(\boldr)$ the scattered field caused by the induced current $\Jms(\boldr)$. 

\begin{figure}[!t]
\centering\includegraphics[width=0.80\columnwidth,trim=0cm 0cm 22cm 0cm,clip]{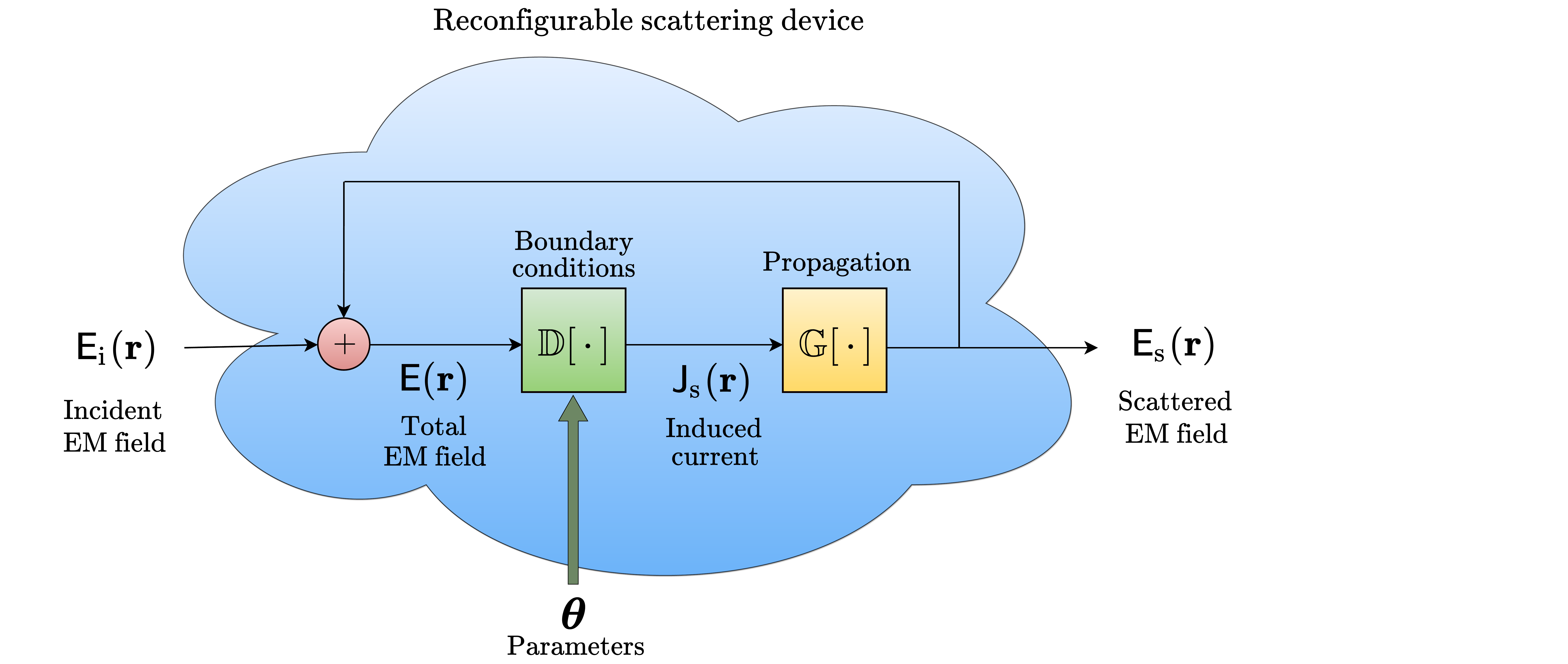}
\caption{The reconfigurable scattering device as a space-variant feedback filter.  } 
\label{Fig:Feedback}
\end{figure}

Suppose we aim to solve the \ac{EM} problem of determining the total electric field $\Em(\boldr)$ and/or its scattered component $\Ems(\boldr)$ when the scattering device is illuminated by an \ac{EM} source. Both the electric field and the induced current must satisfy \eqref{eq:Er}, as well as the constitutive equation \eqref{eq:D}, which imposes a boundary condition, i.e., a constraint on the \ac{EM} field on $\CalS$. Examining the relationship between \eqref{eq:Er} and \eqref{eq:D}, it is noteworthy that any scattering device can be viewed as a feedback system, as illustrated in Fig.~\ref{Fig:Feedback}. In this framework, $\Emi(\boldr)$ represents the input, the propagation operator $\GOperator$ acts as a space-invariant filter, filtering the wavenumber components of the electric field within the support $\CalE$ as defined in \eqref{eq:wavecondition} and providing $\Ems(\boldr)$ as its output, and the scattering device itself operating as a space-variant filter $\DOperator$ (analogous to time-variant filters), with the induced current $\Jms(\boldr)$ as its output. The filter $\DOperator$ can be designed through the parameters $\btheta$, i.e., $\DOperator ( \btheta)$, which means that different boundary conditions, and hence different \ac{EM} wave ``processing" tasks, can be realized through a proper design of parameters $\btheta$, as will be explored in the next sections.

The analysis of the feedback system in Fig.~\ref{Fig:Feedback} leads to integral equations, the solution of which is generally a complex problem that is typically tackled numerically. A more efficient approach is to use the method of moments, which involves expressing the fields and currents as series expansions using appropriately defined basis sets \cite{BalB:24}.
To this end, let $\left \{ \Phim_k(\boldr) \right \}_{k=1}^{K}$ denote a complete orthonormal vector basis set for $\CalS$, which depends solely on the geometry of $\CalS$. In general, $K$ is infinite for the basis set to be complete, but $K$ can be set to a sufficiently large finite value based on the desired level of accuracy. Consequently, the current density $\Jms(\boldr)$, the incident and scattered electric fields $\Emi(\boldr)$, $\Ems(\boldr)$, and the total electric field $\Em(\boldr)$ within the space $\CalS$, i.e., for $\boldr \in \CalS$, can be expressed as a linear combination of these basis functions, namely
\begin{align}
\Jms(\boldr)  \!=\! \sum_{k=1}^K [\js]_k \, \Phim_k(\boldr)\, , \quad    \Emi(\boldr)  \!=\! \sum_{k=1}^K [\ei]_k \, \Phim_k(\boldr)\, , \quad   \Ems(\boldr)  \!=\! \sum_{k=1}^K [\es]_k \, \Phim_k(\boldr)\, , \quad   \Em(\boldr)  \!=\! \sum_{k=1}^K [\e]_k \, \Phim_k(\boldr)\, , 
\end{align}
where the vectors $\js$, $\ei$, $\es$, and $\e$ represent generalized currents and voltages. The complex coefficients $\{ [\js]_k \}$ composing the vector $\js$ are given by $[\js]_k= \left < \Jms(\boldr) , \Phim_k(\boldr)  \right > $, for $ k=1,2, \ldots, K$.  The vectors $\ei$, $\es$, and $\e$ are defined analogously, containing the coefficients of the series expansions of, respectively, $\Emi(\boldr)$, $\Ems(\boldr)$, and $\Em(\boldr)$,  according to the same basis set $\{\Phim_k(\boldr) \}$.   
 When the scattering device consists of a set of radiating elements (e.g., dipoles), each with a programmable loaded port, a suitable choice of the functions $\left \{ \Phim_k (\boldr) \right \}$ allows $\js$ and $\es$ to be directly associated with the actual currents and voltages at the ports. This effectively maps the original problem into an electrical circuit model (see Section~\ref{Sec:Discrete}).  

From Fig.~\ref{Fig:Feedback}, it straightforward to write  $\Em(\boldr)=\Emi(\boldr)+(\GOperator \, \Jms )(\boldr)=\Emi(\boldr)+(\GOperator \, \DOperator \, \Em )(\boldr)$.
As a consequence,  $\e=\ei+\es=\ei+\boldG \, \js=\ei+\boldG \, \boldD(\btheta)   \, \e$, where $\boldG\in \mathbb{C}^{K\times K}$ is a matrix accounting for the self-coupling of the scattering device having elements $[\boldG]_{i,k}=\innerprod{ \Phim_i(\boldr)}{(\GOperator \, \Phim_k) (\boldr) }$ for $i,k=1,2, \ldots , K$,  and $\boldD(\btheta) \in \mathbb{C}^{K\times K}$ is a matrix with elements $[\boldD(\btheta)]_{i,k}=\innerprod{\Phim_i(\boldr) }{(\DOperator \, \Phim_k )(\boldr)}$, for $i,k=1,2, \ldots , K$.
It follows that 
\begin{equation} \label{eq:es}
\es=\boldG \, \js=\boldG \, \boldD(\btheta) (\boldI_K-\boldG \, \boldD(\btheta))^{-1} \, \ei =\boldR(\btheta) \, \ei\, , 
\end{equation}
where $\boldI_K$ is the identity matrix of size $K$, and $\boldR(\btheta) = \boldG \, \boldD(\btheta) (\boldI_K - \boldG \, \boldD(\btheta))^{-1}$ serves as the \emph{reflection matrix} (or \emph{scattering matrix} in the electromagnetic community). Specifically, the $(k,i)$-th element of $\boldR(\btheta)$ represents the reflection coefficient corresponding to the $i$-th component of the impinging electric field and the $k$-th component of the scattered field, as described by the basis set $\{\Phim_k(\boldr)\}$.
Notably, depending on the available technology, the matrix $\boldD(\btheta)$ might be constrained to take a particular structure. For example, if we exclude the presence of non-reciprocal materials in the scatterer, the response of the scattering device must verify the reciprocity theorem, hence resulting in a symmetric matrix \cite{BucIse:97}. 
Another example is the case where the boundary conditions imposed by the reconfigurable scatterer are local, i.e.,  $\Dm(\boldr,\bolds)=\Dm(\boldr) \, \delta(\boldr-\bolds) $ and hence $\Jms(\boldr)=\Dm(\boldr) \, \Em(\boldr)$ in \eqref{eq:D} \cite{Dar:J24}.   
For instance, if we consider the reconfigurable scattering device to be composed of a set of $K$ uncoupled discrete elements each described by the $k$-th function $\Phim_k(\boldr)$ of the basis set, the $\left \{ \Phim_k(\boldr) \right \}$'s are disjoint functions, and hence matrix $\boldD(\btheta)$ is restricted to be diagonal as well as the reflection matrix $\boldR(\btheta)$ being the elements uncoupled by assumption (see Section~\ref{Sec:Technologies} for examples). 

With reference to Fig.~\ref{Fig:Scenario}, let us now consider two distinct coupling problems: from the source space $\St$ to the scattering device space $\CalS$, and from the scattering device space to the receiving space  $\Sr$, both involving the operator $\GOperator$. A direct coupling between the source and receiving spaces may occur; however, the reconfigurable scattering device cannot control this interaction. Therefore, unless otherwise specified, we will focus solely on the \ac{EM} field component received from the scattering device.

Suppose the communication modes associated with the first link (from the source space to the scattering device) and the second link (from the scattering device to the receiving space) have been derived through the coupled eigenfunction problems \eqref{eq:Eigen} by truncating the series to the first $K$ terms, where $K$ is sufficiently large. Denote with $\left ( \left \{ \VarPhimt_n(\boldr) \right \}, \left \{ \Psimt_n(\boldr) \right \},  \sigma_n^{(\text{T})}  \right )$ the $n$-th basis couple and singular value of the first link, for $n=1,2, \ldots , K$, and with  $\left ( \left \{ \VarPhimr_m(\boldr) \right \}, \left \{ \Psimr_m(\boldr) \right \},  \sigma_m^{(\text{R})}  \right )$ the $m$-th basis couple and singular value of the second link, for $m=1,2, \ldots , K$.  
Denote with ${\Gt = \diag{\sigma_1^{(\text{T})}, \sigma_2^{(\text{T})}, \ldots , \sigma_K^{(\text{T})}} \in \mathbb{C}^{K \times K}}$ and $\Gr= \diag{\sigma_1^{(\text{R})}, \sigma_2^{(\text{R})}, \ldots , \sigma_K^{(\text{R})}} \in \mathbb{C}^{K \times K}$  the coupling matrices between, respectively, the source and device spaces, and the device and receiving spaces. The effect of the scattering device can be fully described by the device operator $\boldC(\btheta) \in \mathbb{C}^{K \times K}$, whose generic element $[\boldC(\btheta)]_{m,n}$ represents the coupling coefficient between the input communication mode $n$ and the output communication mode $m$. 
As a consequence, the end-to-end coupling between the source space and the receiving space is given by
\begin{equation}\label{eq:Htheta}
\boldH(\btheta)=\Gr \, \boldC(\btheta) \, \Gt \, . 
\end{equation}
The device operator $\boldC(\btheta)$, also known as the \emph{mode transfer matrix}, is determined by the specific structure and technology of the scattering device, as well as its configuration parameters $\btheta$, which enable the realization of different end-to-end processing functionalities $\boldH(\btheta)$.
In essence, any linear scattering device can be seen as a \emph{mode converter} \cite{Miller:19}.

We now derive the relationship between $\boldC(\btheta)$ and $\boldD(\btheta)$. By defining $\boldU \in \mathbb{C}^{K\times K}$, with elements $[\boldU]_{m,k}= \innerprod{\VarPhimr_m(\boldr)}{\Phim_k(\boldr)} $ as the transformation matrix between the basis $\left \{\VarPhimr_m(\boldr) \right \}$ and $\left \{\Phim_k(\boldr) \right \}$, and $\boldV\in \mathbb{C}^{K\times K}$, with elements $[\boldV]_{n,k}=\innerprod{\Psimt_n(\boldr)}{\Phim_k(\boldr)}$, the transformation matrix between the basis $\left \{ \Phim_k(\boldr) \right \}$  and $ \left \{ \Psimt_n(\boldr)  \right \}$, we can write 
\begin{align}  \label{eq:bT}
 \boldC (\btheta)
 =\boldU \, \boldD(\btheta)\,  (\boldI_K-\boldG \, \boldD(\btheta))^{-1}  \, \boldV^{-1} \, .
\end{align}
It is useful to examine how \eqref{eq:bT} simplifies under the assumption of weak reflections, i.e., when $ |\Ems(\boldr)| \ll |\Emi(\boldr)| $ for $ \boldr \in \CalS $, which corresponds to $ \| \boldG \, \boldD(\btheta) \| \ll 1 $. Specifically, we have
\begin{align} \label{eq:Coupling}
 \boldC(\btheta) \approx \boldU \, \boldD(\btheta) \, \boldV^{-1} \, .
 \end{align}
This corresponds to the Born approximation, which involves using the incident field as the driving field at each point in the scattering device, rather than the total \ac{EM} field. This approximation is valid when the scattered field is small relative to the incident field on the scattering device. Notably, this result is equivalent to removing the feedback loop in Fig.~\ref{Fig:Feedback}.
Depending on the specific technology being analyzed, the use of the approximation \eqref{eq:Coupling} instead of \eqref{eq:bT} should be considered with caution, as it may fail to capture important physical phenomena. An example of this is the Floquet modes in periodic metamaterials, which correspond to spurious reflections in undesired directions when used as a \ac{RIS} \cite{MarMac:22, Dar:J24}.

In conclusion, the \ac{ESP} problem can be framed as finding the optimal configuration parameter set $\btheta$ that best approximates a given objective end-to-end processing functionality $\boldHo$ according to a specific criterion. This can be formulated as the following minimization problem:
 \begin{equation} \label{eq:hattheta}
\hat{\btheta}=\arg \min_{\btheta} \| \Gr \, \boldC(\btheta) \, \Gt -   \boldHo \|_{\mathrm{F}}\, ,    
\end{equation}
subject to a constraint on the radiated power, where $\| \cdot \|_{\mathrm{F}}$ denotes the Frobenius norm. Depending on the adopted technology, additional constraints should be incorporated into the minimization problem to account for the specific structure and symmetries of the matrix $\boldD(\btheta)$ and hence $\boldC(\btheta)$. This underscores the need for tailored optimization algorithms based on physically consistent models. In this regard, specific examples will be provided in Section~\ref{Sec:Reactive}.

\section{Fundamental Limits}

As described in Section~\ref{Sec:CommModes}, the number of \ac{DoF} of the receiver space $\Sr$ represents the minimum number of parameters sufficient to approximate any element $\Em(\boldr)$ of the signal space $\mathcal{Y}={\mathbb{L}}^2(\Sr)$, defined in $\Sr$, up to arbitrary precision. Therefore, when $\Em(\boldr)$ is the result of wave processing by the reconfigurable scattering device, a fundamental question is to determine the signal processing capability of the device in approximating the desired \ac{EM} response, i.e., the number of \ac{DoF} the device can manage.   
The ultimate limit in terms of the number of \ac{DoF} of the scattering device can be obtained supposing that, through a proper configuration of the device and illumination from the source space, it is possible to generate any element $\Jm(\boldr) \in \mathcal{X}$, where $\mathcal{X}={\mathbb{L}}^2(\CalS)$ is the signal space of the induced current distributions on $\CalS$. 
In the following, we show that even in the case of an unbounded space (infinite receiver's size), the number of \ac{DoF} is limited by the size of the scattering device compared to the wavelength. This number of \ac{DoF} is further limited when finite source and receiver spaces are considered. In any case, these theoretical limits have to be intended as upper bounds because the particular technology considered might pose further constraints.

\subsection{DoF in the Unbounded Space} \label{sec:DoFunbounded}

In the following, we revisit the main results regarding the number of \ac{DoF} for the radiating \ac{EM} field in an unbounded space, given a current density $\Jm(\boldr)$ defined within a limited space $\CalS$, using signal processing arguments. 

\subsubsection{DoF for 1D Structures}

In this case, $\CalS$ is a line segment of length $L$, with coordinate $r_x \in [-L/2, L/2]$. 
Without loss of generality, we consider vertical \ac{EM} waves polarization. 
Suppose $\Jm(\boldr)=J(r_x) \, \delta(r_y) \, \delta(r_z)\, \versory$ defined in $\CalS$.
It is $\TildeJm(\wavefreq)=\fourierthree{\Jm(\boldr)}=\fourier{J(r_x)} \versory=\TildeJ (\wavefreqx) \, \versory=\TildeJm(\wavefreqx)$, where ${\TildeJ (\wavefreqx)=\fourier{J(r_x)}}$.
It is well-known that every function $J(r_x)$ limited in $[-L/2,L/2]$ can be represented as an infinite series expansion of Fourier basis functions (i.e., modes) \cite{Dar:J24} 
\begin{equation}
J(r_x)= \sum_{n=-\infty}^{\infty} j_n  \frac{1}{\sqrt{L}} \rect{\frac{r_x}{L}}\, e^{\jmath k_n \, r_x }\, , 
\end{equation}
where $k_n=2\, n\, \pi/L$ and $\rect{x}=1$, for $|x|<0.5$, and zero otherwise. Then it is 
\begin{equation} \label{eq:TildeJ}
\TildeJ(\wavefreqx)=\sum_{n=-\infty}^{\infty} j_n \, \sqrt{L}  \, \sinc{\frac{L}{2\pi} \left (\wavefreqx - k_n \right )  }\, , 
\end{equation}
being $\sinc{x}=\sin(\pi x)/(\pi x)$. Note that, according to \eqref{eq:wavecondition}, only the modes for which ${|k_n|\le \kappazero=2 \pi/\lambda}$ radiate. Then, only $2 L/\lambda$ terms in \eqref{eq:TildeJ} fall in the visible region.\footnote{More precisely, we assume that $L \gg \lambda$; otherwise, the tails of certain sinc functions centered in the non-visible region could significantly contribute to the visible region.}
Being the wavenumber support limited (i.e., $\TildeJ (\wavefreqx)$ band limited in $[-\kappazero,\kappazero]$), we can say that any \ac{EM} radiated field that can be generated by a current line source of length $L$ can be fully represented by sampling it in the wavenumber domain at frequencies $k_n=2n\pi/L$ for $|n|\le L/\lambda$. Thus, the required number of samples leads to the number of \ac{DoF} in the unbounded space $\Nu=\frac{2L}{\lambda}$ (per polarization).
To draw a parallel, when beamforming is operated using $\Narray$ half-wavelength spaced antennas arranged as a \ac{ULA}, $\Nu=\Narray$ orthogonal beams (i.e., $2L/\lambda$ when the antenna spacing is $\lambda/2$) can be realized. These can be adopted to reach users in every possible angular direction (e.g., for initial access). These beams correspond to the set of orthogonal transmission directions that can be spanned by the ULA considering the classical \ac{DFT}-based codebook of beamforming vectors given by \cite{Pre:J24} 
\begin{equation} \label{eq:DFT}
    \frac{1}{\sqrt{\Narray}}
    \begin{bmatrix}
        1 ,\,\, 
        e^{- \frac{\jmath 2 \pi n }{N}} ,\,\, 
        e^{- \frac{ \jmath 2 \pi 2 n}{N}} ,\,\, 
        \ldots ,\,\, 
        e^{- \frac{\jmath 2 \pi (\Narray-1)n} {N}} 
        \end{bmatrix}^{\transpose}
        = 
         \frac{1}{\sqrt{\Narray}}
    \begin{bmatrix}
        1 ,\,\, 
        e^{- \frac{\jmath k_n \lambda}{2}} ,\,\, 
        e^{- \frac{ \jmath 2 k_n \lambda}{2}} ,\,\, 
        \ldots ,\,\, 
        e^{- \frac{\jmath  (\Narray-1) k_n \lambda } {2}} 
        \end{bmatrix}^{\transpose} \, .
\end{equation}

 The different vectors correspond to the transmission directions towards angles $\gamma_n=\arcsin{\frac{2n}{\Narray}}$, for $n=\pm 1, \pm 2, \ldots, \left\lfloor \frac{\Narray}{2} \right\rfloor$. 
The characteristic of these beams is that each one points towards the null directions of all the other beams. Every possible channel vector (i.e., direction) can be realized as a linear combination of the vectors in \eqref{eq:DFT}, which is an orthonormal basis for $\mathbb{C}^{\Narray}$.

\subsubsection{DoF for 2D Structures}
In this second case, $\CalS$ is a surface of size $\Lx \times \Ly$. 
Similarly to the 1D case, the current density can be represented as a \ac{2D} series expansion of Fourier basis functions. In the wavenumber domain, we have \cite{Dar:J24} 
\begin{equation} \label{eq:TildeJ2D}
\TildeJ(\wavefreqx,\wavefreqy)= \sum_{n_x,n_y} j_{n_x,n_y} \,  \sqrt{\Lx \, \Ly}\, \sinc{\frac{\Lx}{2\pi} \left (\wavefreqx - k_{n_x} \right )  } \sinc{\frac{\Ly}{2\pi} \left (\wavefreqy - k_{n_y} \right )  }\, , 
\end{equation}
where $k_{n_x}=2\, n_x\, \pi/\Lx$, and $k_{n_y}=2\, n_y\, \pi/\Ly$.
Note that only the set of indexes $(n_x,n_y) \in \CalP_2$, with $\CalP_2=\left \{ (n_x,n_y) :  \left (\frac{n_x \lambda}{\Lx} \right )^2+\left (\frac{n_y \lambda}{\Ly} \right )^2 \le 1 \right \}$, satisfying $k_{n_x}^2+k_{n_y}^2 \le \kappazero^2$, are not evanescent, i.e., lie in the visible region. For instance, for a square surface of size $\Lx=\Ly=L$, it is $\Nu \simeq |\CalP_2| \simeq \frac{ \pi L^2}{\lambda^2}$ (per polarization) \cite{Dar:J24}.  Interestingly, this is $\pi/4=0.79$ of the square of the \ac{DoF} of a linear segment. 
As a consequence, by drawing a parallel as before with uniform planar arrays, we have that a $\Narray\times \Narray$ planar array will not be able to generate $\Narray^2$ orthogonal beams (i.e., $\frac{4 L^2}{\lambda^2}$), as could be erroneously suggested by intuition, but only a subset $\Nu<\Narray^2$, specifically $\Nu=\frac{\pi}{4}\Narray^2\approx 0.79\, \Narray^2$ beams satisfying the condition of non-evanescent modes.

\subsubsection{DoF for 3D Structures}

When $\CalS$ is a volume of size $\Lx \times \Ly \times \Lz$, in the wavedomain we have 
\begin{equation} \label{eq:TildeJ3D}
\TildeJ(\wavefreqx,\wavefreqy,\wavefreqz)= \hspace{-0.3cm} \sum_{n_x,n_y,n_z} \hspace{-0.3cm} j_{n_x,n_y,n_z}  \sqrt{\Lx \Ly \Lz}  \, \sinc{\frac{\Lx}{2\pi} \left (\wavefreqx - k_{n_x} \right )  } \sinc{\frac{\Ly}{2\pi} \left (\wavefreqy - k_{n_y} \right )  } \sinc{\frac{\Lz}{2\pi} \left (\wavefreqz - k_{n_z} \right )  }\, , 
\end{equation}
where $k_{n_x}=2\, \pi \, n_x/\Lx$, $k_{n_y}=2\, \pi \, n_y/\Ly$, and  $k_{n_z}=2\, \pi \, n_z/\Lz$.

\begin{figure}[!t]
\centering\includegraphics[width=0.75\columnwidth]{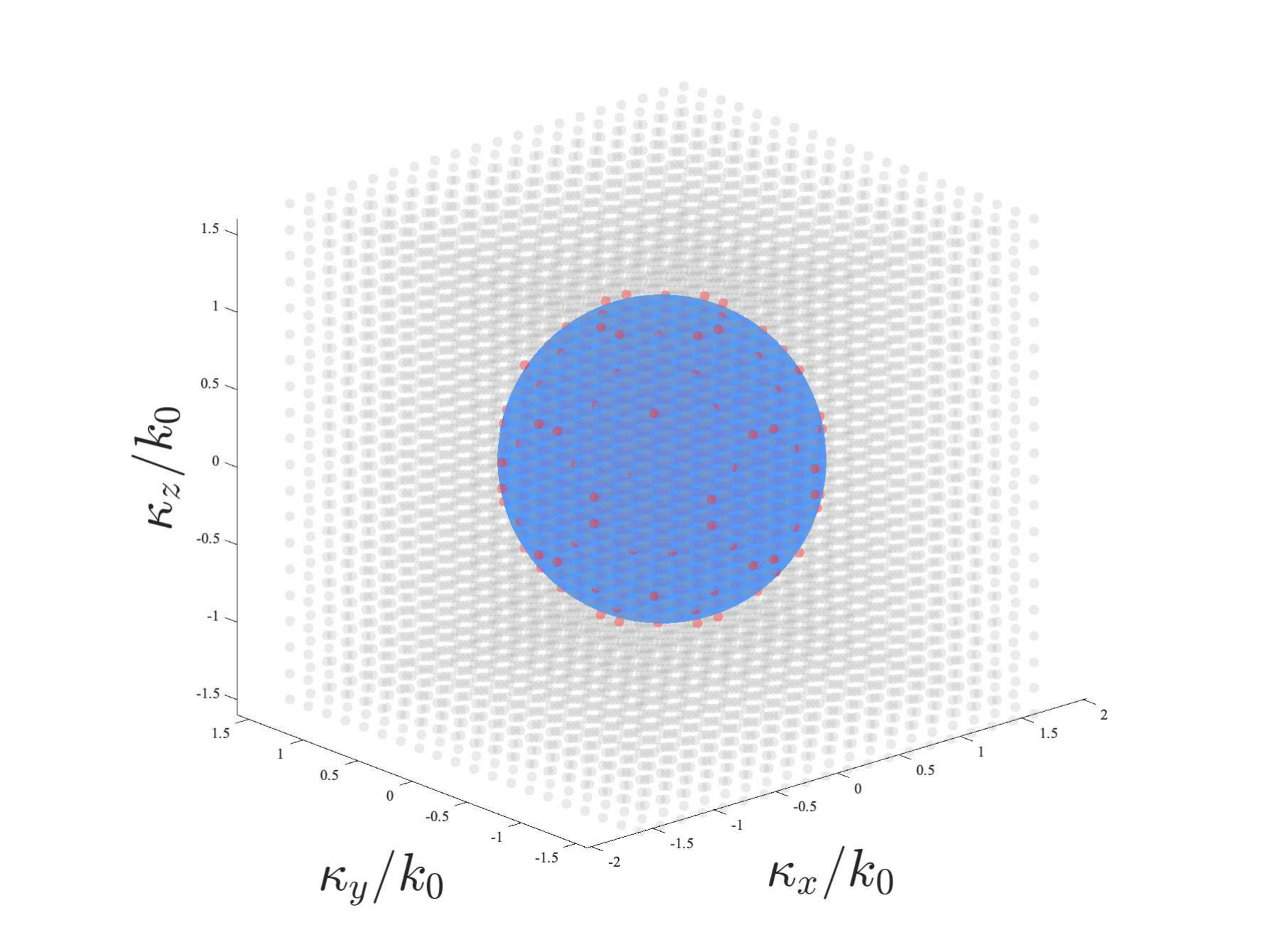}
\caption{Filtering operation of the Green's operator in the wavenumber domain for $L/\lambda=6$. The sphere corresponding to the set $\CalE$ is depicted in blue, while the grey points represent the \ac{3D} sinc functions in \eqref{eq:TildeJ3D}. The red points instead denote the subset of \ac{3D} sinc functions in \eqref{eq:TildeJ3D} having significant intersection with the sphere, hence generating \ac{EM} waves.} 
\label{Fig:Sfera}
\end{figure}

The derivation of the number of \ac{DoF} for a \ac{3D} structure is more involved. The typical approach is to consider the smallest sphere enclosing the space $\CalS$,  define a complete basis set given by the spherical harmonic functions on its surface, and identify the $N$ dominant ones \cite{BalB:24}.  
Here we follow an approximate but simpler approach for the case $\Lx=\Ly=\Lz=L$.  The modes in \eqref{eq:TildeJ3D} contributing to the \ac{EM} field are those with index combinations $(n_x,n_y,n_z)$, corresponding to the wavenumber points $\wavefreq_{n_x,n_y,n_z}=(k_{n_x},k_{n_y},k_{n_z})$, for which the intersection with the surface of radius $\kappazero$ defined in \eqref{eq:wavecondition} is not negligible.  Their number can be approximatively evaluated as the number of wavenumber points $\wavefreq_{n_x,n_y,n_z}$ falling within the shell of radius $\kappa_0$ and thickness $ \delta$ (see Fig.~\ref{Fig:Sfera}). Denote with $\CalP_3$ the set of indexes $(n_x,n_y,n_z)$ whose wavenumber satisfies this condition. For instance, $\delta$ can be taken equal to the first lobe of the sinc function, i.e., $\delta=2\pi /L$. Specifically, it is $\Nu \simeq |\CalP_3|=\nu \cdot V_{\delta}$, where $\nu=\left ( \frac{L}{2 \pi} \right )^3$ is the density of the  wavenumber points and $V_{\delta}$ is the volume of the shell. Therefore, it is
\begin{equation}
\Nu \simeq \nu \cdot V_{\delta}=\frac{\pi}{3} + \frac{4 \pi L^2}{\lambda^2} \simeq \frac{ \pi^2 L^2}{\lambda^2}\, , 
\end{equation} 
for $L \gg \lambda$, which is the same result one can obtain using a more precise but complex approach based on spherical harmonic waves. Note that, when using volumetric antennas instead of planar antennas, a \ac{DoF} gain of at most $\pi$ can be achieved.
It can be pointed out that while for the 1D case sampling the current with spacing $\lambda/2$ is necessary to represent the current components contributing to the radiating \ac{EM}  field, in the \ac{2D} and \ac{3D} cases sampling at $\lambda/2$ is only a sufficient but no longer necessary condition. Readers can refer to \cite{PizTorSanMar:22} for a more detailed discussion.

\subsection{DoF of the Single Link}

Consider a communication system involving a transmitting space $\St$ and a receiving space $\Sr$. Each space allows for a maximum number of \ac{DoF} in the unbounded space, as discussed in Section~\ref{sec:DoFunbounded}. However, only a subset of these available \ac{DoF} can be exploited for communication, corresponding to the $N$ well-coupled communication modes with eigenvalues $\sigma_n^2$ described in Section~\ref{sec:CommModes}. When currents are excited at $\St$ corresponding to these well-coupled modes, the resulting \ac{EM} field is predominantly concentrated on the receiver space $\Sr$. In contrast, exciting modes outside this subset of well-coupled modes leads to an \ac{EM} field largely dispersed away from $\Sr$.

\begin{figure}[!t]
\centering\includegraphics[width=0.95\columnwidth]{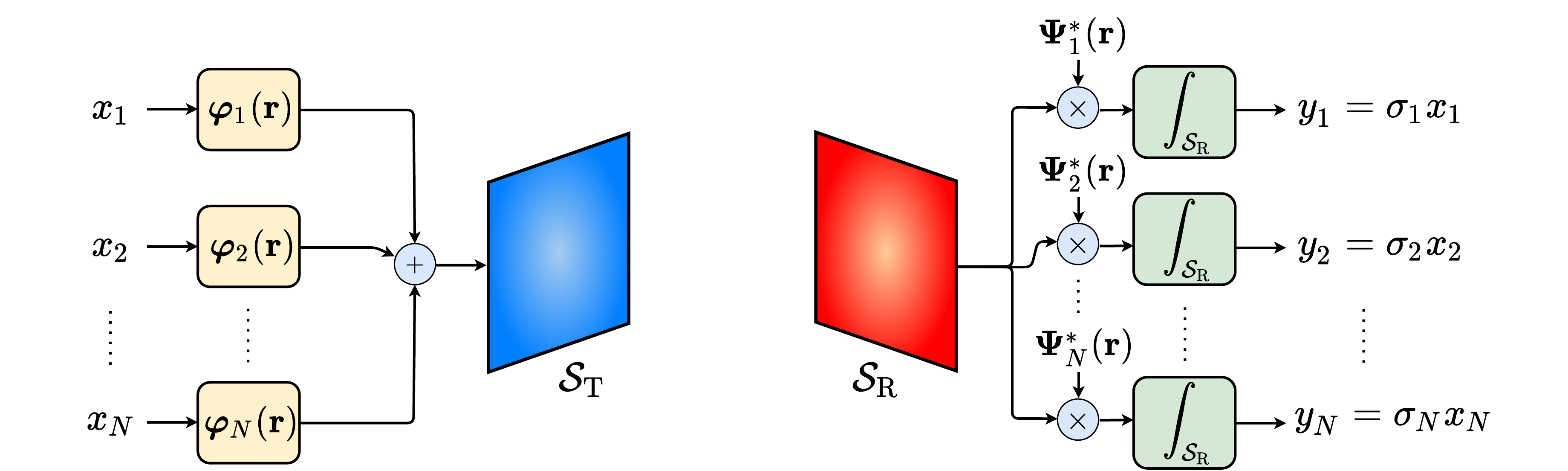}
\caption{Optimal communication scheme between spaces based on communication modes.}
\label{Fig:Communication}
\end{figure}

In the context of communication systems, the $N$ communication modes represent the available \ac{DoF} between the transmitting space $\St$ and the receiving space $\Sr$, which can be modeled using the communication scheme shown in Fig. \ref{Fig:Communication}. Here, $N$ information symbols $\boldx \in \mathbb{C}^N$ are associated with the basis functions $\left\{\VarPhim_n(\boldr)\right\}$, forming the source current density $\Jm(\boldr) = \sum_n x_n \VarPhim_n(\boldr)$ on $\St$. At the receiver, the information is recovered by correlating the received \ac{EM} field $\Em(\boldr)$ with the corresponding basis functions $\left\{\Psim_n(\boldr)\right\}$ on $\Sr$, i.e., $y_n = \innerprod{\Em(\boldr)}{\Psim_n(\boldr)}$, resulting in a set of matched filters operating in the spatial domain. Formally, considering the effect of noise, the received signal $\mathbf{y} \in \mathbb{C}^N$ is given by the following equation
\begin{equation}
    \mathbf{y} = \boldsymbol{\Sigma} \, \mathbf{x}+\mathbf{n}\, , 
\end{equation}
where $\boldsymbol{\Sigma}=\diag{\sigma_n}\in\mathbb{C}^{N\times N}$ , $\mathbf{n}\in\mathbb{C}^{N}$ is the \ac{AWGN} with ${\mathbf{n} \sim {\mathcal{CN}}\left (\mathbf{0}, \sigma_{\mathrm{noise}}^2 \, \mathbf{I}_N \right )}$, where $\sigma_{\mathrm{noise}}^2$ is the noise power and ${\mathcal{CN}}\left (\cdot, \cdot \right)$ denotes a multivariate complex normal distribution. Moreover, let $\mathbf{Q}$ be the positive semi-definite (i.e., ${\mathbf{Q}} \succeq 0$) covariance matrix of the transmitted signal, i.e., $\mathbf{Q}=\mathbb{E}[\mathbf{x}\mathbf{x}^{\ctranspose}]$, where $\mathbb{E}[\cdot]$ is the expectation operator and $(\cdot)^{\ctranspose}$ denotes the complex transpose. Then, the transmit power constraint is $\mathrm{tr} (\mathbf{Q})\leq \Pt$, where $\mathrm{tr}(\cdot)$ is the trace operator and $\Pt$ is the total transmit power. 

As well-known from the MIMO theory, the transmitter must split its power among the $N$ parallel \ac{SISO} channels. For any $\mathbf{Q}$, the overall capacity is the sum of the 
capacities of the individual channels, leading to \cite{TseVis:B05}
\begin{equation}
C=\mathop{\max}_{\substack{p_1\ge0,\ldots,p_N\ge0\\\mathrm{tr} (\mathbf{Q})\leq \Pt}} \sum_{n=1}^N \log_2 \left (1 + \frac{p_n \sigma_n^2}{\sigma_{\mathrm{noise}}^2}\right )\, , 
\end{equation}
where the optimum power allocation follows the water-filling solution as per 
\begin{equation}\label{eq:waterfilling}
p_n=\max\left(\mu - \frac{\sigma_{\mathrm{noise}}^2}{\sigma_n^2},0 \right), \quad\quad n=1,\ldots,N\, , 
\end{equation}
and being the variable $\mu$ selected to fulfill the condition $\sum_{n=1}^N p_n = \Pt$.
The eigenvalues $\sigma_n^2$ represent the channel gains and, being ordered in decreasing magnitude, the power is allocated from the strongest to the weakest channels. Consequently, the number of significant eigenvalues (i.e., the number of \ac{DoF}) and their distribution have a significant impact on the maximum capacity. A few examples are hereby provided.

\subsubsection{DoF for 1D Structures}

Consider two 1D segments as described in Section~\ref{sec:DoFunbounded}, placed parallel to each other and aligned along their centers (i.e., in a paraxial configuration). We focus on a single polarization. The lengths of the segments are denoted as $\Lt$ and $\Lr$, and the distance between their centers is $d$. 
It can be demonstrated that the $N$ most significant eigenvalues in this configuration are nearly identical in magnitude and then rapidly decay. Therefore, according to the water-filling principle in \eqref{eq:waterfilling}, the power can be evenly distributed among the $N$ modes. The number of \ac{DoF} of the link, or the number of significant eigenvalues, can thus be determined for a single polarization as follows \cite{Miller:19}
\begin{equation}\label{DoFlink:1D}
N=\frac{\Lt \Lr}{\lambda \, d}\, .
\end{equation}
This classical expression, while accurately determining the number of \ac{DoF} when $\Lt, \Lr \gg d$ and the sizes of $\Lt$ and $\Lr$ are nearly comparable, becomes inadequate when one of the segments grows excessively large. In fact, \eqref{DoFlink:1D} appears to suggest that the DoF can increase indefinitely if either $\Lt$ or $\Lr$ becomes infinitely large. However, this contradicts the results obtained in Section~\ref{sec:DoFunbounded}, where the maximum number of DoF is bounded and cannot exceed the values derived for an unbounded space. A more accurate expression was derived in \cite{DecDar:J21} by analyzing the scalar Green's operator and modeling the communication modes as focused beams, which leads to 
\begin{align}\label{DoFlink:1Dnew}
N = \frac{2\Lt}{\lambda} \zeta\, ,
\end{align}
with $\zeta=\Lr/\sqrt{4d^2+\Lr^2}$.
From expression \eqref{DoFlink:1Dnew}, it can be observed that as $ \Lr \to \infty $, we have $ \zeta \to 1 $, and thus the limit is $ N = \Nu = \frac{2\Lt}{\lambda} $ which corresponds to the actual limit in the unbounded space presented in Section~\ref{sec:DoFunbounded}. Among the $ \Nu = \frac{2\Lt}{\lambda} $ orthogonal beamsteering directions associated with \eqref{eq:DFT} that can be generated with the segment of length $ \Lt $, \eqref{DoFlink:1Dnew} indicates the number of beams that intersect the parallel segment of length $ \Lr $ at a distance $ d $.

\subsubsection{DoF for \ac{2D} Structures}

Consider now two square surfaces with areas, respectively, $\At=\Lt^2$ and $\Ar = \Lr^2$. These surfaces are located at a distance $d$ in a paraxial configuration and outside the reactive region of the transmitting area. When %
$d \gg \Lt, \Lr$
, the classical expression for the available DoF (per polarization) is \cite{Miller:19}
\begin{equation}\label{DoFlink:2D}
N=\frac{\At \Ar}{\lambda^2 d^2}=\Nu^{(\text{T})}\, \Nu^{(\text{R})}\, \left (\frac{\lambda}{\pi d} \right )^2  
\le \min \left (\Nu^{(\text{T})}, \Nu^{(\text{R})} \right )
\end{equation}
where $\Nu^{(\text{T})}$ and $\Nu^{(\text{R})}$ are the number of \ac{DoF} in the unbounded space of the transmitting and receiving areas, respectively. 
In \cite{Dar:J20}, the following alternative expression is derived for $\At < \Ar$ 
\begin{equation}\label{DoFlink:2Dnew}
N=\frac{4 \At}{\lambda^2} \zeta \tan^{-1}\zeta \, .
\end{equation}
Unlike \eqref{DoFlink:2D}, \eqref{DoFlink:2Dnew} provides the actual number of \ac{DoF} for squared planar paraxial surfaces of any size, also demonstrating a saturation when $ d \ll \Lr $. In the case $ \Lr \rightarrow \infty $, the limit is $ N = \Nu^{(\text{T})} =\frac{\pi \At}{\lambda^2} $, which matches the result obtained for the unbounded space in Section~\ref{sec:DoFunbounded}.

\subsection{DoF of the Cascade Link including a Reconfigurable Scattering Device}

When a reconfigurable scattering device is placed between the transmitter and the receiver, as discussed in Section~\ref{sec:scatterdevice} and shown in Fig. \ref{Fig:Scenario}, we define $ N $ and $ M $ as the maximum number of ``usable" channels (i.e., communication modes) into and out of the scattering device, respectively. Here, $ N $ and $ M $ correspond to the rank of $ \Gt $ and $ \Gr $, respectively, with $ N, M \leq K $. The maximum number of end-to-end communication modes is given by $ \Nc = \min(N, M) $. 
Note that, since $N,M\le \Nu$, where $\Nu$ is the number of \ac{DoF} of the reconfigurable scattering device in the unbounded space derived in Section~\ref{sec:DoFunbounded}, it is $ \Nc \le \Nu$.
The equivalent end-to-end channel can be written as
\begin{equation} \label{eq:y}
    \mathbf{y} = \mathbf{H}(\boldsymbol{\theta}) \, \mathbf{x}+\mathbf{n}\, , 
\end{equation}
where now $\boldy,\boldx, \boldn \in\mathbb{C}^{K}$, and the end-to-end channel matrix $\boldH(\theta)$ defined in \eqref{eq:Htheta} can be written equivalently as $\boldH(\theta)=\Gr\boldU\boldG^{-1}\boldR(\boldsymbol{\theta})\boldV^{-1}\Gt=\boldH_{\mathrm{R}} \boldR(\boldsymbol{\theta}) \boldH_{\mathrm{T}}$, with $\boldH_{\mathrm{R}}=\Gr\boldU \boldG^{-1}$ and $\boldH_{\mathrm{T}}=\boldV^{-1}\Gt$, to explicit the effect of the reflection matrix $\boldR(\btheta)$.

It is of interest to investigate the optimal configuration of the reconfigurable scattering device through the matrix $\boldD(\btheta)$ (thus, $\boldR(\boldsymbol{\theta})$ and $\mathbf{H}(\boldsymbol{\theta})$) maximizing the end-to-end channel capacity given $\Gt$ and $\Gr$ under some constraint on the matrix $\boldR(\boldsymbol{\theta})\approx  \boldG \, \boldD(\boldsymbol{\theta})$ (we make use of the Born approximation), for example, a lossless constraint for the reconfigurable scattering device\footnote{In this case, the power of the incident electric field $\ei$ equals that of the scattered electric field $\es=\boldR(\boldsymbol{\theta})\, \ei$.} leading to ${\boldR(\boldsymbol{\theta}) \, \boldR^{\ctranspose}(\boldsymbol{\theta}) =  \boldI_K}$.

Adopting the approach followed in \cite{BarAbrDecDarDiR:J23}, it can be proved that the end-to-end channel capacity of the reconfigurable scattering device-assisted communication is 
\begin{equation}
C = \sum\limits_{k = 1}^{K} 
\log_2 \left ( 1 + \left (\tilde{\sigma}_k^{(\mathrm{T})} \right )^2 
\left (\tilde{\sigma}_k^{(\mathrm{R})} \right )^2 
\frac{p_k}{\sigma_{\mathrm{noise}}^2} \right ) \, , 
\label{Eq_C}
\end{equation}
where $p_k = \max \left(\mu  - \frac{\sigma_{\mathrm{noise}}^2}
{\left (\tilde{\sigma}_k^{(\mathrm{T})} \right )^2 \left (\tilde{\sigma}_k^{(\mathrm{R})} \right )^2},0\right)$, with $\mu$ obtained by fulfilling the identity $\sum\nolimits_{k = 1}^{K} {p_k}  = \Pt$, and being $\tilde{\sigma}_k^{(\mathrm{T})}$ and $\tilde{\sigma}_k^{(\mathrm{R})}$ the singular values of $\boldH_{\mathrm{T}}$ and $\boldH_{\mathrm{R}}$, respectively.\footnote{We have $ \tilde{\sigma}_k^{(\mathrm{T})} = \sigma_k^{(\mathrm{T})} $, since $ \boldV $ is unitary. In contrast, $ \tilde{\sigma}_k^{(\mathrm{R})} $ may differ from $ \sigma_k^{(\mathrm{R})} $ due to the presence of the matrix $ \boldG^{-1} $ in $\boldH_{\mathrm{R}}$, which accounts for the coupling between the current density and the scattered electric field at the reconfigurable scattering device.} It is important to note that the last $ K - \Nc $ power coefficients are zero due to the maximum rank of $ \mathbf{H}(\boldsymbol{\theta}) $ being $ \Nc $. Specifically, because of the diagonal structure of $ \Gt $ and $ \Gr $, where only the first $ N $ and $ M $ diagonal entries, respectively, correspond to significant singular values, only the first $ \Nc $ elements of the symbol vector $ \boldx $ (out of $ K $) carry ``usable'' information data. In addition, \eqref{Eq_C} is obtained by setting $\mathbf{Q} = \diag{ p_1,p_2, \ldots p_K}$ and the coupling matrix $\boldC =\boldI_K$. Under the Born approximation, from \eqref{eq:Coupling}, the optimal device matrix should satisfy the condition
\begin{align}
  \label{eq:Opt}
  \boldR(\boldsymbol{\theta})=\boldV_{\text{R}} \boldU^{\ctranspose}_{\text{T}} \, , 
\end{align}
where $\boldV_{\text{R}}$ and $\boldU_{\text{T}}$ represent the left and right eigenvector matrices of the \ac{SVD} of $\boldH_{\mathrm{T}}$ and $\boldH_{\mathrm{R}}$, respectively. 
From \eqref{eq:Opt}, it can be inferred that the end-to-end capacity is maximized by designing $\boldR(\boldsymbol{\theta})$ (and consequently $\boldD(\boldsymbol{\theta})$) in a way that aligns with the left eigenvectors of the first link and the right eigenvectors of the second link. This ensures that the end-to-end transmitter-device-receiver channel is diagonalized. Furthermore, the water-filling power allocation is applied to the ordered product of the singular values of the individual transmitter-device and device-receiver channels.

The physical realizability of the optimal device matrix $\mathbf{R}(\boldsymbol{\theta})$ depends on the specific technology being used. As mentioned in Section~\ref{Sec:Model}, when the boundary conditions imposed by the scattering device are local, such as in conventional \acp{RIS}, the matrix $\mathbf{R}(\boldsymbol{\theta})$ is constrained to be diagonal, meaning that \eqref{eq:Opt} can only be approximated \cite{BarAbrDecDarDiR:J23}. Recently, new \ac{RIS} structures have been introduced, which enable reconfigurable coupling between the elements, leading to a non-diagonal matrix (non-diagonal \acp{RIS}) \cite{LiSheNerCle:24}.

\section{Enabling EM Technologies and SP  Methods} 
\label{Sec:Technologies}

In this section, we first provide a connection between \ac{EM} and circuit-based representations, then a comprehensive overview of key enabling \ac{EM} technologies that can be utilized within the \ac{ESP} framework by presenting specific examples of their application in practical processing tasks.

\subsection{Port-controlled Reconfigurable Scattering Devices: Equivalent Circuit Model}
\label{Sec:Discrete}

In most practical implementations, the reconfigurable scattering device consists of a discrete set of $K$ antenna elements positioned at locations $\boldr_k$, whose scattering characteristics can be adjusted by varying the load at their ports. In this section, we illustrate how the general framework outlined in Section~\ref{Sec:Model} can be linked to circuit theory that is typically adopted to model multiport systems.
With a limited loss of generality, for the $k$-th antenna element of the reconfigurable scattering device, let $\boldF_k(\boldr)$ $\, [1/\mathrm{m}^2]$ denote the corresponding current density normalized to an input port current of 1 Ampere.
Here, we are implicitly assuming that the antenna element can be characterized by only one dominant mode, which is a good approximation, for instance, for Hertzian dipoles and resonant structures such as the half-wave dipole.  

The function $\boldF_k(\boldr)$ depends on the particular structure of the element as well as its position $\boldr_k$ and orientation. Such a distribution can be obtained in general through  \ac{EM}-based simulations, but closed-form expressions are available for simple structures such as Hertzian or half-wave dipoles \cite{BalB:24}. 

The entire scattering device can be considered as a linear $K$-port network where we define the complex voltage and current envelopes (in the following denoted simply as voltages and currents) at the ports, respectively, $\boldv$ (open circuit voltage) and $\is$.
The actual current density of the $k$-th element is therefore ${\Jms}_k(\boldr)=[{\is}]_k\cdot \boldF_k(\boldr)$.  The corresponding unitary-energy basis function is $\Phim_k(\boldr)=\boldF_k(\boldr)/\| \boldF_k(\boldr) \| $, where $\|\boldF_k(\boldr) \|=\sqrt{\innerprod{\boldF_k(\boldr)}{\boldF_k(\boldr)}}$. For simplicity, in the following, we consider the same structure for all elements so that $\| \boldF_k(\boldr) \|=\alpha, \forall k$.
As a consequence, the relationship between $\js$ and $\is$ is $\js=\alpha \, \is $, defined in Section~\ref{sec:scatterdevice}, and the total current density can be expressed as a function of the ports' currents   
 \begin{align}
  \Jm(\boldr) =\sum_{k=1}^K {\Jms}_k(\boldr)=\sum_{k=1}^K [\is]_k\, \alpha \, \Phim_k(\boldr) \, .
\end{align}

Exploiting the reciprocity theorem, the open circuit voltage at the $k$-th element's port is $ [\boldv]_k=\alpha \, \innerprod{\Em(\boldr)}{\Phim_k(\boldr) } =\alpha \, [\bolde]_k $.  
 
\begin{figure}[!t]
\centering\includegraphics[width=0.55\columnwidth]{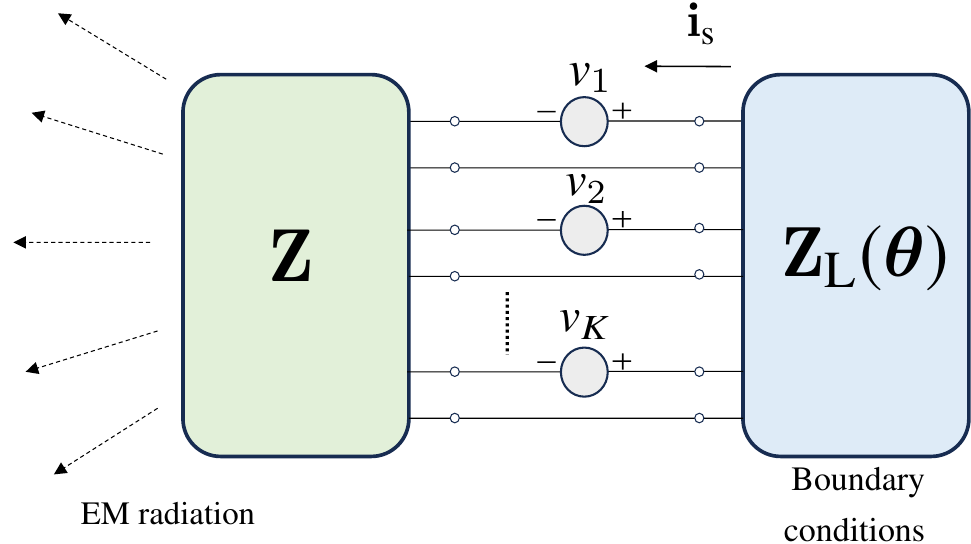}
\caption{Equivalent multiport circuit of the reconfigurable scattering device in Fig. \ref{Fig:Feedback}. The circles represent voltage sources.} 
\label{Fig:Circuit}
\end{figure}

In Section~\ref{sec:scatterdevice}, we have seen that the vectors representing the total, incident, and scattered \ac{EM} fields are related by $\bolde = \ei + \es$, where $\bolde=\boldD^{-1} \js$ and $\es=\boldG \, \js$. Moreover, since $\{ \Phim_k(\boldr) \}$'s are disjoint functions, then $\boldD$ is diagonal.  
By multiplying the first relationship by the constant $\alpha$, it is possible to map it into a relationship between currents and voltages at the ports, i.e., $\alpha^2 \, \boldD^{-1} \is =\vi +\alpha^2 \boldG \, \is$, being ${\vi=[v_1, v_2, \ldots, v_{K}]^{\transpose}}$ the impressed open circuit voltages associated with the incident field $\ei$. After simple steps, we obtain 
 \begin{align} \label{eq:DSA}
 -\left ( \bZl(\btheta)  +\boldZ  \right ) \is =\vi\, , 
\end{align}
where $\boldZ=\alpha^2 \, \boldG$ is the impedance matrix of the antenna structure and $\bZl(\btheta)=-\alpha^2 \, \boldD^{-1}$ is a diagonal impedance matrix accounting for the boundary conditions depending on the loads and possibly mutual coupling between the elements.\footnote{The sign change is a consequence of the convention followed in Fig.~\ref{Fig:Circuit} for current direction.} Equation \eqref{eq:DSA} corresponds to the multiport circuit depicted in Fig.~\ref{Fig:Circuit}, and represents the equivalent voltage/current formulation of \eqref{eq:es}. In fact, after simple substitutions, it is
\begin{equation}
\es=-\boldZ   \left (\bZl(\btheta)+\boldZ \right )^{-1} \, \ei =\boldR(\btheta) \, \ei \, , 
\end{equation}
which gives the reflection matrix $\boldR(\btheta)$ as a function of the circuit's impedances. 
Therefore, Fig.~\ref{Fig:Circuit} can be seen as the multiport circuit version of the scheme in Fig. \ref{Fig:Feedback} that involves \ac{EM} quantities.  All the couplings between the elements of the structure are captured by the impedance matrix $\boldZ$, which does not depend on the reconfigurable loads and relates the open-circuit voltages and currents of the $K$ ports \cite{BalB:24}.
In the case Hertzian dipoles of infinitesimal length $l\ll \lambda$ located at position  $\boldr_k$ with a generic polarization (orientation) $\versorp_k$, it can be computed analytically as 
$[\boldZ]_{n,k}=l^2 \, \versorp_n \scalprod \GreenE{\boldr_n-\boldr_k} \, \versorp_k$, with $\GreenE{\boldr}$ given by \eqref{eq:Gej}.
 The real part of the impedance matrix $\boldZ$ affects the radiated \ac{EM} field whose power is given by $\Pt=\is^{\ctranspose} \Re\{\boldZ\} \, \is $ for a lossless structure, where $\Re\{x\}$ denotes the real part of $x$.

In general, the equivalent multiport circuit is accurate provided that the impedance matrix 
$\boldZ$ is well characterized. This typically requires full-wave simulations or measurements, especially for complex structures. 
Thus, while the equivalent multiport circuit is highly convenient for signal processing and communication theorists, it does not by itself reveal the underlying physical phenomena, even though these are inherently embedded in the circuit model.
 
\begin{figure}[t!]
    \centering
  \subfloat[ ]{
\includegraphics[width=0.5\columnwidth]{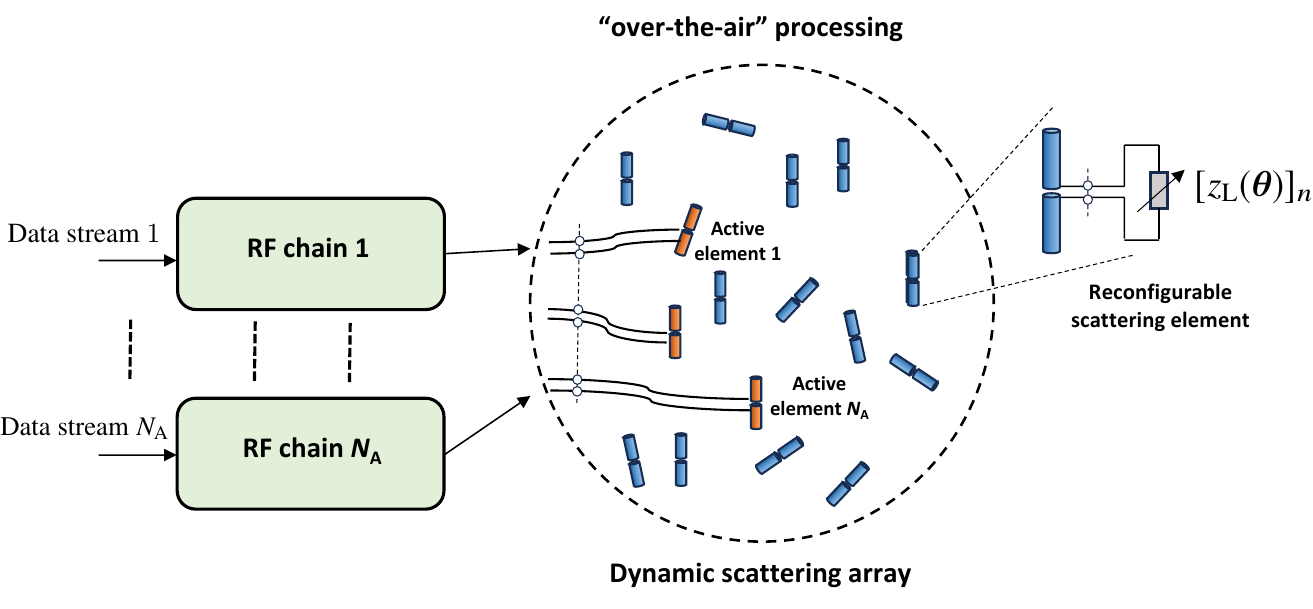}}
  \subfloat[ ]{
\includegraphics[width=0.5\columnwidth]{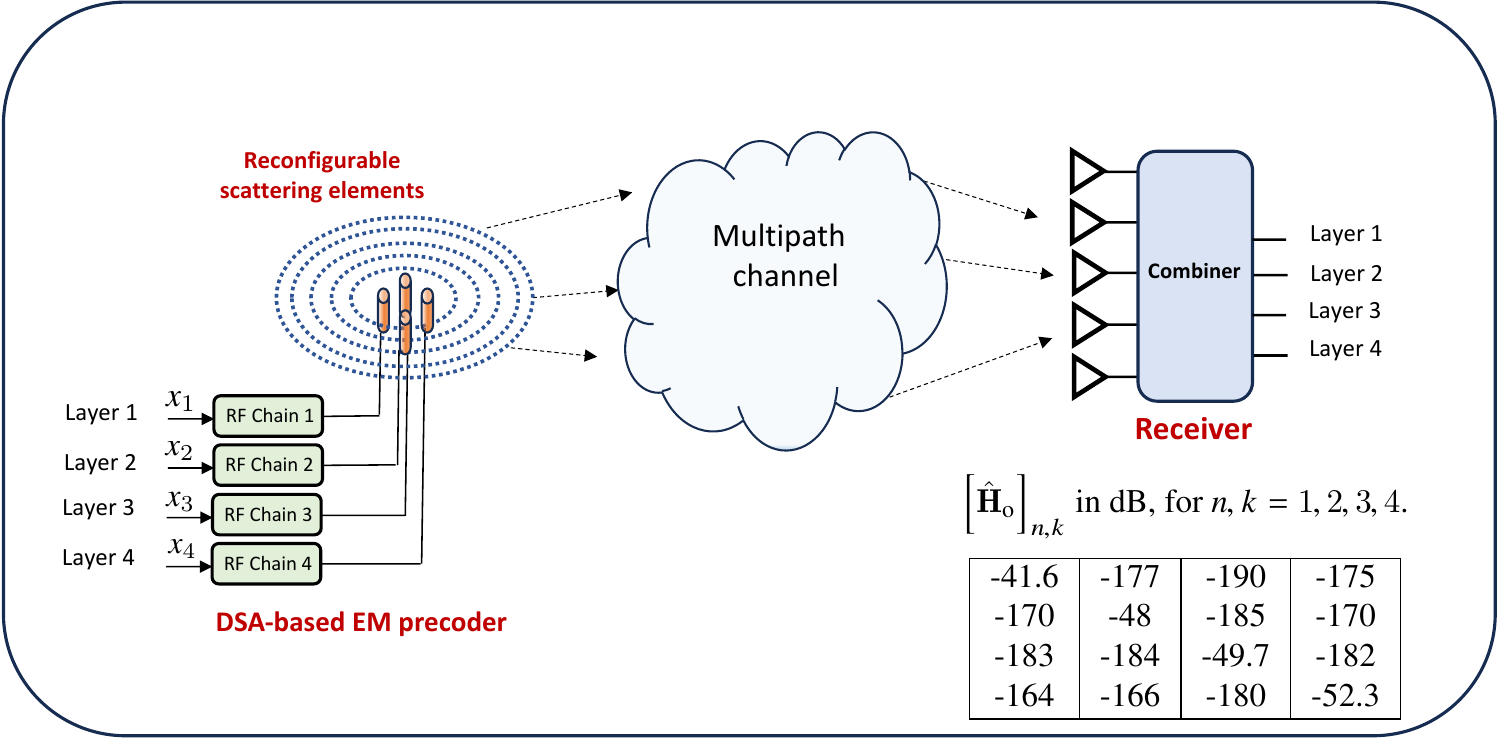}}
\hfill  
        \caption{The generic DSA structure with $\Na$ RF chains and $\Na$ active elements surrounded by $\Ns$ reconfigurable scattering elements. Example of MIMO communication in which the DSA is adopted at the transmitter. The table reports the actual end-to-end matrix  $\hat{\mathbf{H}}_{\text{o}}$ obtained by optimizing \eqref{eq:hattheta1}.}
    \label{Fig:DSA}
\end{figure}

\subsection{EM Signal Processing in the Reactive Near-Field }
\label{Sec:Reactive}

We illustrate here two examples related to the use of the general framework illustrated in the previous sections to model and design a \acf{DSA} and a \acf{SIM} whose aim is to realize a given signal processing task at the \ac{EM} level and minimize the number $\Na$ of RF chains.

\subsubsection{Dynamic Scattering Arrays}
\label{Sec:DSA}

With reference to Fig.~\ref{Fig:DSA}, a \ac{DSA} consists of $K$ antenna elements (e.g., dipoles) located within the space $\CalS$, of which $\Na$ are active and connected to an RF chain, while the remaining $\Ns = K - \Na$ are passive scattering elements \cite{Har:78,BucJuaKamSib:20,HanZhaSheLiChiMuc:21,Dar:C24}.  The response of the \ac{DSA} can be changed by loading each scattering antenna element with reconfigurable loads of impedances $\bzl=\bzl(\btheta)$. Ideally, the loads are designed to be only reactive to avoid power losses, i.e., $\bzl=\jmath \btheta$, with $\btheta \in \mathbb{R}^K$. 
Now, the $\Na$ RF chains introduce impressed open circuit voltages ${\vi=[v_1, v_2, \ldots, v_{\Na}, 0, 0, \ldots, 0]^{\transpose}}$ to the first $\Na$ elements that carry the information and are responsible for the impressed field $\ei$. To account for this, define the identity matrix $\boldQ \in \mathbb{R}^{K \times \Na} $ so $\vi=\boldQ \, \va$, with $\va=[v_1, v_2, \ldots, v_{\Na}]^{\transpose}$. The \ac{DSA} is governed by the relationship \eq{eq:DSA} where the impedance matrix $\boldZ$ embeds also the $\Na$ active sources in this case. To ensure the maximum power transfer between the sources (i.e., the RF chains) and the \ac{DSA}, a power matching network has to be inserted, even though it is not explicitly shown in the figure for simplicity. Since the input impedance of the $\Na$ ports of the \ac{DSA} depends on the \ac{DSA} configuration $\btheta$, the matching network should be configured accordingly \cite{IvrNos:10}.

For a given receiver space, from \eqref{eq:y} and  \eqref{eq:DSA}, by setting $\boldH_{\mathrm{T}}=\boldI_{K}$ since the source and scattering spaces coincides, the end-to-end mapping between the open circuit voltage $\va$ at the $\Na$ ports and the observed field $\boldy$ is 
\begin{align}  \label{eq:by}
\boldy=-\alpha \, \Gr \boldU \,  \left (\bZl(\btheta)  +\boldZ  \right )^{-1} \vi +\boldn =-\alpha \, \Gr \boldU \,  \left (\bZl(\btheta)  +\boldZ  \right )^{-1} \boldQ \, \va +\boldn \,. 
\end{align}
The previous equation can be used as a general model for the \ac{DSA} in an optimization problem once the desired response $\boldy$ is fixed.

One design approach is named \emph{characteristic mode analysis} and consists of designing $\btheta$ such that the current $\is$, needed to obtain the desired response $\boldy$, represents the dominant scattering mode of the structure. This can be simply accomplished by selecting $\btheta$ such that the current $\is$ resonates, i.e., $[\btheta]_k=\frac{1}{[\is]_k} \left [ \Im\{\boldZ\} \is \right ]_k$, for $k=1,2, \ldots, K$, where $\Im\{x\}$ denotes the imaginary part of $x$.  This approach, originally proposed in \cite{MauHar:73}, has been widely exploited for the design of \acp{ESPAR} with a single \ac{RF} chain \cite{BucJuaKamSib:20,HanZhaSheLiChiMuc:21}. Unfortunately, this simple approach is not accurate when applied to large structures in which secondary modes might deviate from the actual response from the desired one, and it is not applicable in the case of multiple active antennas. 

More in general, one might want to solve the following constrained optimization problem of the type in \eqref{eq:hattheta} for a given objective end-to-end channel matrix $ \boldHo \in \mathbb{C}^{K \times \Na}$  
\begin{equation} \label{eq:hattheta1}
\hat{\btheta}=\arg \min_{\btheta} \left\|\, -\alpha \, \Gr \boldU \,  \left (\bZl(\btheta)  +\boldZ  \right )^{-1} \boldQ -   \boldHo \right\|_{\mathrm{F}}\, ,
\end{equation}
constrained to a fixed radiated power $\Pt$ \cite{Dar:C24}. The $n$-th column  of $\boldHo$ represents the desired end-to-end channel response associated with the $n$-th port of the \ac{DSA}, for $n=1,2, \ldots , \Na$. It is worth noticing that with the same configuration $\hat{\btheta}$ of the \ac{DSA}, $\Na$ different responses are obtained simultaneously, each one associated with the input port of one specific \ac{RF} chain. 


Hereafter, we illustrate a numerical example in which \eqref{eq:hattheta1} is optimized to realize a \ac{MIMO} \ac{EM} precoder. 
In particular, we consider a \ac{MIMO} communication in which the \ac{DSA} is adopted at the transmitter as illustrated in Fig.~\ref{Fig:DSA}.
A receiving user equipped with a standard \ac{ULA} consisting of $K=20$ elements spaced at $\lambda/2$ is located at a distance of \unit[10]{m} under \ac{NLOS} conditions.
The simulated \ac{NLOS} channel consists of 5 multi-paths caused by the reflection of 5 scatterers in the environment corresponding to the following strongest singular values of the channel $(\unit[-35.6]{dB},\unit[-42.4]{dB},\unit[-43.7]{dB},\unit[-46.3]{dB}, \unit[-82]{dB}, \ldots )\,$, which has clearly rank $r=4$.
The \ac{DSA} is equipped with $\Ns=121$ Hertzian reconfigurable scattering dipoles deployed in 5 concentric rings and spaced apart of $\lambda/4$. The carrier frequency is $\unit[28]{GHz}$ so that the total size of the \ac{DSA} is $\unit[3.2]{cm}$.
It is well known from \ac{MIMO} theory that, in a generic propagation scenario characterized by rank $r$, up to $r$ parallel orthogonal links (or layers) can be established between the transmitter and the receiver, enabling the transmission of $r$ independent data streams \cite{TseVis:B05}.
To exploit all of them, the number of active antennas (i.e., RF chains)  must be $\Na=r$ and the \ac{DSA} must implement a suitable precoding strategy, i.e., act as an \emph{\ac{EM} precoder}. Specifically, setting the target end-to-end channel to be diagonal and matching the singular values of the link $\boldHo=\alpha\, \Gr \boldQ$ (channel diagonalization), it must be 
\begin{equation}  \label{eq:opt2}
\left (\bZl(\btheta)  +\boldZ  \right )^{-1} \propto \boldW \, ,   
\end{equation}
where $\boldW=\boldU^{-1}$ represents the optimal precoding matrix. 
 The  approximating end-to-end channel matrix  $\hat{\mathbf{H}}_{\text{o}}$, obtained by solving \eqref{eq:hattheta1} numerically while maximizing the radiated power $\Pt$ (see \cite{Dar:C24} for the details), is reported in  Fig.~\ref{Fig:DSA}.
 
 The comparison between the diagonal and off-diagonal values indicates that the coupling between different layers is completely negligible, i.e., the channel is almost perfectly diagonalized. Compared to the actual singular values of the channel, the intensity of the diagonal elements of $\hat{\mathbf{H}}_{\mathrm{o}}$ exhibit the same behavior with a loss of about $6\,$dB that can be ascribed to the fact that the antenna is not ideal. The designed \ac{DSA} implements optimal precoding at the \ac{EM} level using no more than $\Na = r$ RF chains—the minimum possible—making it significantly simpler and more energy-efficient than conventional full-digital or hybrid solutions, especially considering that the reconfigurable scattering elements are passive.
 Further examples of \ac{DSA} design for the realization of superdirective beamforming and multi-user \ac{MISO} can be found in \cite{Dar:C24}.  
The main challenge from the \ac{SP} perspective lies in finding efficient techniques to tackle high-dimensional optimization problems constrained by \ac{EM} laws like that in \eqref{eq:hattheta1} for a large number of reconfigurable scattering elements.

\subsubsection{Stacked Intelligent Metasurfaces}

Recently, a new technology known as \ac{SIM} has gained significant attention within the research community due to its advantages and potential applications \cite{AnXuNgAleHuaYueHan:23}. A \ac{SIM} consists of a sealed vacuum structure that incorporates multiple layers of metasurfaces, each containing many reconfigurable meta-atoms, i.e., small reconfigurable scattering elements, which are interconnected with a controller (e.g., an FPGA) whose purpose is to implement the optimization algorithm (see the example below) and change the response of each meta-atom,  typically consisting in a two-state phase shift.
%
%
Operating on the well-established Huygens-Fresnel principle, when an \ac{EM} wave interacts with a meta-atom in any given layer, it acts as a secondary point source, illuminating all subsequent meta-atoms in the following layer. This architectural approach, which can be seen as a particular case of \acp{DSA}, with cells organized in layers and coupling occurring in only one direction, provides  
greater flexibility in \ac{EM} waveform manipulation and enhanced spatial-domain gain when compared to conventional \ac{MIMO} architectures, e.g., transmitarrays and hybrid \ac{MIMO} antennas \cite{HassanetAl:J24}. It significantly reduces the number of required \ac{RF} chains, hence allowing for improved \ac{EM}-level processing with very limited energy consumption \cite{AnXuNgAleHuaYueHan:23}.

Interestingly, thanks to their peculiar layered structure, several optimization algorithms for determining the optimal \ac{SIM} transmission coefficients have been inspired by the well-known error backpropagation algorithm used for \ac{DNN} training, hence leveraging their particular hardware architecture \cite{LiuetAl:J22}. These algorithms allow for the optimization of the meta-atom configuration and the phase shifts imposed on \ac{EM} waves passing through these structures. Such optimizations enable the implementation of various wave-domain functionalities, such as beamforming, precoding, and combining. For instance, \cite{LiuetAl:J22} applied deep reinforcement learning to jointly optimize the phase shifts of \acp{SIM} and the transmit power allocation. In a complementary approach, \cite{perovic2024mutual} proposed a \ac{PGD} method to iteratively optimize the phase shifts of \acp{SIM} at the transceivers layer-by-layer, aimed at minimizing the channel cutoff rate under practical modulation constraints. Furthermore, \cite{PapaetAl:J24} introduced a hybrid digital and wave-domain \ac{SIM}-based \ac{MIMO} transceiver architecture and proposed a projected gradient ascent (PGA) method that jointly optimizes the phase shifts of the \ac{SIM}, alongside digital transmit precoding and receive combining. Further works, such as \cite{LiEtAl:2021, an2024two}, employed backpropagation-inspired approaches to determine the optimal \ac{SIM} configurations for specific tasks, including channel estimation, image reconstruction, and \ac{DoA} estimation.

\begin{figure}[t!]
    \centering
  \subfloat[ ]{
\includegraphics[width=0.50\columnwidth]{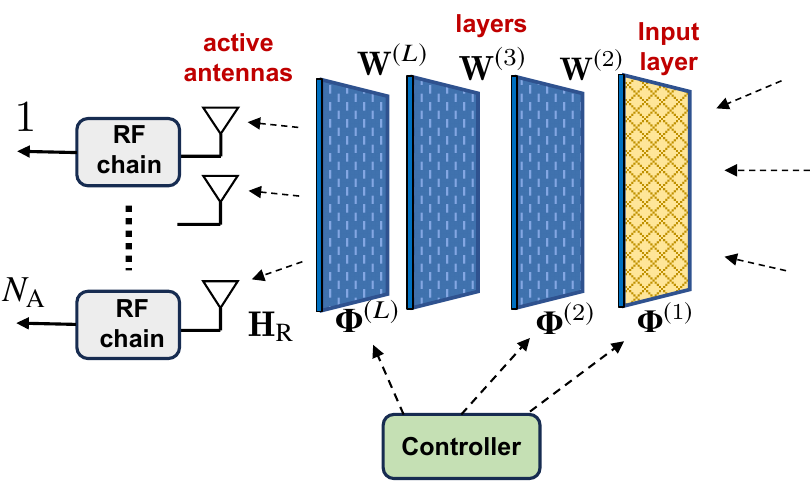}}
  \subfloat[ ]{
\includegraphics[width=0.4\columnwidth]{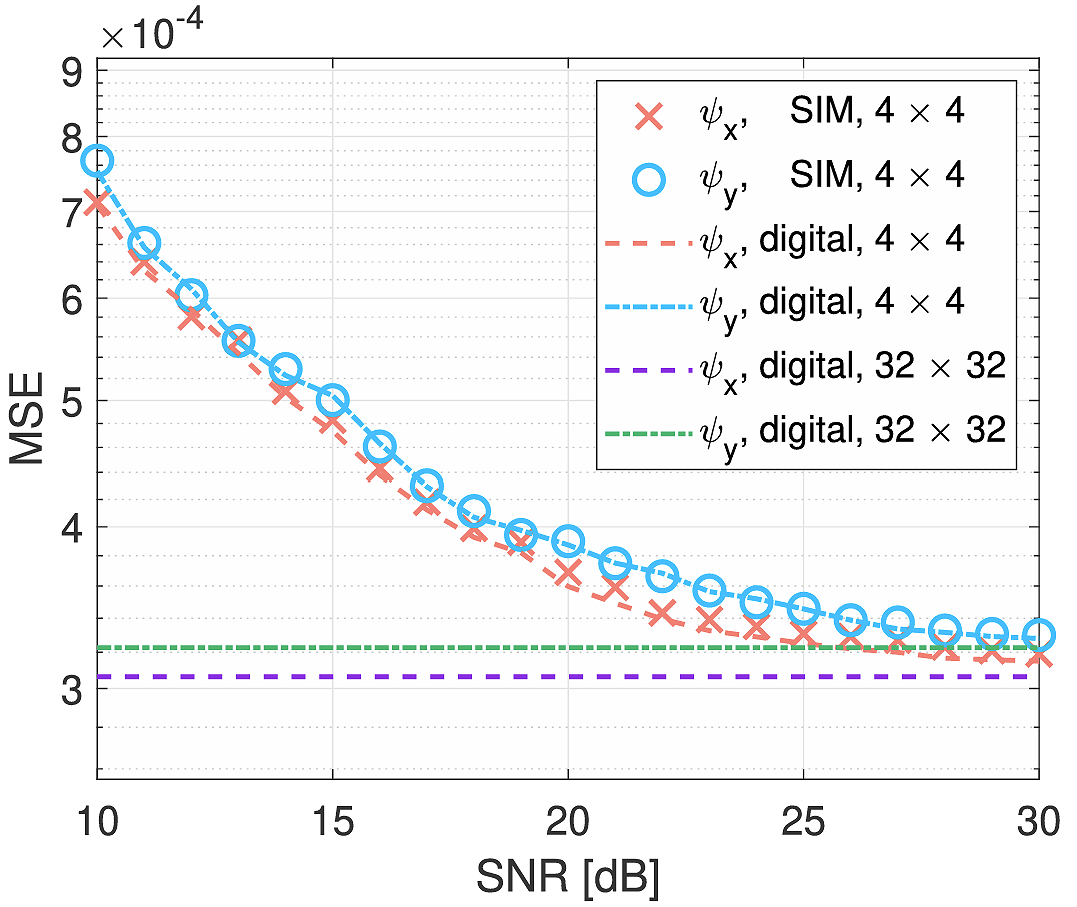}}
        \caption{\ac{SIM}-based antenna architecture, comprising a \ac{ULA}, $L$ metasurface layers, and a controller for reconfigurability. Example of performance comparison of the \ac{SIM}-based \ac{DoA} estimator and the conventional approach via a \ac{2D} digital \ac{DFT} \cite{an2024two}. } 
    \label{fig:SIMsScenario}
\end{figure}

To illustrate, consider a \ac{SIM}-based receiving antenna as shown in Fig.~\ref{fig:SIMsScenario}, where a receiving \ac{SIM} composed of $L$ metasurface layers is followed by a conventional antenna array, such as a \ac{ULA}, consisting of $\Na$ active antennas, with $\Na$ also representing the number of receiving \ac{RF} chains. For simplicity, we assume that every metasurface layer composing the SIM is equipped with the same number $M$ of meta-atoms, where $M > \Na$.\footnote{The condition  $M>\Na$ is not strictly necessary; however, it is generally desirable. In practice, $\Na$ is kept small to limit the number of RF chains, thereby reducing hardware complexity and power consumption, whereas $M$ is chosen to be large to increase the \ac{DoF} available for processing the \ac{EM} wave.} Moreover, we assume that each layer of the \ac{SIM} has an isomorphic square lattice arrangement and we model each metasurface layer as a uniform planar array \cite{LiuetAl:J22}. Specifically, the distance between adjacent meta-atoms is denoted as $d_{\mathrm{atom}}$, the area of each meta-atom is $A$, and the spacing between each metasurface layer is represented by $d_{\mathrm{layer}}$.

The transmission coefficient of the $m$-th meta-atom in the $l$-th metasurface layer is represented by $\phi_m^{(l)} = e^{\jmath \theta_m^{(l)}}$, with $\theta_m^{(l)} \in [0,2\pi)$, for $m = 1,2,  \dots, M$ and $l = 1,2,  \dots, L$. Consequently, the matrix of transmission coefficients for the $l$-th metasurface layer can be expressed as $\boldsymbol{\Phi}^{(l)} = \diag{ \phi_1^{(l)}, \phi_2^{(l)}, \ldots, \phi_M^{(l)}} \in \mathbb{C}^{M \times M}$. 
Moreover, we denote the intra-layer propagation matrix of the SIM, which describes the transmission coefficients from the $(l-1)$-th transmit metasurface layer to the $l$-th layer, as $\mathbf{W}^{(l)} \in \mathbb{C}^{M \times M}$ for $l = 1,2,  \dots, L$. According to Rayleigh-Sommerfeld diffraction theory \cite{BalB:24}, the $(m,i)$-th entry of $\mathbf{W}^{(l)}$ is expressed as
\begin{equation}\label{eq:intralayprop}
w_{m,i}^{(l)} = \frac{A \, d_{\mathrm{layer}}}{d_{m,i}^{(l)}}\left(\frac{1}{2\pi d_{m,i}^{(l)}} - \frac{\jmath}{\lambda}\right) e^{\jmath \kappazero \, d_{m,i}^{(l)}}, \quad l = 1, 2, \dots, L \, , 
\end{equation}
where $d_{m,i}^{(l)}$ is the transmission distance from the $i$-th meta-atom of the $(l-1)$-th metasurface layer to the $m$-th meta-atom of the $l$-th layer. Therefore, the overall effect of the receiving SIM can be represented as
\begin{equation}
\mathbf{R} (\boldsymbol{\theta})= \boldsymbol{\Phi}^{(L)} \mathbf{W}^{(L)} \ldots \boldsymbol{\Phi}^{(2)} \mathbf{W}^{(2)} \boldsymbol{\Phi}^{(1)}  \in \mathbb{C}^{M \times M} \, , 
\end{equation}
where $\boldsymbol{\theta}=\left \{ \theta_m^{(l)}  \right \}$, for $m=1,2, \ldots , M$, $l=1,2, \ldots, L$, collects all the parameters to be optimized.
In addition, we assume that the receiving \ac{ULA} is positioned close to the last SIM layer, that is, within its reactive near-field propagation region. Consequently, the wireless interaction between the receive \ac{ULA} and the last layer of the \ac{SIM} is characterized by the matrix $ \Hr = \left\{ h_{m,s}\right\}\in \mathbb{C}^{\Na \times M }$, whose elements can be derived by substituting $d_{m,i}^{(l)}$ in \eqref{eq:intralayprop} with $r_{m,s}^{(1)}$, i.e., the distance between the $m$-th meta-atom of the last layer of the \ac{SIM} and the generic $s$-th receiving antenna. Note that, despite matrices $\boldsymbol{\Phi}^{(l)}$ being diagonal, the coupling effect at each layer makes the overall reflection matrix $\boldR(\btheta)$ non-diagonal, thus increasing the variables available for the optimization.
When optimizing the phase profiles of the \ac{SIM} for a given task, e.g., to realize an \ac{EM} \ac{MIMO} precoder, perform localization or channel estimation, the optimization problem that needs to be addressed is often a particular case of \eqref{eq:hattheta1}, which typically is in the form \cite{AnXuNgAleHuaYueHan:23} 
\begin{align}\label{eq:SIMoptimization}
\underset{\boldsymbol{\theta}}{\operatorname{minimize}} & \quad \boldsymbol{\Gamma}=\left\| \Hr \, \mathbf{R}(\boldsymbol{\theta})    -\boldHo \right\|_{\mathrm{F}}^2 \, , 
\end{align}
where $\boldHo \in \mathbb{C}^{\Na \times M }$ is a target response matrix that depends on the specific functionality to be achieved. Thanks to the layered structure of the \ac{SIM}, the solution to \eqref{eq:SIMoptimization} can be found thanks to a gradient descent algorithm based on error back-propagation, as outlined in Algorithm~\ref{alg:SIMbackprop}. 

\begin{algorithm}[t]
\caption{Gradient Descent Algorithm for Solving \eqref{eq:SIMoptimization}}
\begin{algorithmic}[1]\label{alg:SIMbackprop}
\STATE \textbf{Input:} $\mathbf{W}^{(l)},\, l = 1, \dots, L,\, \boldHo$
\STATE Randomly initialize the phase shifts $\theta_m^{(l)}, \; m = 1, \dots, M, \; l = 1, \dots, L$
\REPEAT
    \STATE Calculate the partial derivatives of $\boldsymbol{\Gamma}$ w.r.t. $\theta_m^{(l)}, \; m = 1, \dots, M, \; l = 1, \dots, L$
    \STATE Normalize the partial derivatives of $\boldsymbol{\Gamma}$ w.r.t. $\theta_m^{(l)}, \; m = 1, \dots, M, \; l = 1, \dots, L$ 
    \STATE Update the phase shifts $\theta_m^{(l)}, \; m = 1, \dots, M, \; l = 1, \dots, L$
    \STATE Diminish the learning rate $\eta$ 
    \STATE Calculate the objective function value $\boldsymbol{\Gamma}$ 
\UNTIL The decrement of $\boldsymbol{\Gamma}$ is less than a preset threshold or the maximum number of iterations is reached
\STATE \textbf{Output:} $\theta_m^{(l)}, \; m = 1, \dots, M, \; l = 1, \dots, L$ 
\end{algorithmic}
\end{algorithm}

For instance, an interesting application of the algorithm mentioned above can be found in \cite{an2024two}, where the Authors formulate an optimization problem in the form of \eqref{eq:SIMoptimization} to compute the \ac{2D} \ac{DFT} in the wave domain, hence propose a gradient descent algorithm to obtain a near-optimal solution for the \ac{SIM} transmission coefficients, minimizing the Frobenius norm of the approximation error between the \ac{SIM}'s end-to-end transfer function and the ideal \ac{2D} \ac{DFT} matrix $\boldHo$. 
The final objective of \cite{an2024two} is to leverage the \ac{2D} \ac{DFT}, resulting from the propagation of the incident wave through the optimized \ac{SIM}, to estimate the \ac{DoA} of the signal at the receiving antenna array placed after the \ac{SIM}. In the absence of noise, the EM waves propagating through the \ac{SIM} are automatically focused on the specific receiving antenna corresponding to the on-grid \ac{DoA} estimate of the incoming signal thanks to the 2D \ac{DFT} operation. As a result, the \ac{DoA} of the incoming signal can be readily estimated by simply measuring the energy distribution across the receiver antenna array. 

In this regard, in Fig.~\ref{fig:SIMsScenario}-right the performance of such a \ac{DoA} estimator is reported from  \cite{an2024two}, in comparison to the conventional method based on digital beamforming. Specifically, the \ac{MSE} of the electrical \ac{DoA} angles $\psi_\text{x}$ and $\psi_\text{y}$ in the x- and y-directions is reported. 
The results show that the \ac{SIM}-based estimator, operating with $\Na=4 \times 4=16$ antennas and collecting the received signals over $T$ = $64$ snapshots before \ac{DoA} estimation, performs comparably to the digital beamforming-based method with the same number of receiver antennas at large \acp{SNR}. Additionally, Fig.~\ref{fig:SIMsScenario}-right plots the \ac{MSE} of a $32 \times 32$ digital receiver array with a single snapshot, which provides a lower bound for the \ac{SIM}-based scheme, since it directly resolves the same angular grid in a single snapshot that the SIM-based method synthesizes over $T=64$ snapshots with a $4\times 4$ array. The results indicate that the \ac{SIM}-based estimator performs comparably to the digital method, with the benefit of relying on energy detection and wave-based signal processing instead of coherent receivers and complex digital processing. Moreover, despite the superior accuracy, the reference digital approach demands a significantly larger array aperture with many more \ac{RF} chains. In contrast, the \ac{SIM}-based estimator achieves effective \ac{DoA} estimation with a smaller array, demonstrating its practical advantage in resource-constrained scenarios. Details regarding the settings used to obtain the numerical results presented in Fig.~\ref{fig:SIMsScenario}, including the use of different electrical angles for $\psi_x$ and $\psi_y$ estimation, resulting in slightly different \ac{MSE} values, are available in \cite{an2024two}.
\begin{figure}[t!]
    \centering
    \includegraphics[width=0.85\columnwidth, keepaspectratio, trim=0mm 28mm 0mm 0mm, clip]{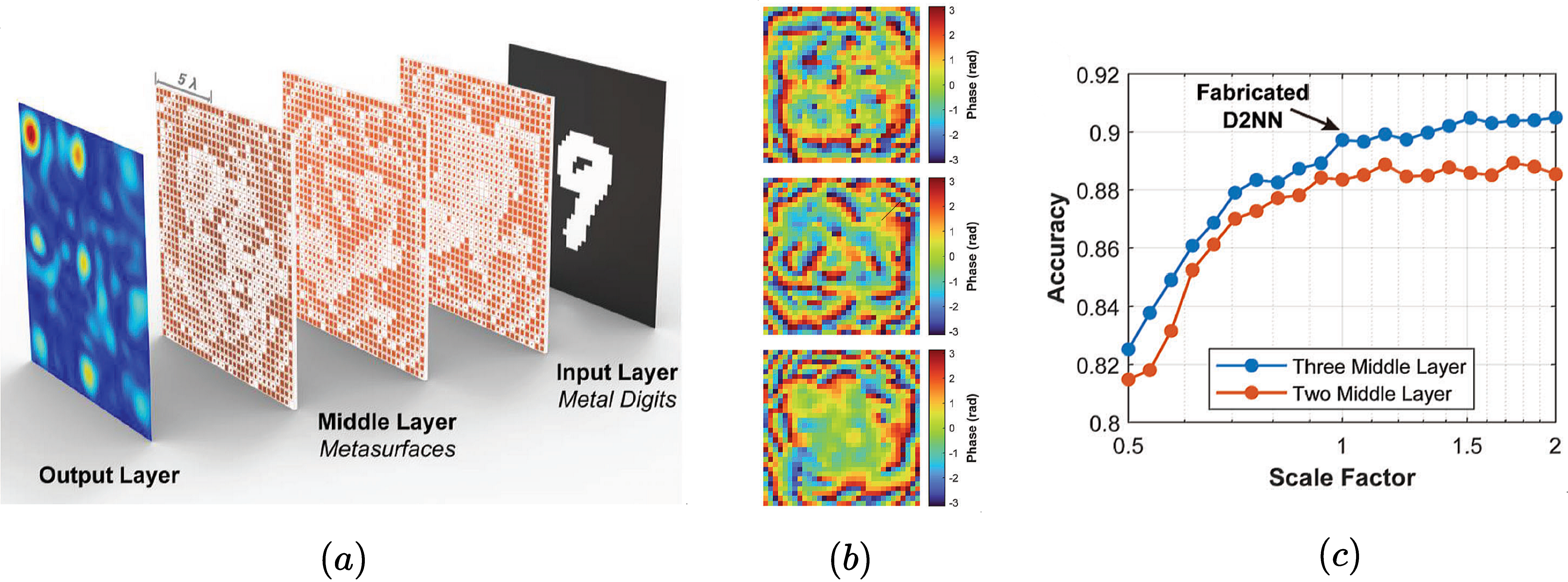}
    \caption{Left: Workflow of the SIM-based $\mathrm{D}^{2}\mathrm{NN}$. Center: Optimized phase distributions for the three metasurface layers. Right: Classification accuracy as a function of system scale and number of metasurface layers  \cite{gu2024classification}.}
    \label{fig:SIMopt}
\end{figure}

Another interesting application can be found in \cite{gu2024classification}, which introduces a novel microwave-based diffractive deep neural network ($\mathrm{D}^{2}\mathrm{NN}$) designed to classify handwritten metallic digits. The proposed architecture consists of a \ac{SIM} comprising $L = 3$ sequential metasurface layers, each of them having $ M = 1024$ meta-atoms capable of introducing phase modulation to the impinging \ac{EM} field. Input data, represented as engraved metal digit patterns, are illuminated by \ac{EM} waves, propagating through the $\mathrm{D}^{2}\mathrm{NN}$ layers, and focusing energy at specific points on the target plane, hence enabling classification. The phase responses of each layer are designed using a stochastic gradient descent algorithm to minimize a cross-entropy loss function based on a training dataset. The system workflow is depicted in Fig.~\ref{fig:SIMopt}-left, which shows the overall \ac{SIM} structure, including the input plane of engraved digits and the output layer with the \ac{EM} field distribution which focuses the energy onto designated points at the target plane for direct classification. The optimized phase distributions of the three diffractive layers, presented in Fig.~\ref{fig:SIMopt}-middle, illustrate how the \ac{SIM} processes input data by modulating the phase of the transmitted \ac{EM} waves. These distributions reveal that the $\mathrm{D}^{2}\mathrm{NN}$ architecture progressively refines the wavefront, enabling precise focusing at the output plane and effectively mapping the input patterns into distinct classification categories. 

Finally, Fig.~\ref{fig:SIMopt}-right examines the classification accuracy as a function of the relative scale of the intermediate diffractive layers, i.e., based on the number of meta-atoms per layer side. The results indicate that the accuracy improves with increasing scale, plateauing at approximately $90.4\%$ for a three-layer configuration, validating the efficiency of the gradient descent algorithm-based optimization in achieving high classification accuracy using a \ac{SIM}.


\subsection{EM Signal Processing in the Radiative Region}
\label{Sec:Radiative}

\subsubsection{Reconfigurable Intelligent Surfaces}

From a structural viewpoint, \acp{RIS} are composed of large arrays comprising a multitude of small, programmable scattering elements that can dynamically control the phase and, in some cases, the amplitude of incident \ac{EM} waves. By carefully configuring these parameters, \acp{RIS} can shape the wireless propagation environment, leading to significant improvements in coverage, capacity, and energy efficiency. Unlike conventional amplify-and-forward or relay nodes, which introduce transmission delays and consume significant power, \acp{RIS} offer low implementation complexity and reduced power consumption, making them a highly attractive solution for various applications and to realize large structures. 
Moreover, \acp{RIS} can be deployed in two main configurations: transmitting and reflecting. Transmitting \acp{RIS}  are designed to allow \ac{EM} waves to pass through while adjusting their peculiar characteristics \cite{demmer2023hybrid}. For instance, they can be used to control the propagation between indoor and outdoor environments or as layers in a \ac{SIM}. Instead, reflecting \acp{RIS} manipulate incoming \ac{EM} waves by directing them via controlled reflections. In this section, we focus exclusively on the latter configuration.

Specifically, by applying specific phase shifts to the incident \ac{EM} waves, \acp{RIS} can reflect signals in directions that deviate from the conventional Snell's law. This phenomenon, referred to as the \textit{universal Snell's law},
allows \acp{RIS} to impose arbitrary directionality on reflected signals, resulting in highly customizable beamforming and beam focusing. This capability enhances signal strength and reception, and depending on the specific hardware structure and reconfigurable elements of the \ac{RIS}, additional functionalities such as beam collimation, beam splitting, polarization control, and various forms of analog processing can also be achieved \cite{DiRDanTre:22,MarMac:22,BjoWymMatPopSanCar:22}.
This versatility positions \acp{RIS} as a promising solution for addressing the complex challenges faced in modern wireless communication systems. Applications of \acp{RIS} span communication, sensing, localization/tracking, and the detection of passive objects.  
Ongoing research is also investigating new avenues for leveraging \acp{RIS} in mapping and imaging to reconstruct high-fidelity and high-resolution maps of the surrounding radio environment \cite{TorrGueZhaGuiYanEldDar:J24}.

\begin{figure*}[t!]
	\centering \includegraphics[trim= {1cm 1cm 1cm 1cm}, clip, width=0.7\linewidth]{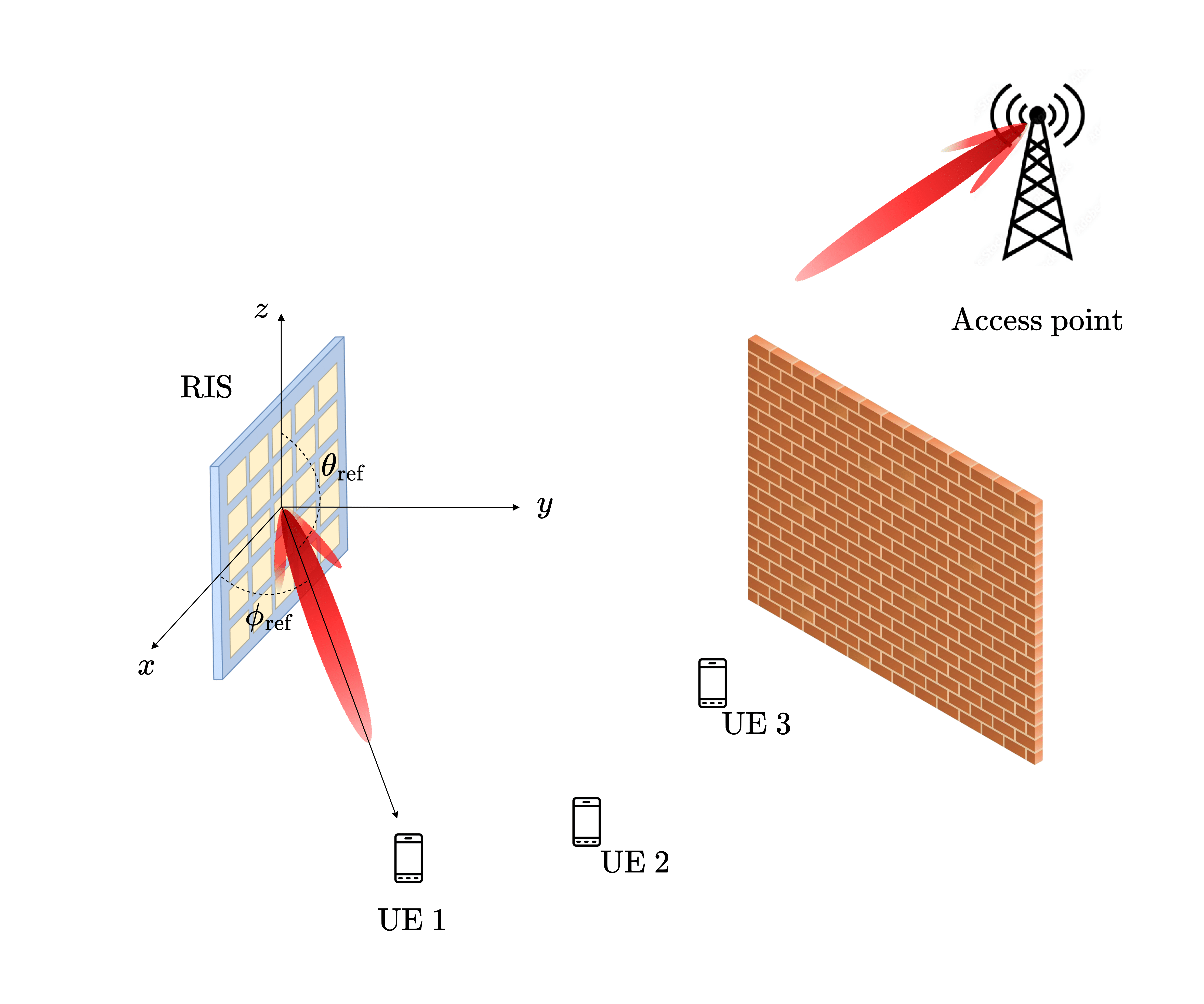}
		\caption{RIS-aided LOS \ac{MIMO} communication.}
		\label{Fig:RIScomm}
\end{figure*}

As an example of the basic functionality of a \ac{RIS}, let us consider a transmitting antenna, e.g., an \ac{AP}, that wants to transmit to a \ac{UE} that is not in direct visibility due to obstruction raised by the presence of an obstacle, e.g., in a configuration similar to Fig.~\ref{Fig:RIScomm}. Thanks to the deployment of the \ac{RIS}, it is possible to create a virtual \ac{LOS} link between the \ac{AP} and the \ac{UE}. 
Specifically, we consider a square \ac{RIS} consisting of $K$ small unit cells (i.e., scattering elements) spaced $\lambda / 2$ apart providing reconfigurable phase shifts, each being characterized by reflection coefficients $\rho_{i,j} = e^{\jmath \theta_{i,j} }$, with $i,j = 1,2, \dots, \sqrt{K}$ being the \ac{2D} position indexes.
According to the discussion in Section~\ref{sec:scatterdevice}, a conventional \ac{RIS} with uncoupled elements is characterized by a diagonal reflection matrix $\boldR(\btheta)=\diag{r_1, r_2, \ldots, r_K}$ (diagonal \ac{RIS}), where $r_k=\rho_{i,j}$ is the reflection coefficient of the $k$-th element at position $(i,j)$, where $k=i+\sqrt{K}(j-1)$,  with $i,j = 1,2,  \dots, \sqrt{K}$.

To perform the so-called \textit{anomalous reflection}, i.e., let the \ac{RIS} reflect the incoming \ac{EM} wave at an arbitrary angle, the reflection coefficient applied at each \ac{RIS} element must be configured such that the phase shift introduced by the $(i,j)$-th element compensates the extra phase shifts due its position with respect to the incoming and reflected wavefronts, that is \cite{DiRDanTre:22}
\begin{equation}
\theta_{i,j} = - \pi i \left(u_x\left(\Theta_{\mathrm{inc}}\right)+u_x(\Theta_{\mathrm{ref}})\right)
- \pi j\left(u_y\left(\Theta_{\mathrm{inc}}\right)+u_y(\Theta_{\mathrm{ref}})\right) \, , 
\end{equation}
where $\Theta_{\mathrm{inc}} = (\theta_{\mathrm{inc}}, \phi_{\mathrm{inc}})$ represents the incident angle (evaluated with respect to the \ac{RIS}'s normal vector), $\Theta_{\mathrm{ref}} = (\theta_{\mathrm{ref}}, \phi_{\mathrm{ref}})$ represents the desired reflection angle, $u_x(\Theta)$ and $u_y(\Theta)$ are functions of the elevation ($\theta$) and azimuth ($\phi$) angles, i.e.,  $u_x(\Theta) = \sin (\theta) \cos (\phi)$ and $u_y(\Theta) = \sin (\theta) \sin (\phi)$. In this manner, the \ac{RIS} can steer the reflected signal towards a specific angle, enabling a high degree of control over the propagation of the \ac{EM} waves. 

One important issue, often overlooked in the literature, arises from reflections in unwanted directions caused by uncontrolled \ac{RF} sources in the environment. These reflections can generate interference, which may be challenging to mitigate. Diagonal \acp{RIS} may be ineffective in addressing this problem due to the limited available number of \ac{DoF} in their reflection matrix, which is equal to $K$.
A potential alternative is the use of non-diagonal \acp{RIS}, which allow controllable coupling between elements. This configuration significantly increases the available number of \ac{DoF}, reaching up to $K\,(K-1)/2$ in reciprocal structures \cite{LiSheNerCle:24}. An even greater number of \ac{DoF} (up to $K^2$) can be achieved with non-reciprocal designs.
Interestingly, a non-diagonal \ac{RIS}, as a \ac{2D} structure, could theoretically achieve the same performance as a \ac{SIM}, which is a \ac{3D} structure but with 2D input and output surfaces. However, as both technologies are still in their early stages, a detailed comparison in terms of implementation complexity is not yet available.

\subsubsection{Self-Conjugating Metasurfaces}

We now present an example of iterative over-the-air \ac{EM} processing involving a transmitting/receiving \ac{MIMO} \ac{AP} and a scattering device located in their respective radiative regions. The \ac{AP} interacts with the scattering device to retrieve information data embedded in the backscattered signal. 
From a \ac{SP} perspective, the challenge is to achieve joint communication and optimal beamforming without requiring \ac{CSI} estimation or signaling, thereby reducing the latency and overhead typically associated with large \ac{MIMO} systems.
The idea is to design the \ac{EM} scattering device to perform the complex conjugation of the signal coming from the \ac{AP}, thereby achieving retrodirectivity. Due to this processing capability, such a device is referred to as \ac{SCM}. 
Additionally, it is envisioned that the \ac{SCM} introduces appropriate phase shifts, uniformly across all cells, into the retransmitted signals, thereby incorporating information through phase modulation.

This dual-function device, referred to as a \emph{modulating \ac{SCM}} in the following, can be considered an evolution of retrodirective antenna arrays, which are systems capable of reflecting incoming signals back toward their direction of arrival \cite{MiyIto:J02}. 
While retrodirective antennas have been studied for decades, the feasibility of retrodirective metasurfaces has only recently been demonstrated in the literature \cite{KaSe:19}.
By leveraging \emph{modulating} \acp{SCM}, it is possible to devise a communication scheme that involves an iterative exchange of signals between a full-duplex \ac{AP} equipped with a conventional antenna array with $N$ elements, and a device equipped with a \emph{modulating} \ac{SCM} with $M$ elements, such as a sensor, which has queued data intended for the \ac{AP}. The former continuously transmits an interrogation signal while the latter retro-directs the received signal after modulating it according to the data addressed to the interrogation device.

\begin{figure*}[t!]
	\centering \includegraphics[trim= {0 0 0 0}, clip, width=0.85\linewidth]{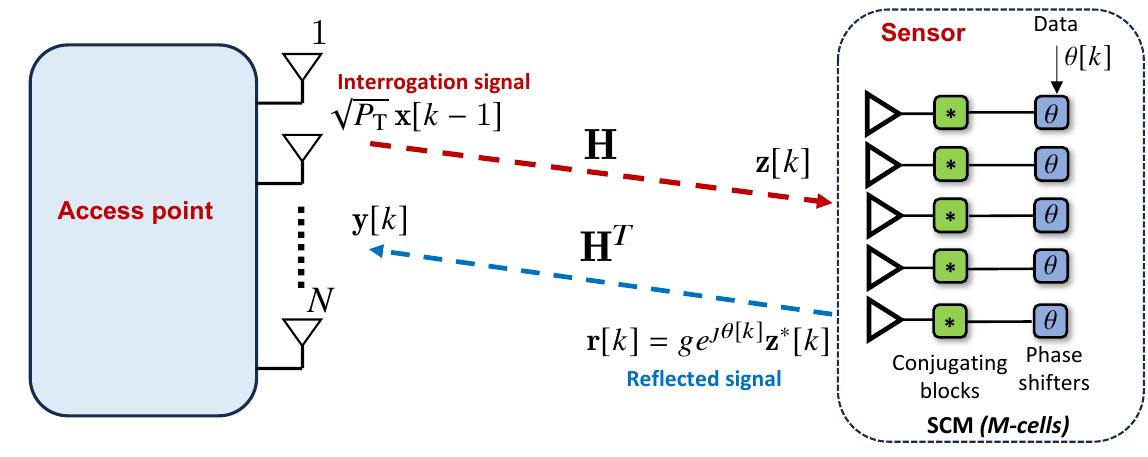}
		\caption{Principle scheme of a modulating \ac{SCM}-based \ac{MIMO} communication.  }
		\label{Fig:SCM}
\end{figure*}

An algorithm designed to establish such an uplink \ac{MIMO} communication with joint beamforming and data transmission, proposed in \cite{DarLotDecPas:J23}, is presented in pseudo-code form in Algorithm~\ref{Algorithm_1} and is discussed below. The key quantities referenced in Algorithm~\ref{Algorithm_1} are highlighted in Fig.~\ref{Fig:SCM}.
\begin{algorithm*}[t]
\setstretch{1.1}
\SetAlgoLined
0: \textit{Initialization}: generate a guess unit norm beamforming vector $\boldx[0]$ ;\\
1: \For{$k=1,\dots,K$}  {
2: transmit: $\sqrt{\Ptx} \, \boldx[k-1]$ \tcp*{signal transmitted by the \ac{AP}}
3: $\boldz[k]=\sqrt{\Ptx}\,  \boldH \, \boldx[k-1] + \boldeta[k]$ \tcp*{signal received by the sensor}
4: $ \boldr[k]=g e^{\jmath \theta[k]} \boldz^*[k]$  \tcp*{signal retro-directed by the sensor}
5: receive: $\boldy[k]=e^{\jmath \theta[k]} \, \boldA^*\,  \boldx^*[k-1] + \boldn^*[k]$    \tcp*{signal received by the \ac{AP}}
6: $\boldx[k]={\boldy^*[k]}/{\left\| \boldy[k] \right\|}$ \tcp*{beamforming vector update}
7: $u[k]=\boldx^{\ctranspose}[k-1] \,  \boldx[k]$ \tcp*{decision variable}
8: $\widehat{\theta}[k]= \mathsf{detection} \left (-\arg \left \{u[k] \right \} \right ) $ \tcp*{data detection}
  }
  \caption{Modified \textit{Power  Method} for
joint communication and beamforming between sensor - \ac{AP}
}
  \label{Algorithm_1}
\end{algorithm*}

\begin{figure}[t!]
	\centering 
 \subfloat[]{\raisebox{10mm}{
       \includegraphics[trim={0 0 0 0},width=0.18\columnwidth]{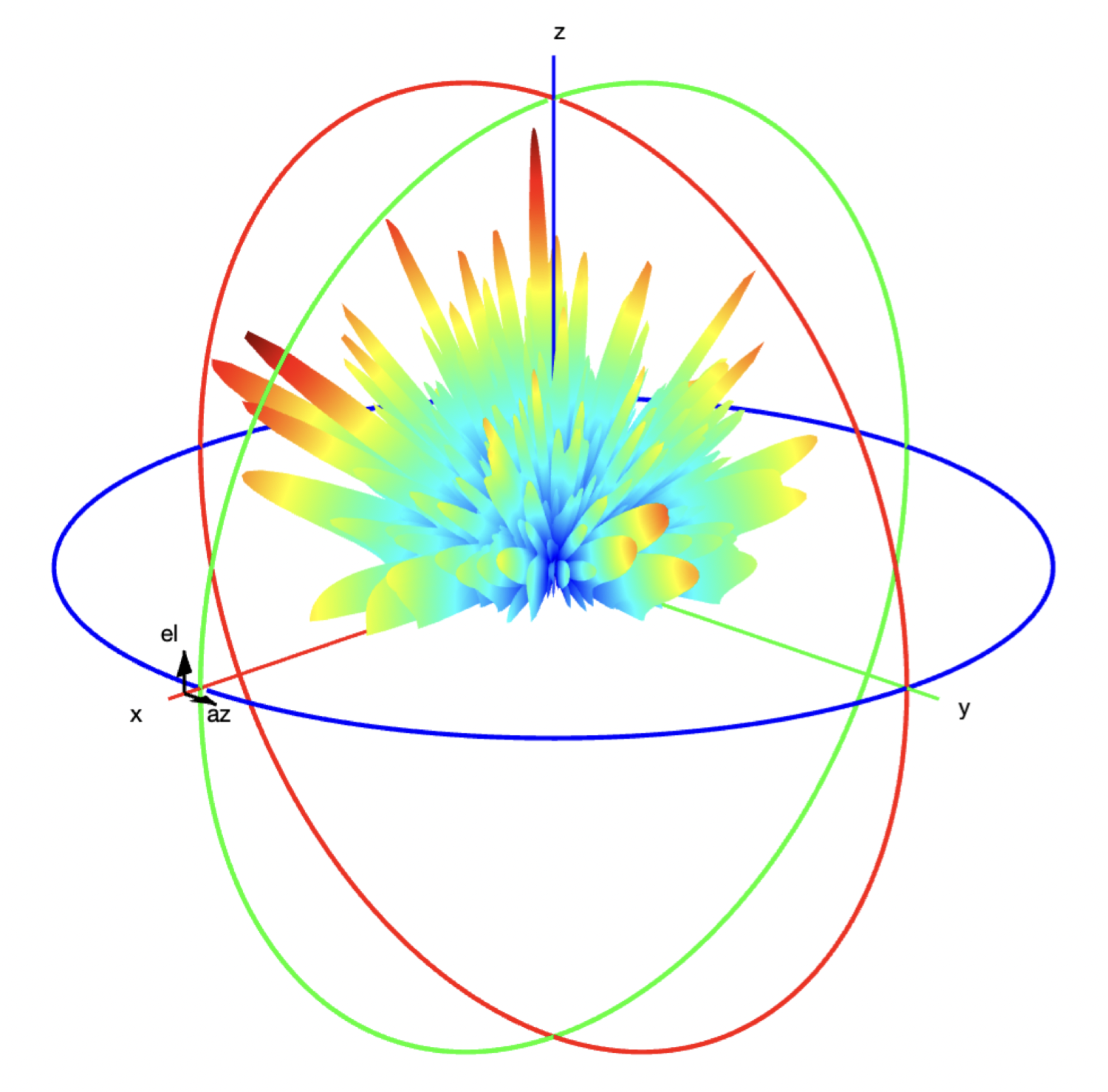}}}
    \hfill
 \subfloat[]{\raisebox{10mm}{
       \includegraphics[trim={0 0 0 0},width=0.17\columnwidth]{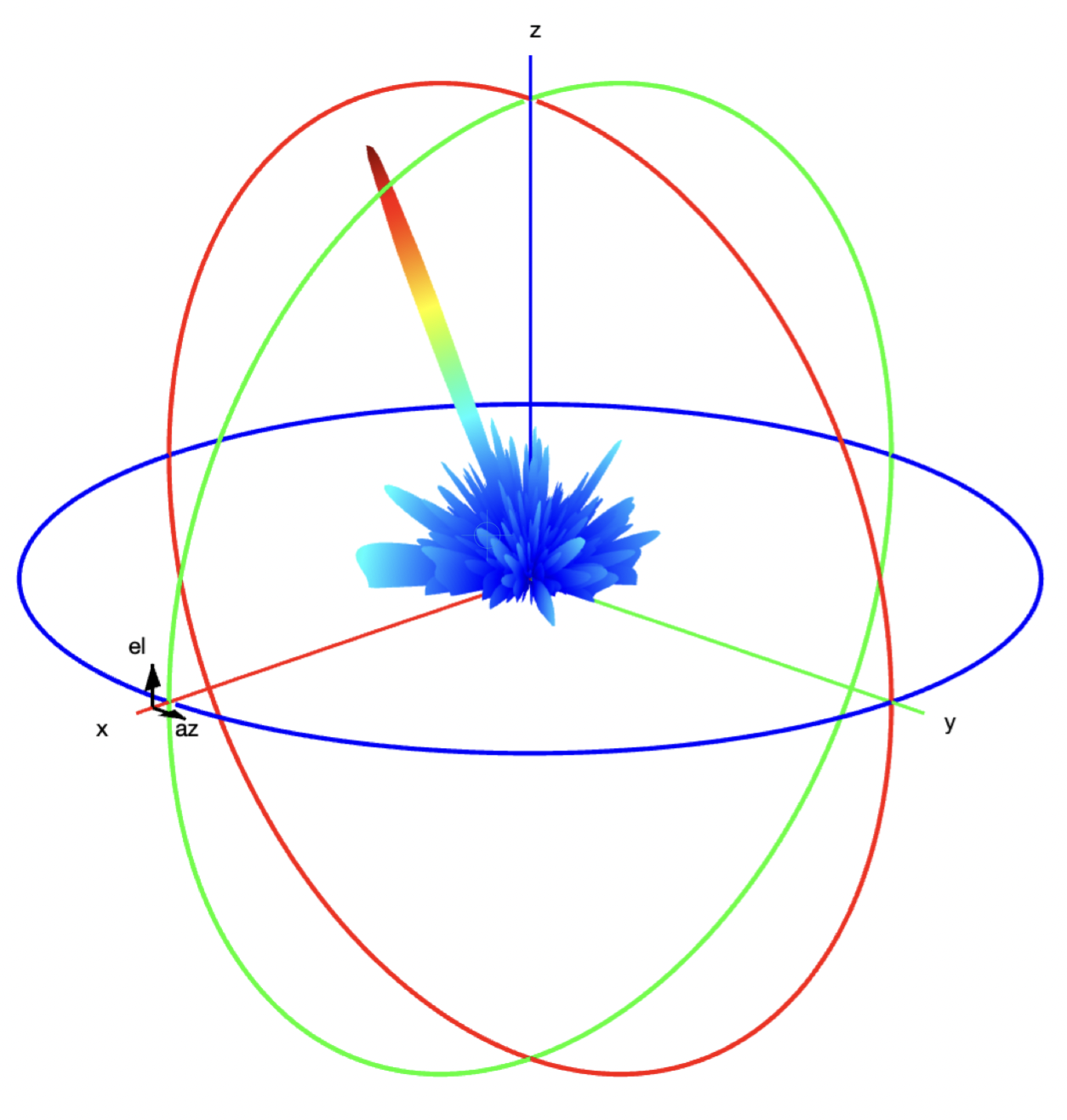}}}
    \hfill
    \subfloat[]{\raisebox{12mm}{
       \includegraphics[trim={0 0 0 0},width=0.17\columnwidth]{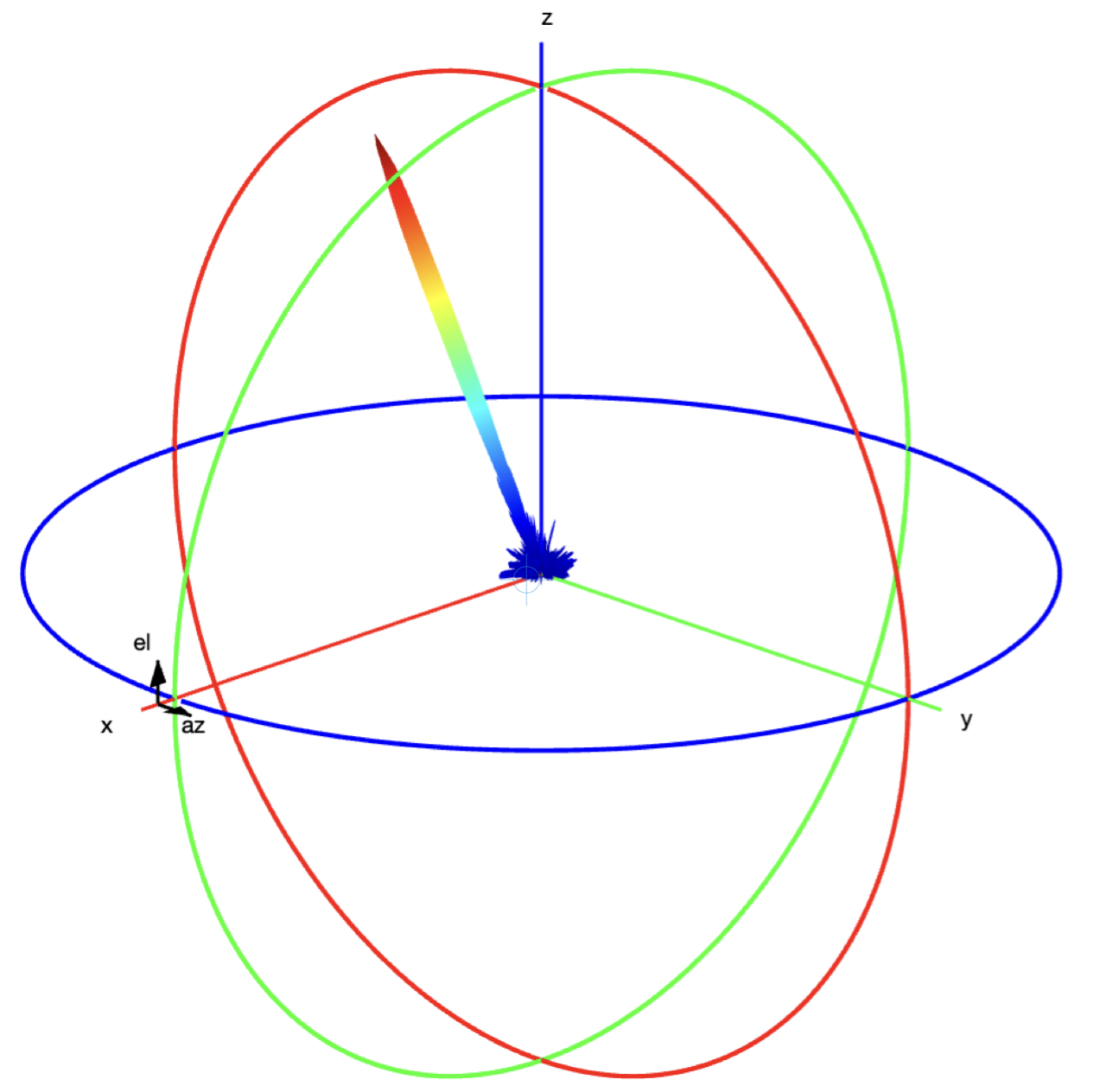}}}
    \hfill
      \subfloat[]{
       \includegraphics[trim= {0 0 0 0}, clip, width=0.4\linewidth]{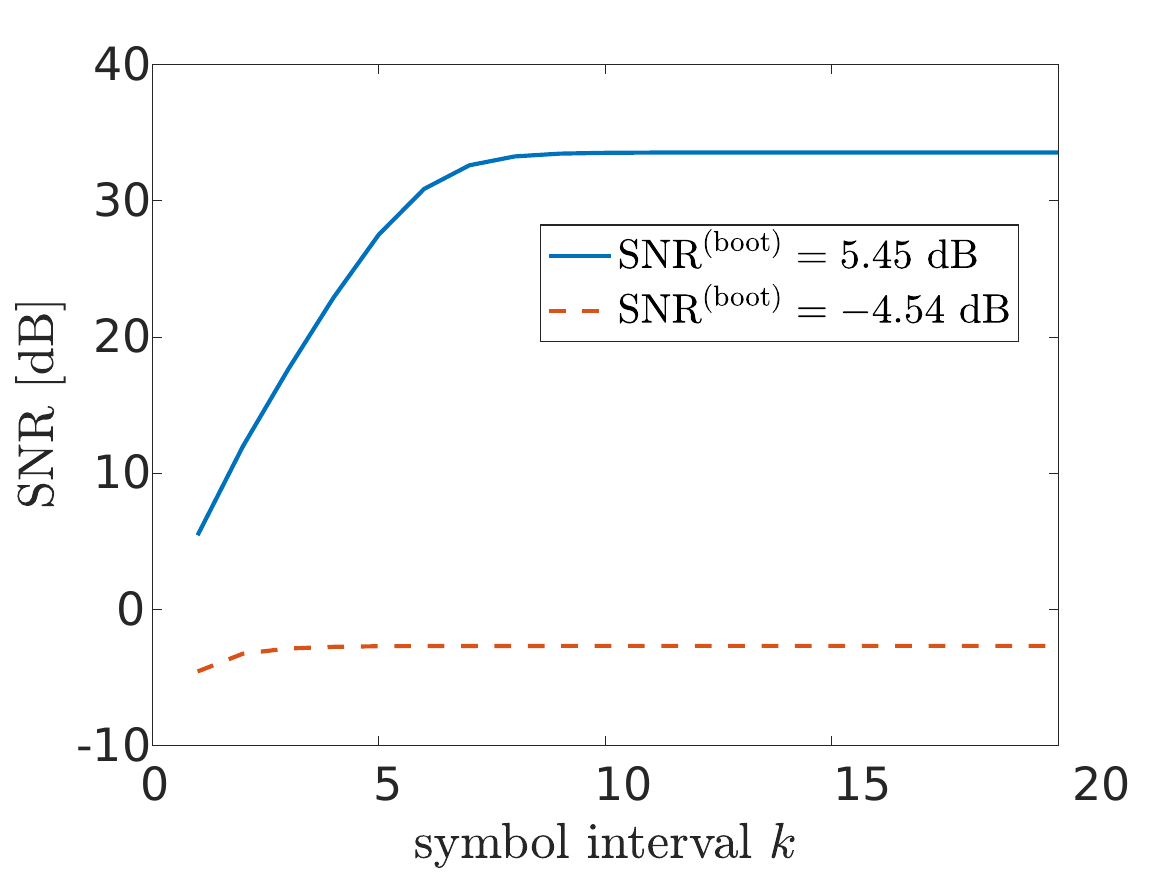}}
    \hfill
 \caption{Evolution of the beam's shape at time $k=0$, $k=1$, and $k=2$, and evolution of the SNR for different bootstrap conditions in the case of rank-1 channel.}
 \label{Fig:Diagrams}
\end{figure}

The process begins (step 0 of the pseudo-code) with the \ac{AP} generating a beamforming vector, $\boldx[0] \in \mathbb{C}^{1 \times N}$. Since the position of the sensor is unknown at startup, this initial beamforming vector is chosen randomly (with norm one) to allow the signal transmitted using $\boldx[0]$ to cover multiple directions,  as shown in Fig.~\ref{Fig:Diagrams}, first beam pattern.  
An iterative process is then initiated (see Algorithm~\ref{Algorithm_1}) to progressively steer the downlink beam toward the active sensor based on the signal it has retro-directed. 
More precisely, at the $(k-1)$-th iteration, with $k \ge 1$, the \ac{AP} transmits the interrogation signal with power $\Ptx$ using the current beamforming vector $\boldx[k-1]$ (step~2). 

This signal, after passing through the \ac{MIMO} channel represented by the matrix $\boldH \in \mathbb{C}^{M \times N}$, is then received by the sensor (step~3), along with the  \ac{AWGN} noise  $\boldeta[k] \in \mathbb{C}^{1 \times M}$.
Thanks to the \emph{modulating} \ac{SCM}, the sensor reflects the received signal along the direction(s) of arrival, also changing its phase based on the data intended for the \ac{AP} (step~4). Specifically, at the sensor side, information data is associated with the phase sequence $\left \{ \theta[k] \right \}_{k=1}^K$, forming a packet of length $K$ symbols, according to any phase-based signaling scheme (e.g., BPSK). When reflecting the signal, the \emph{modulating} \ac{SCM} may also introduce a gain $g$, with $g<1$ if it is passive.
In step~5, the \ac{AP} receives the response $\boldy[k] \in \mathbb{C}^{1 \times N}$ from the sensor, which incorporates the total \ac{AWGN} at the receiver, $\boldn[k] \in \mathbb{C}^{1 \times N}$, and accounts for the round-trip path experienced by the signal and the conjugation performed by the \ac{SCM} through the matrix $\boldA=\sqrt{\Ptx} \, g\,  \boldH^{\ctranspose} \boldH \in \mathbb{C}^{N \times N}$.
Then, a normalized and conjugated version of the received vector $\boldy[k]$ is computed (step~6) and used as the updated beamforming vector $\boldx[k]$ in the subsequent iteration (see Fig.~\ref{Fig:Diagrams}, second beam pattern). 
Iteration after iteration, the beamforming vector is progressively refined over the air, ideally converging toward the optimal configuration (see Fig.~\ref{Fig:Diagrams}, third beam pattern). 
Information data is also extracted by the \ac{AP} at each iteration by correlating the current received vector with the previous beamforming vector $\boldx^{\ctranspose}[k-1]$, thus forming the scalar decision variable $u[k]$ (step~7).
The decision on the modulation symbol conveyed by $\widehat{\theta}[k]$ at the $k$-th time instant is obtained by means of the function $\mathsf{detection(\cdot)}$, according to the adopted modulation scheme (step~8). 
It is worth noting that data demodulation occurs while the \ac{AP} transmits the interrogation signal using the current beamforming vector, thanks to the full-duplex capability of the \ac{AP}.

In the absence of noise and data, the processing operated in  Algorithm~\ref{Algorithm_1} corresponds to the well-known  \emph{Power Method}, which allows the estimate of the strongest eigenvector (top eigenvector) of a square matrix $\boldA$,  described by the recurrence relation
\begin{equation}
    \boldx[k]=\frac{\boldA \, \boldx[k-1]}{ \left\| \boldA \, \boldx[k-1]  \right\|}\, , 
\end{equation}
being $\boldx[0]$ typically an initial random guess of the top eigenvector. For $k\rightarrow \infty$, the direction of $\boldx[k]$ converges to that of the top eigenvector. 
This means that $\boldx[k]$ tends to the direction of the top left eigenvector of the \ac{MIMO} channel matrix $\boldH$, i.e., the optimal beamforming vector.
In the presence of noise, new noise samples enter the loop at each iteration, and the convergence to the top eigenvector of the channel is no longer guaranteed. In this regard, the analysis in \cite{DarLotDecPas:J23} shows that the iterative scheme still converges to the top eigenvector if the \emph{bootstrap} \ac{SNR}, defined as ${\mathsf{SNR}^{(\mathrm{boot})}=\mathsf{SNR}^{(\mathrm{max})}/N}$, is greater than one, where $\mathsf{SNR}^{(\mathrm{max})}$ denotes the maximum \ac{SNR} achievable along the direction of the top eigenvector, which occurs when the \ac{AP} beam is perfectly aligned with the sensor. When the algorithm operates successfully, the \ac{SNR} experienced by the \ac{AP} at the $k$-th iteration, denoted $\mathsf{SNR}[k]$, converges iteratively to $\mathsf{SNR}^{(\mathrm{max})}$ independently of the initial guess $\boldx[0]$.

In Fig.~\ref{Fig:Diagrams}, we also report some simulation results to investigate the performance of the scheme proposed in \cite{DarLotDecPas:J23}.
The scenario considered consists of an \ac{AP} operating at 28~GHz and equipped with a uniform planar array of $N=20 \times 20$ antennas deployed along the $xy$-plane and a sensor with a \emph{modulating} \ac{SCM} comprising $M=10 \times 10$ cells assumed to lie on the same plane.
Specifically, Fig.~\ref{Fig:Diagrams} shows the time evolution of $\mathsf{SNR}[k]$  under different conditions: The blue curve corresponds to the case ${\mathsf{SNR}^{(\mathrm{max})}} = \unit[35]{dB}$, associated with a \emph{bootstrap} \ac{SNR} of $\unit[5.45]{dB}$, while the red curve corresponds to the case  ${\mathsf{SNR}^{(\mathrm{max})}} = \unit[25]{dB}$, resulting in a \emph{bootstrap} \ac{SNR} of \unit[-5.45]{dB}. As shown, when ${\mathsf{SNR}^{(\mathrm{boot})}}$ exceeds  $\unit[0]{dB}$, $\mathsf{SNR}[k]$ converges to $\mathsf{SNR}^{(\mathrm{max})}$ after a few iterations, indicating that perfect alignment has been achieved and the packet can be successfully detected. Conversely, when ${\mathsf{SNR}^{(\mathrm{boot})}}$ is below $\unit[0]{dB}$, $\mathsf{SNR}[k]$ converges to a much lower value and the link cannot be established.

It is worth noticing that the convergence time is below 10 time intervals, as confirmed by a more extensive investigation available in \cite{DarLotDecPas:J23}. The algorithm is able to dynamically adjust the beam orientation in response to any change in the position of the sensor during ongoing communication, thereby exhibiting tracking capabilities. 
Notably, this is achieved without requiring \ac{ADC} chains at the sensor, explicit channel estimation, and time-consuming beamforming or alignment processes. 
Application of this solution to grant-free random access schemes can be found in \cite{DarLotDecPas:J24}.

\section{Discussion and Future Research Directions}

In previous sections, we explored the significant advancements in the development of \ac{ESP} in recent years. By processing the signal directly at the \ac{EM} level through the interaction of passive scatterers, \ac{ESP} offers considerable benefits, including reduced power consumption, lower latency, and smaller device size compared to fully digital or hybrid alternatives. These features make \ac{ESP}-based solutions highly promising for addressing scalability and sustainability challenges. 
However, these advantages are accompanied by trade-offs. New challenges must be considered, as they may limit the processing flexibility and applicability of these solutions compared to traditional digital-based implementations. Specifically, signal processing schemes must account for constraints imposed by fundamental \ac{EM} laws, and the characteristics of the specific technology employed can further restrict the \ac{DoF} available for implementing certain functionalities, as demonstrated in the previous sections. Therefore, it is crucial to adopt physically consistent models, such as those based on multi-port circuits, and design tailored \ac{EM}-compliant signal processing methods.

Despite the recent progress discussed in this article, significant work remains to be done to address various issues at the intersection of \ac{EM} and signal processing theory. Here, we highlight some promising research directions: 

\begin{figure}[t!]
    \centering
  \subfloat[ ]{
\includegraphics[width=0.5\columnwidth]{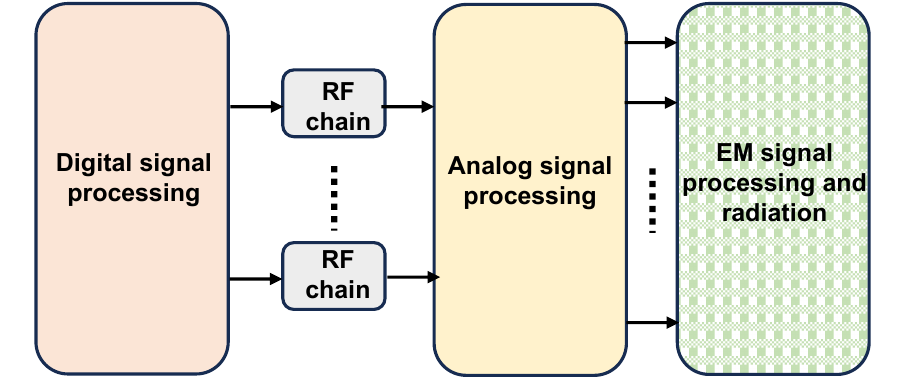}}
  \subfloat[ ]{
\includegraphics[width=0.3\columnwidth]{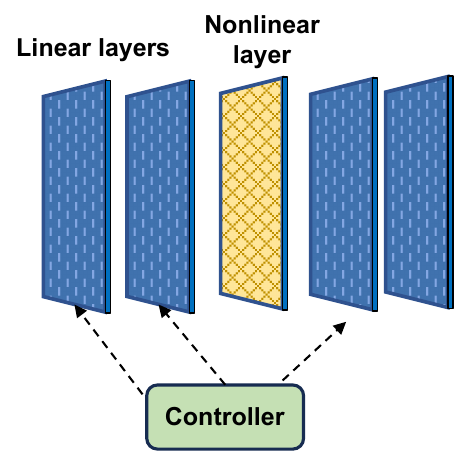}}
        \caption{Left; The tri-hybrid architecture for multi-domain processing. Right: Example of a nonlinear SIM. }
    \label{fig:Perpectives}
\end{figure}

\begin{itemize}
\item \emph{Optimization algorithms and novel hybrid \ac{SP} architectures}: 
From the \ac{SP} perspective, an important issue is the development of 
efficient techniques to address high-dimensional optimization problems constrained by \ac{EM} laws, such as that expressed in \eqref{eq:hattheta1}, particularly when dealing with a large number of reconfigurable scattering elements. Evidently, \ac{ESP} alone cannot address all issues. A promising approach is the so-called tri-hybrid architecture \cite{CasYanChaHea:25} shown in Fig. \ref{fig:Perpectives}-left, where \ac{SP} is distributed across digital, analog (e.g., through phase shifters and adders), and \ac{EM} domains. From the \ac{SP} algorithm design perspective, the key task is to determine how to partition processing among these domains to achieve a given trade-off between power consumption, scalability, latency, and flexibility.
Another open challenge involves designing optimization techniques capable of determining the optimal geometric configuration of scatterers to achieve a desired class of signal processing functionalities.

\item \emph{Space-time-frequency \ac{EM} processing}: 
Most of the existing literature has focused on linear, time-invariant \ac{ESP} solutions that operate primarily in the spatial domain. Consequently, our analysis in the previous sections has been largely confined to narrowband and time-invariant systems.
However, emerging research is exploring the extension of these concepts to wideband and time-varying systems, addressing processing across the space, time, and frequency domains. For instance, \ac{EM} devices operating with wideband signals often exhibit frequency-dependent effects, such as beam squinting or unwanted phase variations. While traditionally viewed as limitations, these phenomena can be strategically harnessed to enhance \ac{EM}-level functionalities. By deliberately exploiting frequency-dependent behaviors, it becomes possible to design systems capable of tailored responses for signals at different frequencies.
This approach is particularly relevant in the study of \ac{STM}-\ac{RIS}, also known as \ac{STMM}, which integrates phase and temporal phase gradients across metasurfaces to enable anomalous, non-reciprocal reflections \cite{MizTagSpa:J24}. This working principle generalizes the design concepts of traditional \ac{RIS}. 
%

\item \emph{\ac{CSI} estimation}:  
The examples of scattering device optimization presented in the previous sections implicitly assume the availability of a perfect \ac{CSI} estimate or precise knowledge of the geometry, including the position and orientation of devices. However, in some scenarios, the scattering device may be located far from the active system components (e.g., a \ac{RIS}) and lack integrated sensors. In dynamic environments, estimating the cascade channel can become time-consuming, leading to high latency and overhead.
Potential solutions could involve long-term statistical optimization schemes that leverage statistical priors, such as device positioning uncertainty and behavioral patterns \cite{JiaAbrKeyWymDarDiR:J23}.

\item \emph{\ac{AI}-aided \ac{ESP} and \ac{ESP}-aided \ac{AI}}: To address the so-called curse of dimensionality in reconfigurable scatterer optimization, \ac{AI}-aided schemes that incorporate physically consistent models can be employed. Additionally, \ac{AI}-based methods can be leveraged to develop hybrid solutions, where AI-driven optimization techniques are combined with traditional model-based approaches. This integration can lead to the creation of more efficient and adaptive \ac{ESP} systems, capable of handling complex and dynamic signal processing tasks. Conversely, passive non-linear elements can be incorporated into reconfigurable scattering devices, such as \acp{SIM}, to emulate the behavior of a \ac{DNN} through the \ac{EM} device. For instance, in Fig. \ref{fig:Perpectives}-right, a layer containing meta-atoms with nonlinear scattering properties is inserted between the linear layers of the \ac{SIM} \cite{FabTorDar:C25}. However, unlike conventional \acp{DNN}, where weights and activation functions can be freely chosen, \ac{EM} physical and technological constraints impose additional limitations. Consequently, models and training schemes must be reimagined to accommodate these constraints. Morover, the fundamental question of under which conditions the \ac{EM} device can act as a universal approximator has still to be answered.

\end{itemize}

In summary, while existing research has offered invaluable insights into the practical and theoretical aspects of \ac{EM} signal processing, there is still much to explore. Advancing this field will require the seamless integration of cutting-edge technologies, such as \ac{AI} and innovative materials, with established physical principles from electromagnetism and related disciplines. This synergy will be pivotal in driving the next wave of innovation and positioning \ac{ESP} as a key enabler for future wireless networks.

\section*{Acknowledgment}
This work was supported by the European Union under the Italian National Recovery and Resilience Plan (NRRP) of NextGeneration EU, partnership on ``Telecommunications of the Future" (PE00000001 - program ``RESTART"), and by the HORIZON-JU-SNS-2022-STREAM-B-01-03 6G-SHINE project (Grant Agreement No. 101095738). Giulia Torcolacci was funded by an NRRP Ph.D. grant.

\bibliographystyle{IEEEtran}
\bibliography{Feature_Paper_ESIT_SPM_REVISED_2nd_ARXIV.bbl}

\end{document}

%% file: acronyms.tex
\acrodef{$P_{D}$}{probability of detection}
\acrodef{$P_{EM}$}{probability of emulation, or false alarm}
\acrodef{$P_{FA}$}{probability of false alarm}
\acrodef{$P_{MD}$}{probability of missed detection}
\acrodef{2D}{bi-dimensional}
\acrodef{3D}{three-dimensional}
\acrodef{5G}{5th generation}
\acrodef{6G}{6th Generation}
\acrodef{ACF}{autocorrelation function}
\acrodef{ACI}{adjacent channel interference}
\acrodef{ACK}{acknowledge}
\acrodef{AcR}{autocorrelation receiver}
\acrodef{ADC}{analog-to-digital converter}
\acrodef{AF}{amplify \& forward}
\acrodef{AFL}{anchor-free localization}
\acrodef{AGNSS}{assisted-GNSS}
\acrodef{AGPS}{assisted GPS}
\acrodef{AI}{Artificial intelligence}
\acrodef{AIC}{Akaike information criterion}
\acrodef{AO}{alternating optimization}
\acrodef{AOA}{angle-of-arrival}
\acrodef{AOD}{angle-of-departure}
\acrodef{AOT}{approximate optimum threshold}
\acrodef{AP}{access point}
\acrodef{API}{application programming interface}
\acrodef{ASK}{amplitude shift keying}
\acrodef{ASNR}{accumulated signal-to-noise ratio}
\acrodef{AUB}{asymptotic union bound}
\acrodef{AWGN}{additive white Gaussian noise}
\acrodef{BAN}{body area network}
\acrodef{BAV}{balanced antipodal Vivaldi}
\acrodef{BCH}{Bose Chaudhuri Hocquenghem}
\acrodef{BEP}{bit error probability}
\acrodef{BER}{bit error rate}
\acrodef{BF}{brute force}
\acrodef{BFC}{block fading channel}
\acrodef{BIC}{Bayesian information criterion}
\acrodef{BLUE}{best linear unbiased estimator}
\acrodef{BPAM}{binary pulse amplitude modulation}
\acrodef{BPF}{bandpass filter}
\acrodef{BPPM}{binary pulse position modulation}
\acrodef{bps}{bits per second}
\acrodef{BPSK}{binary phase shift keying}
\acrodef{BPZF}{band-pass zonal filter}
\acrodef{BS}{base station}
\acrodef{BSC}{binary symmetric channel}
\acrodef{BTB}{Bellini-Tartara bound}
\acrodef{c.c.d.f.}{complementary cumulative distribution function}
\acrodef{c.d.f.}{cumulative distribution function}
\acrodef{CAD}{computer-aided design}
\acrodef{CAIC}{consistent Akaike information criterion}
\acrodef{CAP}{continuous aperture phased}
\acrodef{CCF}{cross correlation function}
\acrodef{CCI}{co-channel interference}
\acrodef{CD}{cooperative diversity}
\acrodef{CDF}{cumulative distribution function}
\acrodef{CDMA}{code division multiple access}
\acrodef{CED}{channel excess delay}
\acrodef{CEOT}{channel ensemble optimum threshold}
\acrodef{CEP}{codeword error probability}
\acrodef{ch.f.}{characteristic function}
\acrodef{CH}{cluster head}
\acrodef{CIR}{channel impulse response}
\acrodef{CIR}{channel impulse response}
\acrodef{CL}{centroid localization}
\acrodef{CM}{channel model}
\acrodef{CNR}{clutter-to-noise ratio}
\acrodef{CP}{ciclic prefix}
\acrodef{CPR}{channel pulse response}
\acrodef{CR}{channel response}
\acrodef{CR}{channel response}
\acrodef{CRB}{Cram\'{e}r-Rao bound}
\acrodef{CRB}{Cram\'{e}r-Rao bound}
\acrodef{CRC}{cyclic redundancy check}
\acrodef{CRLB}{Cram\'{e}r-Rao lower bound}
\acrodef{CS}{clock skew}
\acrodef{CSCG}{circularly symmetric complex Gaussian}
\acrodef{CSI}{channel state information}
\acrodef{CSMA}{carrier sense multiple access}
\acrodef{CSS}{chirp spread spectrum}
\acrodef{CTS}{clear-to-send}
\acrodef{CW}{continuous wave}
\acrodef{CW}{continuous wave}
\acrodef{DAA}{detect and avoid}
\acrodef{DAB}{digital audio broadcasting}
\acrodef{DAC}{digital-to-analog converter}
\acrodef{DBB}{digital base band}
\acrodef{DBPSK}{differential binary phase shift keying}
\acrodef{DCM}{dual-carrier modulation}
\acrodef{DDP}{detected direct path}
\acrodef{DF}{detect \& forward}
\acrodef{DF}{detect \& forward}
\acrodef{DFMS}{monopole dual feed stripline antenna}
\acrodef{DFT}{discrete Fourier transform}
\acrodef{DGPS}{differential GPS}
\acrodef{DL}{deep learning}
\acrodef{DLL}{delay-locked loop}
\acrodef{DMA}{dynamic metasurface antenna}
\acrodef{DNN}{deep neural network}
\acrodef{DoA}{direction-of-arrival}
\acrodef{DoD}{Department of Defense}
\acrodef{DOF}{degrees of freedom}
\acrodef{DoF}{degrees-of-freedom}
\acrodef{DP}{direct path}
\acrodef{DP}{direct path}
\acrodef{DR}{detection rate}
\acrodef{DRT}{distance ratio test}
\acrodef{DS}{delay spread}
\acrodef{DS}{delay spread}
\acrodef{DSA}{dynamic scattering array}
\acrodef{DS-SS}{direct-sequence spread-spectrum}
\acrodef{DTR}{differential transmitted-reference}
\acrodef{DTT}{Diffraction Tomography Theory}
\acrodef{DVB-H}{digital video broadcasting\,--\,handheld}
\acrodef{DVB-T}{digital video broadcasting\,--\,terrestrial}
\acrodef{e.m.}{electromagnetic}
\acrodef{ECC}{European Community Commission}
\acrodef{ED}{energy detector}
\acrodef{EDR}{energy detector receiver}
\acrodef{EFIM}{equivalent Fisher information matrix}
\acrodef{EIRP}{effective radiated isotropic power}
\acrodef{EIT}{electromagnetic information theory} 
\acrodef{EKF}{extended Kalman filter}
\acrodef{ELAA}{extremely large aperture array} 
\acrodef{ELP}{equivalent low-pass}
\acrodef{EM}{electromagnetic}
\acrodef{EMCB}{extended Miller Chang bound}
\acrodef{EME}{minimum eigenvalue ratio detector}
\acrodef{EMI}{electromagnetic interference}
\acrodef{EMO}{electromagnetic object}
\acrodef{ENP}{estimated noise power}
\acrodef{ESA}{European Space Agency}
\acrodef{ESIT}{electromagnetic signal and information theory} 
\acrodef{ESP}{over-the-air electromagnetic signal processing}
\acrodef{ESPAR}{electrically steerable passive array radiator}
\acrodef{ESPRIT}{Estimation of signal parameters via rotational invariant techniques}
\acrodef{EU}{European Union}
\acrodef{EVD}{eigenvalue decomposition}
\acrodef{FAR}{false alarm rate}
\acrodef{FCC}{Federal Communications Commission}
\acrodef{FCC}{Federal Communications Commission}
\acrodef{FCC}{Federal Communications Commission}
\acrodef{FDMA}{frequency division multiple access}
\acrodef{FDMA}{frequency division multiple access}
\acrodef{FEC}{forward error correction}
\acrodef{FEC}{forward error correction}
\acrodef{FF}{far-field}
\acrodef{FFD}{full function device}
\acrodef{FFR}{full function reader}
\acrodef{FFT}{fast Fourier transform}
\acrodef{FG}{factor graph}
\acrodef{FH}{frequency-hopping}
\acrodef{FH-SS}{frequency-hopping spread-spectrum}
\acrodef{FIM}{Fisher information matrix}
\acrodef{FIM}{Fisher information matrix}
\acrodef{FLL}{Frequency-locked loop}
\acrodef{FPGA}{field programmable gate array}
\acrodef{FS}{frame synchronization}
\acrodef{FT}{Fourier Transform}
\acrodef{GA}{Gaussian approximation}
\acrodef{GD}{gradient descent}
\acrodef{GDA}{gradient descent algorithm}
\acrodef{GDOP}{geometric dilution of precision}
\acrodef{GDOP}{geometric dilution of precision}
\acrodef{GLR}{generalized likelihood ratio}
\acrodef{GLRT}{generalized likelihood ratio test}
\acrodef{GLRT}{generalized likelihood ratio test}
\acrodef{GML}{generalized maximum likelihood}
\acrodef{GML}{generalized maximum likelihood}
\acrodef{GP}{Gaussian process}
\acrodef{GPRS}{general packet radio service}
\acrodef{GPS}{Global Positioning System}
\acrodef{GPS}{global positioning system}
\acrodef{HAP}{high altitude platform}
\acrodef{HCRB}{hybrid Cram\'{e}r-Rao bound}
\acrodef{HDSA}{high-definition situation-aware}
\acrodef{HDSA}{high-definition situation-aware}
\acrodef{Hi-RADIAL}{High-accuracy RAdio Detection, Identification, And Localization}
\acrodef{HMIMO}{holographic multiple-input multiple-output}
\acrodef{HMIMO}{holographic multiple-input multiple-output}
\acrodef{HMM}{hidden Markov model}
\acrodef{HPA}{high-power amplifier}
\acrodef{HPBW}{half power beam width}
\acrodef{HPBW}{half power beam width}
\acrodef{HW}{hardware}
\acrodef{i.i.d.}{independent, identically distributed}
\acrodef{i.i.d.}{independent, identically distributed}
\acrodef{ICT}{information and communication technologies}
\acrodef{IE}{informative element}
\acrodef{IEEE}{Institute of Electrical and Electronics Engineers}
\acrodef{IF}{intermediate frequency}
\acrodef{IF}{intermediate frequency}
\acrodef{IFFT}{inverse fast Fourier transform}
\acrodef{IIoT}{industrial Internet of things}
\acrodef{IIoT}{industrial Internet-of-things}
\acrodef{IMF}{Ideal Matched Filter}
\acrodef{IMF}{ideal matched filter}
\acrodef{IMU}{inertial measurement unit}
\acrodef{IMU}{inertial measurement unit}
\acrodef{INR}{interference-to-noise ratio}
\acrodef{INR}{interference-to-noise ratio}
\acrodef{INS}{inertial navigation system}
\acrodef{INS}{inertial navigation system}
\acrodef{IoT}{Internet of things}
\acrodef{IR}{impulse radio}
\acrodef{IRI}{inter-reader interference}
\acrodef{IRS}{intelligent reconfigurable surface}
\acrodef{IRS}{intelligent reflecting surface} 
\acrodef{IR-UWB}{impulse radio UWB}
\acrodef{IR-UWB}{impulse radio UWB}
\acrodef{ISAC}{integrated sensing and communications}
\acrodef{ISI}{inter-symbol interference}
\acrodef{ISI}{inter-symbol interference} 
\acrodef{isi}{intra-symbol interference} 
\acrodef{ISM}{industrial, scientific and medical}
\acrodef{ISNR}{interference-plus-signal-to-noise-ratio}
\acrodef{ISP}{inverse scattering problem}
\acrodef{IT}{information theory}
\acrodef{IT}{interference temperature}
\acrodef{ITC}{information theoretic criteria}
\acrodef{JBSF}{jump back and search forward}
\acrodef{JF}{just forward}
\acrodef{JF}{just forward}
\acrodef{KF}{Kalman filter}
\acrodef{KKT}{Karush–Kuhn–Tucker}
\acrodef{KKT}{Karush–Kuhn–Tucker}
\acrodef{LDC}{low duty cycle}
\acrodef{LDPC}{low density parity check}
\acrodef{LEO}{localization error outage}
\acrodef{LEO}{localization error outage}
\acrodef{LG}{Laguerre-Gaussian}
\acrodef{LIS}{large intelligent surface}
\acrodef{LIS}{large intelligent surface}
\acrodef{LLR}{log-likelihood ratio}
\acrodef{LLRT}{log-likelihood ratio test}
\acrodef{LNA}{low-noise amplifier}
\acrodef{LOS}{line-of-sight}
\acrodef{LOS}{line-of-sight}
\acrodef{LOS}{line-of-sight}
\acrodef{LRT}{likelihood ratio test}
\acrodef{LRT}{likelihood ratio test}
\acrodef{LRT}{likelihood ratio test}
\acrodef{LRT}{likelihood ratio test}
\acrodef{LS}{least square}
\acrodef{LS}{least squares}
\acrodef{LS}{least squares}
\acrodef{m.g.f.}{moment generating function}
\acrodef{MAC}{medium access control}
\acrodef{MAC}{medium access control}
\acrodef{MAC}{medium access control}
\acrodef{MAE}{mean absolute error}
\acrodef{MAI}{multiple access interference}
\acrodef{MAN}{metropolitan area network}
\acrodef{MAP}{maximum a posteriori}
\acrodef{MB}{multi-band}
\acrodef{MB-OFDM}{multi-band OFDM}
\acrodef{MB-UWB}{multi-band UWB}
\acrodef{MC}{multi-carrier}
\acrodef{MCB}{Miller Chang bound}
\acrodef{MCRB}{modified Cram\'{e}r-Rao bound}
\acrodef{MDD}{minimum distance distribution}
\acrodef{MDL}{minimum description length}
\acrodef{MF}{matched filter}
\acrodef{MF}{matched filter}
\acrodef{MF}{matched filter}
\acrodef{MF}{matched filter}
\acrodef{MGF}{moment generating function}
\acrodef{MI}{mutual information}
\acrodef{MIMO}{multiple-input multiple-output}
\acrodef{MIMO}{multiple-input multiple-output}
\acrodef{MIMO}{multiple-input multiple-output}
\acrodef{MIS}{medium intelligent surface}
\acrodef{MISO}{multiple-input single-output}
\acrodef{MISO}{multiple-input single-output}
\acrodef{ML}{machine learning}
\acrodef{ML}{maximum likelihood}
\acrodef{ML}{maximum likelihood}
\acrodef{ML}{maximum likelihood}
\acrodef{MM}{min-max}
\acrodef{MME}{maximum-minimum eigenvalue ratio detector}
\acrodef{MMSE}{minimum mean-square error}
\acrodef{MMSE}{minimum-mean-square-error}
\acrodef{mmWave}{millimiter wave}
\acrodef{mm-waves}{Millimeter Waves}
\acrodef{MPC}{multipath component}
\acrodef{M-PSK}{$M$-ary phase shift keying}
\acrodef{M-QAM}{$M$-ary quadrature amplitude modulation}
\acrodef{MRC}{maximal ratio combiner}
\acrodef{MS}{mobile station}
\acrodef{MS}{mobile station}
\acrodef{MSB}{most significant bit}
\acrodef{MSE}{mean square error}
\acrodef{MSE}{mean squared error}
\acrodef{MSK}{minimum shift keying}
\acrodef{MUI}{multi-user interference}
\acrodef{MUI}{multi-user interference}
\acrodef{MUI}{multi-user interference}
\acrodef{MUR}{Multistatic radar}
\acrodef{MUR}{multistatic radar}
\acrodef{MVU}{minimum variance unbiased}
\acrodef{MZZB}{modified Ziv-Zakai bound}
\acrodef{NB}{narrowband}
\acrodef{NBI}{narrowband interference}
\acrodef{NBI}{narrowband interference}
\acrodef{NEO}{navigation error outage}
\acrodef{NF}{near-field}
\acrodef{NFER}{near-Þeld electromagnetic ranging}
\acrodef{NFF}{near-field focused}
\acrodef{NL}{nonlinear}
\acrodef{NL}{nonlinear}
\acrodef{NLOS}{non-line-of-sight}
\acrodef{NLOS}{non-line-of-sight}
\acrodef{NMSE}{normalized mean squared error}
\acrodef{NP}{Neyman-Pearson}
\acrodef{NTIA}{National Telecommunications and Information Administration}
\acrodef{NTP}{network time protocol}
\acrodef{OAM}{orbital angular momentum} 
\acrodef{OC}{optimum combining}
\acrodef{OFDM}{orthogonal frequency division multiplexing}
\acrodef{OFDM}{orthogonal frequency division multiplexing}
\acrodef{OOK}{on-off keying}
\acrodef{OP}{outage probability}
\acrodef{OT}{optimum threshold}
\acrodef{p.d.f.}{probability density function}
\acrodef{p.m.f.}{probability mass function}
\acrodef{PA}{power amplifier}
\acrodef{PAM}{pulse amplitude modulation}
\acrodef{PAM}{pulse amplitude modulation}
\acrodef{PAN}{personal area network}
\acrodef{PAR}{peak-to-average ratio}
\acrodef{PCA}{principal component analysis}
\acrodef{P-CRLB}{Posterior Cramer-Rao Lower Bound}
\acrodef{PD}{probability of detection}
\acrodef{PDF}{probability distribution function}
\acrodef{PDP}{power delay profile}
\acrodef{PDP}{power delay profile}
\acrodef{PE}{probability of emulation}
\acrodef{PEB}{position error bound}
\acrodef{PEB}{position error bound}
\acrodef{PEC}{perfect electric conductor}
\acrodef{PEP}{packet error probability}
\acrodef{PF}{particle filter}
\acrodef{PFA}{probability of false alarm}
\acrodef{PGD}{projected gradient descent}
\acrodef{PHY}{physical layer}
\acrodef{PHY}{physical layer}
\acrodef{PL}{path-loss}
\acrodef{PLL}{phase-locked loop}
\acrodef{P-Max}{$P$-Max}  
\acrodef{PMD}{probability of missed detection}
\acrodef{PN}{pseudo-noise}
\acrodef{ppm}{part-per-million}
\acrodef{ppm}{part-per-million}
\acrodef{PPM}{pulse position modulation}
\acrodef{PPM}{pulse position modulation}
\acrodef{PPP}{Poisson point process}
\acrodef{PR}{pseudo-random}
\acrodef{PR}{pseudo-random}
\acrodef{PRake}{partial Rake}
\acrodef{PRake}{partial rake}
\acrodef{PRF}{pulse repetition frequency}
\acrodef{PRP}{pulse repetition period}
\acrodef{PRP}{pulse repetition period}
\acrodef{PSD}{power spectral density}
\acrodef{PSD}{power spectral density}
\acrodef{PSEP}{pairwise synchronization error probability}
\acrodef{PSF}{point spread function}
\acrodef{PSK}{phase shift keying}
\acrodef{PSNR}{peak signal to noise ratio}
\acrodef{PSVD}{product singular value decomposition}
\acrodef{PSWF}{prolate spheroidal wave function}
\acrodef{PSWF}{prolate spheroidal wave function}
\acrodef{PU}{primary user}
\acrodef{QAM}{quadrature amplitude modulation}
\acrodef{QoS}{quality of service}
\acrodef{QPSK}{quadrature phase shift keying}
\acrodef{R.V.}{random variable}
\acrodef{RAA}{retro-directive antenna array}
\acrodef{RADAR}{radar}
\acrodef{RCS}{radar cross section}
\acrodef{RCS}{radar cross section}
\acrodef{RDL}{"random data limit"}
\acrodef{REM}{radio environment map}
\acrodef{REO}{ranging error outage}
\acrodef{RF}{radiofrequency}
\acrodef{RF}{radiofrequency}
\acrodef{RF}{radio-frequency}
\acrodef{RFID}{radio frequency identification}
\acrodef{RFID}{radio-frequency identification}
\acrodef{RFR}{reduced function reader}
\acrodef{RFT}{reduced function tag}
\acrodef{RII}{ranging information intensity}
\acrodef{RIS}{reconfigurable intelligent surface}
\acrodef{rms}{root mean square}
\acrodef{RMSE}{root mean square error}
\acrodef{RMSE}{root-mean-square error}
\acrodef{ROC}{receiver operating characteristic}
\acrodef{ROI}{region of interest}
\acrodef{RRC}{root raised cosine}
\acrodef{RSN}{radar sensor network}
\acrodef{RSS}{received signal strength}
\acrodef{RSSI}{received signal strength indicator}
\acrodef{RSSI}{received signal strength indicator}
\acrodef{RTLS}{real time locating system}
\acrodef{RTLS}{real time locating systems}
\acrodef{RTT}{round-trip time}
\acrodef{RTT}{round-trip time}
\acrodef{RTT}{round-trip time}
\acrodef{RV}{random variable}
\acrodef{SA}{simulated annealing}
\acrodef{SaG}{stop-and-go}
\acrodef{SAR}{synthetic aperture radar}
\acrodef{SBS}{serial backward search}
\acrodef{SBS}{serial backward search}
\acrodef{SBSMC}{serial backward search for multiple clusters}
\acrodef{SCM}{self-conjugating metasurface}
\acrodef{SCR}{signal-to-clutter ratio}
\acrodef{SCR}{signal-to-clutter ratio}
\acrodef{SDR}{software-defined radio}
\acrodef{SEP}{symbol error probability}
\acrodef{SFD}{start frame delimiter}
\acrodef{SIM}{stacked intelligent metasurface}
\acrodef{SIM}{stacked intelligent metasurface}
\acrodef{SIMO}{single-input multiple-output}
\acrodef{SINR}{signal-to-interference noise ratio}
\acrodef{SINR}{signal-to-interference plus noise ratio}
\acrodef{SIR}{signal-to-interference ratio}
\acrodef{SIR}{signal-to-interference ratio}
\acrodef{SIS}{small intelligent surface}
\acrodef{SIS}{small intelligent surface}
\acrodef{SISO}{single-input single-output}
\acrodef{SNR}{signal-to-noise ratio}
\acrodef{SNR}{signal-to-noise ratio}
\acrodef{SoC}{System on Chip}
\acrodef{SoC}{system on chip}
\acrodef{SoO}{signal of opportunity}
\acrodef{SoP}{System on Package}
\acrodef{SoP}{system on package}
\acrodef{SOT}{sub-optimum threshold}
\acrodef{SP}{signal processing}
\acrodef{SPAWN}{sum-product algorithm over a wireless network}
\acrodef{SPEB}{squared position error bound}
\acrodef{SPMF}{Single-Path Matched Filter}
\acrodef{SPMF}{single-path matched filter}
\acrodef{SPP}{spiral phase plate}
\acrodef{SQNR}{signal-to-quantization-noise ratio}
\acrodef{SRE}{smart radio environment}
\acrodef{SRE}{smart radio environment}
\acrodef{SS}{spread spectrum}
\acrodef{ST}{simple thresholding}
\acrodef{STCM}{space-time coded metasurface}
\acrodef{STM}{space-time modulated}
\acrodef{STMM}{space-time modulated metasurfaces}
\acrodef{SU}{secondary user}
\acrodef{S-V}{Saleh-Valenzuela}
\acrodef{S-V}{Saleh-Valenzuela}
\acrodef{SVD}{singular value decomposition}
\acrodef{SVD}{singular value decomposition}
\acrodef{SW}{software}
\acrodef{SW}{sync word}
\acrodef{TDE}{time delay estimation}
\acrodef{TDL}{tapped delay line}
\acrodef{TDMA}{time division multiple access}
\acrodef{TDOA}{time difference-of-arrival}
\acrodef{TDOA}{time difference-of-arrival}
\acrodef{TH}{time-hopping}
\acrodef{TH}{time-hopping}
\acrodef{THz}{terahertz}
\acrodef{TNR}{threshold-to-noise ratio}
\acrodef{TNR}{threshold-to-noise ratio}
\acrodef{TOA}{time-of-arrival}
\acrodef{TOA}{Time-of-arrival}
\acrodef{TOF}{time-of-flight}
\acrodef{TOF}{time-of-flight}
\acrodef{TPC}{transmit power control}
\acrodef{TR}{time-reversal}
\acrodef{TR}{transmitted-reference}
\acrodef{TS}{tabu search}
\acrodef{TSVD}{truncated singular value decomposition}
\acrodef{TV}{total variation denoising}
\acrodef{UAV}{unmanned aerial vehicle}
\acrodef{UAV}{unmanned aerial vehicle}
\acrodef{UB}{union bound}
\acrodef{UCA}{uniform circolar array}
\acrodef{UCA}{uniform circular array}
\acrodef{UDP}{undetected direct path}
\acrodef{UE}{user equipment}
\acrodef{UE}{user equipment}
\acrodef{UHF}{ultra-high frequency}
\acrodef{UHF}{ultra-high frequency}
\acrodef{ULA}{uniform linear array}
\acrodef{ULA}{uniform linear array}
\acrodef{ULP}{user location protocol}
\acrodef{UMP}{uniformly most powerful}
\acrodef{UMPI}{uniformly most powerful invariant}
\acrodef{URA}{uniform rectangular array}
\acrodef{UT}{user terminal}
\acrodef{UTC}{coordinated universal time}
\acrodef{UTM}{universal transverse Mercator}
\acrodef{UTRA}{UMTS terrestrial radio access}
\acrodef{UUV}{unmanned underwater vehicle}
\acrodef{UWB}{ultrawide bandwidth}
\acrodef{UWB}{ultrawide-band}
\acrodef{UWBcap}[UWB]{Ultrawide band}
\acrodef{V2X}{vehicle-to-everything}
\acrodef{VFIL}{virtual force iterative localization}
\acrodef{VGA}{variable-gain amplifier}
\acrodef{VNA}{vector network analyzer}
\acrodef{WAF}{wall attenuation factor}
\acrodef{WB}{wideband}
\acrodef{WBI}{wideband interference}
\acrodef{WBI}{wideband interference}
\acrodef{WCL}{weighted centroid localization}
\acrodef{WED}{wall extra delay}
\acrodef{WiMAX} {worldwide interoperability for microwave access}
\acrodef{WLAN}{wireless local area network}
\acrodef{WLS}{weighted least squares}
\acrodef{WMAN}{wireless metropolitan area network}
\acrodef{WPAN}{wireless personal area networks}
\acrodef{WPAN}{wireless personal area networks}
\acrodef{WRAPI}{wireless research application programming interface}
\acrodef{WSN}{Wireless Sensor Network}
\acrodef{WSN}{wireless sensor network}
\acrodef{WSN}{wireless sensor network}
\acrodef{WSR}{wireless sensor radar}
\acrodef{WSR}{wireless sensor radar}
\acrodef{WSS}{wide-sense stationary}
\acrodef{WSS}{wide-sense stationary}
\acrodef{WSSUS}{WSS with uncorrelated scattering}
\acrodef{WWB}{Weiss-Weinstein bound}
\acrodef{WWLB}{Weiss-Weinstein lower bound}
\acrodef{WWLB}{Weiss-Weinstein lower bound}
\acrodef{XL-MIMO}{extremely large-scale multiple-input multiple-output}
\acrodef{ZZB}{Ziv-Zakai bound}
\acrodef{ZZB}{Ziv-Zakai bound}
\acrodef{ZZLB}{Ziv-Zakai lower bound}

%% file: def_EM_SP.tex
\newcommand{\Am} {\EMsymb A}

\newcommand{\Bm} {\EMsymb B}

\newcommand{\boldA} {{\mathbf{A}}}

\newcommand{\boldC} {{\mathbf{C}}}
\newcommand{\boldD} {{\mathbf{D}}}
\newcommand{\bolde} {{\mathbf{e}}}

\newcommand{\boldeta} {{\boldsymbol{\eta}}}

\newcommand{\boldF} {{\mathbf{F}}}

\newcommand{\boldG} {{\mathbf{G}}}

\newcommand{\boldHo} {{\mathbf{H}_{\text{o}}}}
\newcommand{\boldH} {{\mathbf{H}}}

\newcommand{\boldI} {{\mathbf{I}}}
\newcommand{\boldi} {{\mathbf{i}}}
\newcommand{\boldj} {{\mathbf{j}}}

\newcommand{\boldn} {{\mathbf{n}}}

\newcommand{\boldQ} {{\mathbf{Q}}}
\newcommand{\boldr} {{\mathbf{r}}}
\newcommand{\boldR} {{\mathbf{R}}}

\newcommand{\bolds} {{\mathbf{s}}}

\newcommand{\boldU} {{\mathbf{U}}}
\newcommand{\boldv} {{\mathbf{v}}}
\newcommand{\boldV} {{\mathbf{V}}}

\newcommand{\boldW} {{\mathbf{W}}}
\newcommand{\boldx} {{\mathbf{x}}}

\newcommand{\boldy} {{\mathbf{y}}}

\newcommand{\boldZ} {{\mathbf{Z}}}
\newcommand{\boldz} {{\mathbf{z}}}

\newcommand{\btheta} {\boldsymbol{\theta}}
\newcommand{\bZl} {\mathbf{Z}_{\text{L}}}
\newcommand{\bzl} {\mathbf{z}_{\text{L}}}

\newcommand{\CalE} {{\cal{E}}}
\newcommand{\CalP} {{\cal{P}}}
\newcommand{\CalS} {{\cal{S}}}

\newcommand{\crossprod}[2] {{#1}\, { \scriptstyle \mathsf{x}}\, {#2} }
\newcommand{\ctranspose}{{\mathrm{H}}}

\newcommand{\diag}[1]{{\rm diag} \left ( #1 \right )}

\newcommand{\Dm} {\EMsymb D}

\newcommand{\DOperator}{\mathbb{D} }
\newcommand{\e} {\bolde}
\newcommand{\ei} {\bolde_{\mathrm{i}}}
\newcommand{\Em} {{\EMsymb E}}
\newcommand{\Emi} {{\EMsymb E}_{\mathrm{i}}}

\newcommand{\Ems} {{\EMsymb E}_{\mathrm{s}}}
\newcommand{\EMsymb} {\mathsf}

\newcommand{\eq}[1]{(\ref{#1})}

\newcommand{\es} {\bolde_{\mathrm{s}}}

\newcommand{\fourier}[1]{\mathcal{F} \left [ #1\right ]}
\newcommand{\fourierthree}[1]{\mathcal{F}_3 \left [ #1\right ]}

\newcommand{\GOperator} {\mathbb{G} }
\newcommand{\Gr} {\mathbf{G}_{\text{R}}}

\newcommand{\GreenE}[1] {\EMsymb{G}\left ( #1 \right ) }

\newcommand{\Gt} {\mathbf{G}_{\text{T}}}

\newcommand{\IIm} {{\EMsymb I}}

\newcommand{\innerprod}[2] {\left < {#1}\, , {#2} \right >}
\newcommand{\intthree} {\int_{{\mathbb{R}}^3}}

\newcommand{\invfourierthree}[1]{\mathcal{F}^{-1}_3 \left [ #1\right ]}

\newcommand{\is} {\boldi_{\mathrm{s}}}

\newcommand{\Jm} {{\EMsymb{J}}}

\newcommand{\Jms} {{\EMsymb J}_{\mathrm{s}}}
\newcommand{\js} {\boldj_{\mathrm{s}}}

\newcommand{\kappazero} {k_0}

\newcommand{\Lx} {L_{\text{x}}}
\newcommand{\Ly} {L_{\text{y}}}
\newcommand{\Lz} {L_{\text{z}}}

\newcommand{\Na} {N_{\mathrm{A}}}
\newcommand{\Nc} {{N_{\text{c}}}}

\newcommand{\Ns} {N_{\text{S}}}

\newcommand{\Nu} {{N_{\text{u}}}}
\newcommand{\Narray} {{N_{\text{a}}}}

\newcommand{\Phim} {{\mathbf{\Phi}}}

\newcommand{\Psim} {{\mathbf{\Psi}}}
\newcommand{\Psimr} {\mathbf{\Psi}^{(\text{R})}}
\newcommand{\Psimt} {\mathbf{\Psi}^{(\text{T})}}
\newcommand{\Pt}{P_{\text{T}}}
\newcommand{\Ptx} {P_{\text{T}}}

\newcommand{\rect}[1] {\text{rect} \left ({#1} \right )}

\newcommand{\scalprod} {\boldsymbol{\cdot}}

\newcommand{\sinc}[1] {\text{sinc} \left ({#1} \right )}

\newcommand{\Sr} {\mathcal{S}_{\text{R}}}

\newcommand{\St} {\mathcal{S}_{\text{T}}}

\newcommand{\TildeAm} { \widetilde{\EMsymb A}}

\newcommand{\TildeEm} { \widetilde{\EMsymb E}}

\newcommand{\TildeGm} { \widetilde{\EMsymb G}}

\newcommand{\TildeJ} {{ \widetilde{ J}}}
\newcommand{\TildeJm} {{ \widetilde{\EMsymb J}}}

\newcommand{\transpose}{{\mathrm{T}}}

\newcommand{\VarPhim} {{\boldsymbol{\varphi}}}
\newcommand{\VarPhimr} {\VarPhim^{(\text{R})}}
\newcommand{\VarPhimt} {\VarPhim^{(\text{T})}}

\newcommand{\versorp} {\hat{\mathbf{p}}}
\newcommand{\versorr} {\hat{\mathbf{r}}}

\newcommand{\versorx} {{\hat{\mathbf{x}}}}
\newcommand{\versory} {{\hat{\mathbf{y}}}}
\newcommand{\versorz} {{\hat{\mathbf{z}}}}
\newcommand{\vi} {\boldv_{\mathrm{i}}}
\newcommand{\va} {\boldv_{\mathrm{a}}}

\newcommand{\wavefreq} {{\boldsymbol{\kappa}}}

\newcommand{\wavefreqx} {\kappa_{x}}

\newcommand{\wavefreqy} {\kappa_{y}}

\newcommand{\wavefreqz} {\kappa_{z}}

\newcommand{\Hr}{\boldH_{\text{R}}}

\newcommand{\Lt} {L_{\mathrm{T}}}
\newcommand{\Lr} {L_{\mathrm{R}}}
\newcommand{\At} {A_{\mathrm{T}}}
\newcommand{\Ar} {A_{\mathrm{R}}}